\documentclass[11pt,a4paper]{article}
\usepackage{jheppub}
\usepackage{amsthm,bm,mathrsfs,stmaryrd}
\usepackage{slashed,tikz}
\usepackage{enumerate}
\usepackage{youngtab,simplewick}

\usepackage{longtable}

\newcommand{\be}{\begin{equation}}
\newcommand{\ee}{\end{equation}}
\newcommand{\ba}{\begin{eqnarray}}
\newcommand{\ea}{\end{eqnarray}}

\newcommand{\mt}[1]{$\mathop{#1}$}

\newcommand{\st}{\scriptstyle}
\newcommand{\sst}{\scriptscriptstyle}

\newcommand{\nn}{\nonumber\\}
\newcommand{\eq}{&=&}

\def\a{\alpha}
\def\b{\beta}

\def\d{\delta}

\def\m{\mu}
\def\n{\nu}

\def\cK{{\cal K}}

\def\cO{{\cal O}}

\author[]{Euihun JOUNG\quad }
\author[]{Luca LOPEZ\quad }
\author[]{Massimo TARONNA}

\affiliation[]{Scuola Normale Superiore and INFN\\
Piazza dei Cavalieri 7, 56126 Pisa, Italy}

\emailAdd{euihun.joung@sns.it}
\emailAdd{luca.lopez@sns.it}
\emailAdd{massimo.taronna@sns.it}

\title{\center On the cubic interactions of  massive and
partially-massless
 higher spins  in (A)dS}

\abstract{Cubic interactions of massive and partially-massless totally-symmetric
higher-spin fields in any constant-curvature background of dimension greater than three
are investigated. Making use of the ambient-space formalism, the consistency condition for the traceless and transverse parts of the parity-invariant interactions is recast into a system of partial differential equations.
The latter can be explicitly solved for given \mt{s_{1}\!-\!s_{2}\!-\!s_{3}} couplings
and the \mt{2\!-\!2\!-\!2} and \mt{3\!-\!3\!-\!2} examples are provided in detail for general choices of the masses.
On the other hand, the general solutions for the interactions involving massive and massless fields
are expressed in a compact form as generating functions of all the consistent couplings. The St\"uckelberg formulation of the cubic interactions
as well as their massless limits are also analyzed.
}

\begin{document}

\maketitle

\section{Introduction}

One of the key lessons of intense studies in higher-spin (HS) field theories is the need to abandon many of the beliefs inherited by years of extraordinary results devoted to understanding their lower-spin counterparts. For instance, the higher-derivative nature of the couplings as well as the need of introducing infinitely many HS fields are clear signals that the standard frameworks are not sufficient.
These features naturally surface in String Theory (ST), where the presence of an infinite tower of massive higher-spin excitations bring about most of its
remarkable properties.
Besides being responsible for planar duality, open-closed duality and modular invariance,
the plethora of
massive HS particles is what makes the
high-energy behavior of string amplitudes softer than in any local quantum field theory. To wit,
although its lower-spin truncations are in general non-renormalizable,
ST is believed to be finite.
The field-theoretical reason for this difference is the contribution of an infinite number of massive HS fields to quantum corrections.
Moreover, since the massive HS spectrum becomes massless in the tensionless limit,
it has long been conjectured that ST may describe a broken phase of an underlying HS gauge theory.
Therefore, in order to better understand the quantum properties of ST as well as other of its remarkable features,
it would be important to
investigate  the dynamics of HS gauge fields and their links to massive counterparts on more general field-theoretical grounds.

Over the years, finding consistent interactions of HS gauge fields
has proven to be a very challenging task.\footnote{
For recent reviews on the subject,
see e.g. the proceeding \cite{SolvayHS} (which includes \cite{
Bianchi:2005yh, Francia:2005bv, Bouatta:2004kk, Bekaert:2005vh, Sagnotti:2005ns})
and \cite{Sorokin:2004ie, Francia:2006hp,Bekaert:2010hw,Sagnotti:2011qp}.}
A long-recognized difficulty concerns the inconsistency of the
gravitational minimal couplings in flat space-time \cite{Aragone:1979hx}.
As shown in \cite{Fradkin:1986qy,Fradkin:1987ks},
this problem can be solved in (anti) de Sitter space-time ((A)dS).
There HS gauge invariance, which is broken when one replaces ordinary partial derivatives by the gravitational covariant ones,
is restored by adding a chain of higher-derivative interactions sized by negative powers of the cosmological constant.
Interestingly, this way of solving the minimal interaction problem is similar to the one used for massive HS fields in the St\"uckelberg formulation.
More precisely, one can restore the St\"uckelberg gauge invariance of the HS fields
by adding higher-derivative interactions sized by inverse powers of the mass.\footnote{
See e.g. \cite{Zinoviev:2009hu} for the EM interaction of spin 2 and
\cite{Zinoviev:2008ck} for the gravitational interaction of spin 3.}
This analogy between the roles of cosmological constant and masses suggests that a
systematic study of massive HS theories in (A)dS
can provide new insights on both
Vasiliev's HS gauge theory\footnote{
Vasiliev's equation provides at present the only known fully non-linear consistent description
of an infinite number of HS gauge fields of all spins \cite{Vasiliev:1988sa,Vasiliev:2003ev}.} in (A)dS
and ST, and eventually shed some light on their relations. However, although both of them have been known for many years, extracting their interaction vertices remains a very difficult program that only recently have been pushed forward by some new but yet not conclusive steps. The present work aims to constructing all consistent cubic interactions of totally symmetric HS relying on the Noether procedure. The cubic results are expected to be further constrained by the higher-order consistency leading eventually to ST and Vasiliev's system and possibly to other consistent theories. We hope our work to be a first step towards HS systematics. Let us mention as well that the construction of consistent interacting massive HS theories is also relevent from  a phenomenological prespective. Indeed, they provide an effective description for hadronic resonances in certain regimes.       


Free massive HS particles
can be described by the Fierz system \cite{Fierz:1939zz} consisting
of dynamical field equations together with the traceless and transverse (TT) constraints. The latter constraints guarantee
the propagation of the correct number
of physical degrees of freedom (DoF).
The Lagrangian reproducing the Fierz system was first
obtained in \cite{Singh:1974qz,Singh:1974rc} in flat space, and further studied in \cite{Zinoviev:2001dt,Biswas:2002nk,Metsaev:2003cu, Hallowell:2005np,
Buchbinder:2005ua,Fotopoulos:2006ci,Buchbinder:2006nu,Buchbinder:2006ge,Buchbinder:2007ix,Horvathy:2007pm,Francia:2007ee, Francia:2008ac,
Buchbinder:2008ss,Zinoviev:2008ze,Metsaev:2009hp,
Bekaert:2009pt,Ponomarev:2010st,Grigoriev:2011gp,Metsaev:2011uy,Metsaev:2011iz}
in flat or (A)dS background.
In dS, the mass spectrum in the unitary region presents a discrete series of mass values, called partially-massless points \cite{Deser:1983mm,Deser:1983tm,Higuchi:1986py,Higuchi:1986wu,Higuchi:1989gz,Deser:2001pe,Deser:2001wx,Deser:2001us,Deser:2000de,Deser:2001xr,Zinoviev:2001dt,Deser:2003gw,Skvortsov:2006at,Francia:2008hd,Gover:2008pt,Alkalaev:2009vm,Alkalaev:2011zv},
for which the fields acquire gauge symmetries and the corresponding representations become shorter. It is worth noticing that the interactions of these partially-massless fields might play some important role in the inflationary cosmology. 

As noticed some time ago, the introduction of interactions for massive HS fields
might either spoil the TT constraints, thus leading to the
appearance of unphysical DoF \cite{Fierz:1939ix}, or violate causality \cite{Velo:1969bt,Velo:1970ur,Velo:1972rt}.
See e.g. \cite{Porrati:1993in, Doria:1994cz,
Cucchieri:1994tx,Giannakis:1998wi,Klishevich:1998yt,
Deser:2000dz, Deser:2001dt,
Porrati:2008gv,Porrati:2008rm,Porrati:2008ha,Zinoviev:2008ck,Zinoviev:2009hu,
Porrati:2009bs,Zinoviev:2010av, Zinoviev:2011fv,Rahman:2011ik}
for some recent works on the consistency of the
electromagnetic (EM) and gravitational couplings to massive HS fields.\footnote{See also \cite{Deser:2006zx} for the study of EM interactions of partially-massless spin 2 fields.}
It is worth noticing that,
as shown for spin 2 in \cite{Argyres:1989cu,Porrati:2011uu} and for arbitrary spins in \cite{Porrati:2010hm},
ST provides a solution for the case of constant  EM background.
See also
 \cite{Taronna:2010qq,Sagnotti:2010at} for an analysis of HS interactions in the open bosonic string and
\cite{Bianchi:2010es,Polyakov:2010sk,Schlotterer:2010kk,Bianchi:2011se,Lee:2012ku} for studies on scattering amplitudes of HS states in superstring and heterotic string theories.
Other works on cubic interactions of massive HS fields in (A)dS
can be found in \cite{Zinoviev:2006im,Zinoviev:2008jz}. 

\paragraph{Traceless and transverse part of the interactions}

The aforementioned difficulties in finding consistent interactions
manifest themselves only at the full \emph{off-shell} level,\footnote{
By \emph{off-shell} we mean the entire Lagrangian including traces and divergences of fields,
as opposed to its TT part.} while they can be circumvented restricting the attention to the physical DoF.
Indeed, relying on the light-cone formalism, Metsaev constructed all consistent cubic interactions
involving massive and massless HS fields in flat space-time \cite{Metsaev:2005ar,Metsaev:2007rn}.
In this approach, what is left is to find the \emph{complete} expressions associated with those vertices.
Starting from the TT parts of the interactions, that can be viewed as the covariant versions of Metsaev's lightcone vertices,
the corresponding complete forms within the Fronsdal setting were obtained recently in \cite{Manvelyan:2010jr,Sagnotti:2010at}.
Moreover, the computation of (tree-level) correlation functions does not require the full vertices but only their TT parts.\footnote{See \emph{e.g} \cite{Sagnotti:2010at,Taronna:2011kt} for the analysis of higher-order interactions of massless particles in flat space.}
Therefore, although they ought to be completed, the TT parts of the vertices are also interesting in their own right.
Motivated by this observation, recently the TT parts of the cubic interactions
of massless HS fields in (A)dS were identified in \cite{Joung:2011ww}.\footnote{See \cite{Vasiliev:2011xf} for the frame-like approach to the same problem.}
In the present paper, we extend this approach to the cases of massive
and partially-massless fields in (A)dS.

\paragraph{Radial reduction with delta  function}

A way of obtaining massive theories
is via dimensional reduction of a
$(d+1)$-dimensional massless theory \cite{Aragone:1988yx,Rindani:1988gb,Rindani:1989ym, Bianchi:2005ze,Francia:2008hd}.\footnote{
The Singh-Hagen massive HS Lagrangian
\cite{Singh:1974qz,Singh:1974rc}
can be obtained through dimensional reduction of  Fronsdal's massless one \cite{Fronsdal:1978rb,Fang:1978wz} after gauge fixing. However, the gauge fixing procedure is non-trivial if one starts with the Fronsdal action and a more convenient one can be found in \cite{Metsaev:2008ks}. Let us also mention that the analysis in \cite{Francia:2008hd} is carried out within the unconstrained setting of \cite{Francia:2005bu,Francia:2007qt} bypassing all the problems related to the constrained Fronsdal formulation. Furthermore, let us mention that similar results can be also recovered starting from the tractor approach \cite{Gover:2008pt,Gover:2008sw,Grigoriev:2011gp}.}
However, when applied to cubic interactions, the conventional Kaluza-Klein (KK) reduction method imposes some restrictions.
In the case of flat-space interactions, these rule out the possibility of reproducing most of the known examples of massive HS interactions,
notably those appearing in ST \cite{Taronna:2010qq,Sagnotti:2010at}.
Notice that, after all, the consistency of the KK reduction does not hold if one considers
only a part of the KK spectrum.\footnote{The only consistent truncation is the massless one
which is not the main goal of the present paper.}
In this paper we avoid this restriction
working within the ambient-space formulation of (A)dS fields
\cite{Fronsdal:1978vb,
Metsaev:1995re,Metsaev:1997nj, Biswas:2002nk,Hallowell:2005np}
with an insertion of a delta function of the radial coordinate
into the $(d+1)$-dimensional action.\footnote{A similar delta-function calculus has been used in the framework of 2T-physics (see \cite{Bars:2010xi} and references therein).}
This means that we are actually dealing with a $d$-dimensional action but
in a $(d+1)$-dimensional representation.
On the other hand, the gauge consistency requires particular attention
in treating the total-derivative terms that, because of
the insertion of the delta function, do not vanish any longer. 

\medskip

Taking into account the aforementioned subtleties,
we translate the consistency condition for the vertices into a set of differential equations.
The latter can be explicitly solved for given \mt{s_{1}\!-\!s_{2}\!-\!s_{3}} couplings
and the \mt{2\!-\!2\!-\!2} and \mt{3\!-\!3\!-\!2} examples are provided in detail for all different combinations of the masses. In the following we summarize our results for arbitrary spins. Let us stress that our analysis is independent of space-time dimensionality, however subtleties arise in three and four dimensions due to the appearance of some identities. More precisely, in three dimensions our analysis is not complete while in four dimensions some parts of the vertices can vanish identically.\footnote{For instance, in four dimensions the Gauss-Bonnet identity allows to rewrite the coupling of three partially-massless spin 2 fields with at most four derivatives as a coupling with at most two derivatives.}

\begin{description}

\item[Massive and massless interactions]

Cubic interactions involving massive and massless fields can be expressed in a compact form via generating functions of all consistent couplings. Depending on the number of massless fields entering the latter,
the corresponding vertices are given by functions $\cK$ of subsets of the following building blocks:
\ba
\tilde{Y}_i \eq \partial_{U_i}\!\cdot\partial_{X_{i+1}}+\alpha_i\,\partial_{U_i}\!\cdot\partial_X\,,\nn
 Z_i\eq\partial_{U_{i+1}}\!\cdot\partial_{U_{i-1}}\,,\nn
\tilde{G} \eq (\partial_{U_1}\!\cdot\partial_{X_{2}}+\beta_1\,\partial_{U_1}\!\cdot\partial_X)\, \partial_{U_2}\!\cdot\partial_{U_3}+(\partial_{U_2}\!\cdot\partial_{X_{3}}+\beta_2\,\partial_{U_2}\!\cdot\partial_X) \,\partial_{U_3}\!\cdot\partial_{U_1}\nn
&&+\,(\partial_{U_3}\!\cdot\partial_{X_{1}}+\beta_3\,\partial_{U_3}\!\cdot\partial_X)\,\partial_{U_1}\!\cdot\partial_{U_2}\,,\nn
\tilde H_{i} \eq
	\partial_{X_{i+1}}\!\!\cdot\partial_{X_{i-1}}\,\partial_{U_{i-1}}\!\!\cdot\partial_{U_{i+1}}
	-\partial_{X_{i+1}}\!\!\cdot\partial_{U_{i-1}}\,\partial_{X_{i-1}}\!\!\cdot\partial_{U_{i+1}}\,,
\ea
that are differential operators acting on ambient-space HS fields
\be
	\Phi(X_{i},U_{i})=\sum_{s=0}^{\infty}\,\frac1{s!}\,
	\Phi^{\sst (s)}_{\sst M_{1}\ldots M_{s}}(X_{i})\,U_{i}^{\sst M_{1}}\cdots U_{i}^{\sst M_{s}}\,.
\ee
Here \mt{\partial_{X^M}=\partial_{X_1^M}+\partial_{X_2^M}+\partial_{X_3^M}} denotes total derivatives,
while the $\a_{i}$'s and the $\b_{i}$'s are parameterized as
\ba
&&\alpha_1=\alpha\,, \qquad\alpha_2=-\tfrac{1}{\alpha+1}\,,\qquad
\alpha_3=-\tfrac{\alpha+1}{\alpha}\,,\nn
&&\beta_1=\beta\,, \qquad\beta_2=-\tfrac{\beta+1}{\alpha+1}\,,\qquad \beta_3=-\tfrac{\alpha-\beta}{\alpha}\,.
\ea
Finally, the TT parts of the cubic interactions for massive and massless HS fields in (A)dS can be expressed as
\be
	\int d^{d+1}X\,\delta\big(\sqrt{\epsilon\,X^{2}}-L\big)\ \cK\ \Phi(X_1,U_1)\, \Phi(X_2,U_2)\, \Phi(X_3,U_3)\, \Big|_{\overset{X_i=X}{\sst U_i=0}}\,,
\ee
where $\epsilon$ is a sign, positive for dS and negative for AdS. The
flat-space interactions can be smoothly recovered as limits of the AdS ones.

\item[Partially-massless interactions]
	
Although at present we are not able to derive a generating function encompassing all possible interactions of partially-massless fields (which are unitary only in dS), this can be done for a class of highest-derivative couplings.
As a result, whenever the $i$-th field is at one of its partially-massless points $\m_i\in\{0\,,\,\ldots\,,\,s_i\!-\!1\}$, the corresponding vertices are consistent provided the condition
\be
\label{rmmcond}
\mu_{i}-|\mu_{i+1}-\mu_{i-1}|\in 2\,\mathbb{N}_{\sst \ge 0}\,,\quad [i\simeq i+3]\,,
\ee
holds. Here, the $\mu_{i}$'s are numbers parameterizing the mass-squared values
\be
	M_i^{2}=-\frac{1}{L^{2}}\,\big[(\mu_i-s_i+2)
	(\mu_i-s_i-d+3)-s_i\big]\,,
\ee
of the spin $s_i$ fields.
It is conceivable that the aforementioned pattern does not change in the general case, giving rise to an enhancement of the number of consistent couplings whenever \eqref{rmmcond} is satisfied. This is indeed the case for all the examples that we have analyzed explicitly, although arriving at a definite conclusion on this issue would require more efforts so that we leave this problem for future work.

\end{description}

\paragraph{St\"uckelberg-field formulation}

For the purpose of getting the full vertices, one would need to implement
gauge symmetry also for massive fields.
Then, as in the massless case, the remaining parts of the interactions could be recursively determined
relying on the gauge invariance of the vertices.
Massive HS fields acquire gauge symmetries in the St\"uckelberg formulation,
wherein one introduces new fields and gauge symmetries
into the massive theory in such a way not to alter it.
The advantage of such a formulation is that it allows to properly
analyze the massless limit of a massive theory that, in general,
turns out to be very delicate. A renowned example is the vDVZ discontinuity \cite{vanDam:1970vg,Zakharov:1970cc},
 related to the fact that the massless limit of a massive spin 2 is not simply a massless spin 2 but involves
 a massless vector and a massless scalar too.\footnote{Let us mention that the vDVZ discontinuity is absent in (A)dS \cite{Porrati:2000cp,Kogan:2000uy,Deser:2000de,Porrati:2002cp,Francia:2008hd}.}
The analysis preserving the number of DoF in the massless limit
can be carried out within the St\"uckelberg formulation. Let us mention here a key difference between the massless limit in flat and in AdS space.
While in flat space a massive spin $s$ splits into a collection of massless fields
of spin from $s$ down to $0$\,, in AdS it gives rise to a massless spin $s$
and a massive spin $s-1$ field \cite{Deser:2001pe,Deser:2001us,Deser:2001xr,Zinoviev:2001dt,Deser:2003gw,Bianchi:2005ze}.
With the aim of extending the analysis of the massless limit to the cubic level, we also provide the St\"uckelberg formulation
of the cubic interactions.
The latter can be obtained making use of the St\"uckelberg shift
encoded in the following generating functions:
\ba
	\cK(\bm Y,\bm Z)
		\,P(w_{1}\,X_{1}\!\cdot\partial_{U_{1}})\,P(w_{2}\,X_{2}\!\cdot\partial_{U_{2}})\,
		P(w_{3}\,X_{3}\!\cdot\partial_{U_{3}})\, \Big|_{w_{i}=0}\,,
\ea
where the $\bm Y_{i}$'s and the $\bm Z_i$'s are given by
\ba
\label{YZshifted}
	&& \bm Y_{i}=Y_{i}+
	\partial_{X_{i}}\!\cdot\partial_{X_{i+1}}\,\partial_{w_{i}}\,, \nn
	&& \bm Z_{i}=
	Z_{i}+
	\partial_{U_{i+1}}\!\!\cdot\partial_{X_{i-1}}\,\partial_{w_{i-1}}
	+\partial_{U_{i-1}}\!\!\cdot\partial_{X_{i+1}}\,\partial_{w_{i+1}}
	+\partial_{X_{i+1}}\!\!\cdot\partial_{X_{i-1}}\,\partial_{w_{i+1}}\,
	\partial_{w_{i-1}}\,,
\ea
and $P(z)={}_{\sst 0}F_{\sst 1}(-\mu\,;\,-L\,z)$\, is a hypergeometric function.
Under the assumption that all mass parameters of the theory scale uniformly in the massless limit, we find that in AdS the leading terms of the interactions are massive couplings involving all the massive spin $s\!-\!1$ components of the original spin $s$ fields. On the other hand, when some of the leading parts are absent, then the new dominant ones start to involve the massless spin $s$ components. Finally, performing the massless limit in flat space one recovers consistent massless vertices in agreement with the aforementioned pattern.

\subsection*{Organization of the paper}

Section \ref{sec: free unitary} is devoted to 
the formulation of the free theories of massive and (partially-) massless fields
in the ambient-space formalism.
In Section \ref{sec: cubic unitary}
we provide the solutions to the Noether procedure for the corresponding cubic interactions.
We then extend the previous results to the St\"uckelberg formulation
and study the massless limit of the massive couplings in Section \ref{sec: Stuckelberg}.
Our results as well as some outlook are summarized and discussed in Section \ref{sec: discussion}.
Appendix \ref{sec: identities} contains
some identities and mathematical tools used in our construction.
In Appendix \ref{sec: pm ex} we provide the detailed examples of \mt{2\!-\!2\!-\!2} and \mt{3\!-\!3\!-\!2} interactions, while in Appendix \ref{sec: hd pm}
we discuss a class of interactions containing the highest number of derivatives.
Finally, Appendices \ref{sec: mless limit} and \ref{sec: ST coup} include further details
on the massless limit in flat space and on the ST interactions, respectively.

\section{Free HS fields in (A)dS}
\label{sec: free unitary}

In this section we present the free theories of massive and
(partially-)massless totally-symmetric HS fields in (A)dS.\footnote{
Throughout this paper, by (A)dS we refer to any constant-curvature background
including flat space.
}
After providing an intrinsic formulation, we introduce the ambient-space formalism
 in which the construction of the cubic vertices becomes considerably simpler.

A massive spin-$s$ boson in (A)dS can be described
in terms of a totally-symmetric rank-$s$ tensor field
$\varphi^{\sst (s)}_{\mu_{1}\ldots \mu_{s}}$\,. In the following,
 we use the generating functions of such fields:
\be
	\varphi^{\sst A}(x,u):=
	\sum_{s=0}^{\infty}\,\frac1{s!}\ \varphi^{\sst A\,(s)}_{\mu_{1}\ldots\mu_{s}}(x)\
	u\cdot e^{\mu_{1}}(x)\,\cdots\, u\cdot e^{\mu_{s}}(x)\,,
\ee
where the contraction with the flat auxiliary variables $u^{\alpha}$ is via
the inverse (A)dS vielbein $e^{\ \mu}_{\alpha}(x)$\,:
$u\cdot e^{\mu}(x)=u^{\alpha}\,e^{\ \mu}_{\alpha}(x)$\,,
and $\st A$ is a \emph{color} index associated with the Chan-Paton factors.
The massive representations of the (A)dS isometry group
correspond to  HS fields satisfying the Fierz system:
\be
\label{Fierz}
	(D^2-M^2)\,\varphi^{\sst A}=0\,,\qquad
	\partial_{u}\cdot e^{\mu}\,D_{\mu}\,\varphi^{\sst A}=0\,,\qquad
	\partial_{u}^{\,2}\,\varphi^{\sst A}=0\,,
\ee
where $M$ is the mass operator defined by
$M^{2}\,\varphi^{\sst (s)}:=m_{\sst s}^{2}\,\varphi^{\sst (s)}$\,,
and $D_{\mu}$ is the covariant derivative:
\be
	D_{\mu} := \nabla_{\mu}+\tfrac12\,\omega_{\mu}^{\alpha\beta}(x)
	\,u_{[\alpha}\partial_{u^{\beta]}}\,.
	\label{cov}
\ee
Here $\nabla_{\mu}$ is the usual (A)dS covariant derivative and
$\omega_{\mu}^{\alpha\beta}$ is the (A)dS spin connection, so that
the (A)dS Laplacian operator
is given simply by $D^{2}$\,.

The quadratic action for HS fields
reproducing the Fierz system \eqref{Fierz} can be written as
\be
	S^{\,\sst (2)}= \frac12\int d^{d}x\sqrt{-g}\,\bigg[\delta_{\sst A_{1}A_{2}}
	\,e^{\partial_{u_{1}}\!\cdot\,\partial_{u_{2}}}\,
	\varphi^{\sst A_{1}}(x_{1},u_{1})
	\left(D^{\,2}_{\sst 2}-M_{\sst 2}^{2}\right)
 	\varphi^{\sst A_{2}}(x_{2},u_{2})+\,\ldots
	\bigg]_{\overset{x_{i}=x}{\sst u_{i}=0}}\,,
	\label{AdS act}
\ee
where the ellipsis denote, henceforth, terms proportional to divergences and traces of the fields as well as possible auxiliary fields.
Since we focus on the TT parts of the cubic interactions, such terms
are not relevant for our discussion although they must be taken into account in order to construct the full theory.\footnote{
See \cite{Singh:1974qz,Singh:1974rc,Hallowell:2005np,
Buchbinder:2005ua,Fotopoulos:2006ci,Buchbinder:2006nu,Buchbinder:2007ix,Francia:2007ee, Francia:2008ac,
Buchbinder:2008ss,Zinoviev:2008ze,Ponomarev:2010st,Grigoriev:2011gp}
for the precise forms of the free action.}
The Lagrangian equations are
\be
	(D^{2}-M^{2})\,\varphi^{\sst A}+ \,\ldots\,  \approx 0\,,
\label{LagE}
\ee
together with possible equations for the auxiliary fields.

A massless spin-$s$ boson in (A)dS corresponds to the mass-squared value:
\be
	m_{s}^{2}=\frac{(-\epsilon)}{L^{2}}\,\big[(s-2)(s+d-3)-s\big]\,,
	\label{masslessAdsmass}
\ee
where $L$ is the (A)dS radius and $\epsilon$ is a sign, negative for AdS
and positive for dS.
For this value of the mass, the action \eqref{AdS act}   admits the gauge symmetries:
\be
	\delta^{\sst (0)}\,\varphi(x,u)=u\cdot e^{\mu}\,D_{\mu}\,\varepsilon(x,u)\,,
	\label{dS gt}
\ee
with the gauge parameter $\varepsilon$ traceless in the Fronsdal's formulation \cite{Fronsdal:1978vb}
and traceful in the unconstrained ones \cite{Francia:2007qt,Buchbinder:2007ak}.
For simplicity, in this paper we disregard the issue of trace constraints
keeping the unconstrained formulation in mind.
However, since we focus on the TT parts of the Lagrangian
such a distinction is irrelevant.

\subsection{Ambient-space formalism}
\label{Ambient space formalism}

It is well known that the $d$-dimensional Euclidean AdS
or Lorentzian dS space can be embedded in the $(d+1)$-dimensional flat space with metric:
\be
	ds^{2}_{\rm Amb}=\eta_{\sst MN}\,dX^{\sst M}\,dX^{\sst N}\,,
	\qquad
	\eta=(-,+,\,\ldots,+)\,.
\ee
The (A)dS space  is then defined as the hyper-surface $X^{2}=\epsilon\,L^{2}$\,,
where, as before, $\epsilon$ is a sign, negative for AdS and positive for dS.
We concentrate on the region of the ambient space with $\epsilon\,X^{2}>0$\,, and consider the generating function of totally-symmetric tensor fields $\Phi_{\sst M_{1}\ldots M_{s}}$ given by
\be
	\Phi(X,U)=\sum_{s=0}^{\infty}\frac1{s!}\,
	\Phi^{\sst (s)}_{\sst M_{1}\ldots M_{s}}(X)\,U^{\sst M_{1}}\cdots U^{\sst M_{s}}\,.
\ee
These fields are equivalent to totally-symmetric tensor fields in (A)dS if they are homogeneous in $X^{\sst M}$ and tangent to constant $X^{2}$ surfaces.
At the level of the generating function, the latter
conditions translate  into
\ba
	&{\rm Homogeneity}: \qquad
	&(X\cdot\partial_{X}-U\cdot\partial_{U}+2+\mu)\,\Phi(X,U)=0\,,
	\label{h con} \\
	&{\rm Tangentiality}: \qquad
	&X\cdot \partial_{U}\,\Phi(X,U)=0\,,	
	\label{t con}
\ea
where $\mu$ is a parameter related to the (A)dS mass.
In order to identify the ambient-space fields with the (A)dS ones,
we parameterize the $\epsilon\,X^{2}>0$ region with the radial coordinates $(R,x)$ given by
\be
	X^{\sst M}=R\,\hat X^{\sst M}(x)\,,\qquad \hat X^{2}(x)=\epsilon\,,
\ee
and rotate the auxiliary $U^{\sst M}$-variables as
\be
	U^{\sst M}= \hat X^{\sst M}(x)\,v+ L\,\tfrac{\partial \hat X^{M}}{\partial x^{\mu}}(x)\,
	e_{\alpha}^{\ \mu}(x)\,u^{\alpha}\,.
\ee
With this change of variables from $(X,U)$ to $(R,x;v,u)$\,, the homogeneity and tangentiality conditions
(\ref{h con}\,,\,\ref{t con}) are solved by the (A)dS intrinsic generating functions as
\be
	\Phi(R,x;v,u)= \left(\tfrac RL\right)^{u\cdot\partial_{u}-2-\mu}\,\varphi(x,u)\,,
	\label{doh}
\ee
and the action \eqref{AdS act} can be written as
\be
	S^{\sst (2)} = \frac1{2}
	\int d^{d+1}X\ \delta\Big(\sqrt{\epsilon\,X^{2}}-L\Big)\,
	\Big[\ \delta_{\sst A_{1}A_{2}}\,e^{\partial_{U_{1}}\!\cdot\,\partial_{U_{2}}}\,
	\Phi^{\sst A_{1}}(X_{1},U_{1})\,\partial_{X_{2}}^{\,2}\,\Phi^{\sst A_{2}}(X_{2},U_{2})+\,\ldots
	\Big]_{\overset{X_{i}=X}{\sst U_{i}=0}}\,.\label{freeAdS}
\ee
In the ambient space,
the Lagrangian equation \eqref{LagE} reads
\be
	\partial_{X}^{2}\,\Phi +\ldots\approx 0\,,
	\label{Amb eom}
\ee
where the ambient-space d'Alembertian is related to the (A)dS one as
\be
	\partial_{X}^{2}\,\Phi=
	\left(\tfrac RL\right)^{u\cdot\partial_{u}-4-\mu}\,(D^{2}-M^{2})\,\varphi\,.
\ee
Here, the mass-squared is given in terms of $\mu$ by
\be
	M^{2}=\frac{(-\epsilon)}{L^{2}}\,\big[(\mu-u\cdot\partial_{u}+2)
	(\mu-u\cdot\partial_{u}-d+3)-u\cdot\partial_{u}\big]\,.
	\label{mass}
\ee
Notice that for dS, where $\epsilon=1$\,, the parameter $\mu$ is in general a complex number, hence, in order for the fields to be
real one has to add the complex conjugate in eq.~\eqref{doh}.
Making a comparison with \eqref{masslessAdsmass},
one can also see that $\mu=0$ corresponds to the massless case.

\subsubsection*{Flat-space limit}

The flat-space limit $L\to \infty$ can be considered
keeping the ambient-space point of view.
In order to do that, we first need to place the origin of the ambient space
in a point on the hyper-surface $X^{2}=\epsilon\,L^{2}$ by translating the coordinate system as
\be
	X^{\sst M}\ \ \to\ \ X^{\sst M}+L\,\hat N^{\sst M}\,.
	\label{flat shift}
\ee
Here, $\hat N$ is a constant vector in the ambient space satisfying $\hat N^2=\epsilon$\,. After this shift,
taking the $L\to \infty$ limit one gets
\be
	\delta\Big(\sqrt{\epsilon\,X^{2}}-L\Big)\quad
	\underset{L\to \infty}{\longrightarrow} \quad
	\epsilon\,\delta(\hat N\cdot X)\,,
\ee
so that the hyper-surface $X^{2}=\epsilon\,L^{2}$ becomes
the hyperplane $\hat N\cdot X=0$\,, defining the $d$-dimensional flat space embedded in the ambient space.
Moreover, the homogeneity and tangentiality conditions (\ref{h con}\,,\,\ref{t con})
 admit a well-defined limit:
\be
\label{rad dep1}
	\left(\hat N\cdot\partial_X-\sqrt{-\epsilon}\,M\right)\,\Phi(X,U)=0\,,\qquad
	\hat N\cdot\partial_U\,\Phi(X,U)=0\,,
\ee
provided one first divides them by $L$ and redefines
$\mu$ in terms of $M$ according to \eqref{mass}.
The latter equations are solved by
\be
\label{genfuncd+1}
\Phi(X,U)=e^{-\sqrt{-\epsilon}\,M\,\rho}\,\varphi(x,u)\,,
\ee
where $\rho:=\hat N\cdot X$ and $(x,u)$ are coordinates
on the hyper-surface of constant $\hat N\cdot X$ and $\hat N\cdot U$\,.
Since the flat limit from dS presents some issues related to the partially-massless
points, in the following we only consider the limit starting from AdS ($\epsilon=-1$).

Let us conclude this section with a few remarks about the role of the delta function.
Notice that without the latter
the ambient-space action \eqref{freeAdS} would contain a diverging factor coming from the radial integral.
The insertion of the delta function precisely cures this divergence.
On the other hand, one may wonder whether we could have avoided such insertion by
taking the extra dimension to be compact.
For instance, in flat space one can consider a compact coordinate \mt{\rho \sim \rho + L}
together with a harmonic $\rho$-dependence of the fields:
\mt{\Phi=e^{i\,\frac{2\pi}L\,m\,\rho}\,{\varphi}}\,. Then, the $\rho$-integral gives
an orthogonality condition:
\be
	\int_0^L d\rho\ e^{i\,\frac{2\pi}L\,m_1\,\rho}\ e^{-i\,\frac{2\pi}L\,m_2\,\rho}
	=L\,\delta_{m_1\,,\,m_2}\,.
\ee
Although this KK reduction works well at the free level,
it turns out to be problematic or at least too restrictive at the cubic level
since one gets in this case an undesired mass equality:
\be
	\int_0^L d\rho\ e^{i\,\frac{2\pi}L\,m_1\,\rho}\ e^{i\,\frac{2\pi}L\,m_2\,\rho}
	\ e^{-i\,\frac{2\pi}L\,m_3\,\rho}
	=L\,\delta_{m_1+m_2\,,\,m_3}\,.
\ee
The latter forbids many interactions, notably those arising in ST,
and can be avoided
via the insertion of a delta function $\delta(\hat N\cdot X)$\,.

\subsection{Gauge symmetries in the ambient-space formalism}

As we have seen, in the intrinsic formulation, HS fields whose mass-squared is given by \eqref{masslessAdsmass}, \emph{i.e.} $\mu=0$\,,
admit the gauge symmetries \eqref{dS gt}.
This gauge invariance of the massless theory can be seen also at the ambient space level. We first consider the linearized gauge symmetries:
\be
	\delta^{\sst (0)}\,\Phi(X,U)=U\cdot\partial_{X}\,E(X,U)\,,
	\label{amb gt}
\ee
where $E$ is the generating function of the gauge parameters.
Since the action \eqref{freeAdS} does not contain any explicit mass term,
the gauge invariance seems to be unrelated to the value of $\mu$\,.
This cannot be the case
since it would imply the presence of gauge symmetries for massive theories in the absence of the St\"uckelberg fields.
Indeed, as we show in the following, the homogeneity and tangentiality
conditions (\ref{h con}\,,\,\ref{t con}) are compatible with the gauge symmetry  \eqref{amb gt}
only for particular values of $\mu$.

\subsubsection{Massless fields}

Starting from eqs.~\eqref{h con} and
 \eqref{amb gt}\,, one can first derive the homogeneity degree of the gauge parameters:
 \be
(X\cdot\partial_X-U\cdot\partial_U+\mu)\,E(X,U)=0\,.
\label{h con E}
\ee
Then, one has to impose the compatibility of the tangentiality condition \eqref{t con} with the gauge transformations \eqref{amb gt}:
\be
X\cdot\partial_{U}\ \delta^{\sst (0)}\,\Phi(X,U)
=(U\cdot\partial_X\, X\cdot\partial_U-\mu)\,E(X,U)=0\,,
\label{t consy}
\ee
where we used eq.~\eqref{h con E}.
When \mt{\mu=0}\,, any gauge parameter satisfying
the tangentiality condition:
\be
	X\cdot\partial_U\,E(X,U)=0\,,
	\label{t con E}
\ee
is a solution of \eqref{t consy}.
Therefore, the ambient-space gauge parameter $E$ is related to the intrinsic (A)dS one $\varepsilon$ as
\be
\label{Ereduc}
	E(R,x;v,u)=\left(\tfrac{R}{L}\right)^{u\cdot\partial_{u}}\,\varepsilon(x,u)\,,
\ee
and the ambient-space gauge transformations \eqref{amb gt} reduce to the (A)dS ones \eqref{dS gt}.

\paragraph{Massive fields}

In the $\mu\neq 0$ case, eq.~\eqref{t consy} implies
\be
E(X,U) = \tfrac{1}{\mu}\, U\cdot\partial_X\, X\cdot\partial_U\,E(X,U)\,,
\ee
that in turn is compatible with the tangent condition provided
\be
\big[\,(U\cdot\partial_X)^2\, (X\cdot\partial_U)^2-2\,\mu\,(\mu-1)\,\big]\,E(X,U)=0\,.
\ee
 If $\mu\neq 1$, the latter gives
\be
E(X,U) = \tfrac{1}{2\,\mu\,(\mu-1)}\, (U\cdot\partial_X)^2\, (X\cdot\partial_U)^2\,E(X,U)\,.
\ee
Hence, when $[\mu]_r:=\mu\,(\mu-1)\cdots(\mu-r+1)\neq0$\,,
one can iterate $r$ times this procedure ending up with
\be
\label{PartMassiter}
\big[\,(U\cdot\partial_X)^r\, (X\cdot\partial_U)^r-r!\,[\mu]_r\,\big]\,E(X,U)=0\,.
\ee
Since $(X\cdot\partial_{U})^{s}\,E^{\sst (s-1)}=0$, whenever $[\mu]_{s}\neq 0$, the spin $s\!-\!1$ component of this equation implies
that eqs.~(\ref{h con}\,,\,\ref{t con}) are compatible with the gauge symmetry only for vanishing $E^{\sst (s-1)}$.
In AdS all unitary representations have non-positive values of $\mu$ \cite{Siegel:1988gd},
therefore the gauge symmetry is allowed only in the massless case.

\subsubsection{Partially-massless fields}
\label{sec: pmf}

In dS, the unitary representations \cite{Dixmier:1961zz,Thielekerr:1974zz,Higuchi:1989gz,Deser:2003gw} include all positive integer values
\mt{\mu=r\in \mathbb N_{\ge 0}}\,. In those cases
 the iteration procedure stops whenever \mt{r<s}\,.
 Therefore, non-vanishing solutions exist for the gauge parameters satisfying\footnote{Similar constraints have been also exploited in \cite{Alkalaev:2009vm,Alkalaev:2011zv} keeping the necessary auxiliary fields in order to achieve an off-shell description.}
\be
	(X\cdot\partial_{U})^{r+1}\,E(X,U)=0\,.
\ee
Inverting \eqref{PartMassiter}, the initial gauge parameter $E$
can be solved in terms of a new gauge parameter $\Omega$ as
\be
E(X,U)=(U\cdot\partial_X)^r\,\Omega(X,U)\,,
\ee
where $\Omega$ satisfies the homogeneity and tangentiality conditions:
\be
	(X\cdot\partial_X-U\cdot\partial_U-r)\,\Omega(X,U)=0\,,
	\qquad X\cdot\partial_U\,\Omega(X,U)=0\,.
\ee
Thus, $\Omega$ can be reduced to the intrinsic dS gauge parameter $\omega$ as
\be
	\Omega(R,x;v,u)=\left(\tfrac{R}{L}\right)^{u\cdot\partial_{u}+r}\,
	\omega(x,u)\,.
\ee
Finally, the gauge transformations\footnote{An anolagous form of the gauge transformations has been obtained in the tractor approach \cite{Gover:2008pt}.}
\be
\delta^{\sst (0)}\,\Phi=(U\cdot\partial_X)^{r+1}\,\Omega(X,U)\,,
\label{pm gt}
\ee
become the dS intrinsic ones:
\be
	\delta^{\sst (0)}\,\varphi(x,u)=
	\left[(u\cdot D)^{r+1}+\ldots \right] \omega(x,u)\,,
\ee
whose form has been obtained recursively in \cite{Zinoviev:2001dt,Francia:2008hd}.

\section{Cubic interactions of  HS fields in (A)dS}
\label{sec: cubic unitary}

In this section we construct the consistent parity-invariant cubic interactions
of massive and partially-massless HS fields in (A)dS.
More precisely, we focus on those pieces which do not contain divergences and traces of the fields (TT parts).
 We begin with the most general expression for the cubic vertices:\footnote{
The dependence on the $X^{\sst M}$ in the ansatz can be neglected
(see \cite{Joung:2011ww}).}
\ba
\label{cubicact1}
S^{\sst (3)} \eq \frac1{3!}\int d^{d+1}X\ \delta\Big(\sqrt{\epsilon\,X^2}-L\Big)\
C_{\sst A_{1}A_{2}A_{3}}(L^{-1}\,;\,\partial_{X_1},\partial_{X_2},\partial_{X_3}\,;\,
\partial_{U_1},\partial_{U_2},\partial_{U_3}) \times \nn
&& \hspace{50pt} \times\ \Phi^{\sst A_{1}}(X_1,U_1)\ \Phi^{\sst A_{2}}(X_2,U_2)\ \Phi^{\sst A_{3}}(X_3,U_3)\ \Big|_{^{X_i=X}_{U_i=0}}+\,\ldots\,.
\ea
Here $C_{\sst A_{1}A_{2}A_{3}}$ denotes the TT part of the vertices.
The cubic interactions in (A)dS
are in general inhomogeneous in the number of derivatives,
the lower-derivative parts being dressed by negative powers of $L$
compared to the highest-derivative one.
Hence, the TT parts of the vertices can be expanded as
\be
\label{seriesEx}
C_{\sst A_{1}A_{2}A_{3}}(L^{-1}\,;\,\partial_{X}\,,\,\partial_{U})
=\sum_{n=0}^\infty\, L^{-n}\,
C^{\sst [n]}_{\sst A_{1}A_{2}A_{3}}(Y,Z)\,,
\ee
where we have introduced the parity-preserving Lorentz invariants:
\be
\label{Y and Z}
Y_i=\partial_{U_i}\!\cdot\partial_{X_{i+1}}\,, \qquad
Z_i=\partial_{U_{i+1}}\!\!\cdot\partial_{U_{i-1}}\,,
\qquad [i\simeq i+3]\,.
\ee
Notice that we have dropped divergences, $\partial_{U_{i}}\!\cdot\partial_{X_{i}}$\,, traces, $\partial_{U_{i}}^{\,2}$ as well as
terms proportional to $\partial_{X_i}\!\cdot\partial_{X_j}$'s.
Indeed, being proportional to the field equations \eqref{Amb eom}
up to total derivatives, the latter can be removed by proper field redefinitions.
Moreover, since we have chosen a particular set of $Y_{i}$'s,
any ambiguity related to the total derivatives has been fixed.

In order to simplify the analysis, it is convenient to
recast the expansion \eqref{seriesEx} in a slightly different, though equivalent form.
First, let us notice that negative powers of $L$ can be absorbed into derivatives of the delta function:
\be\label{deltafunc}
	\delta^{\sst (n)}(R-L)\,\,R^{\lambda}
	=\frac{(-2)^{n}\,[(\lambda-1)/2]_{n}}{L^{n}}\,\delta(R-L)\,R^{\lambda}\,.
\ee
where $\delta^{\sst (n)}(R-L)=\left(\tfrac{L}{R}\,\tfrac{d}{dR}\right)^n\,\delta(R-L)$\,.
Then, introducing $\hat{\delta}$ with the following prescription:
\be
	\d^{\sst (n)}(R-L)\equiv\d(R-L)\,(\epsilon\,\hat\delta)^{n}\,,
	\label{d rel}
\ee
each coefficient of \eqref{seriesEx} can be redefined as
\be
	L^{-n}\,C^{\sst [n]}_{\sst A_{1}A_{2}A_{3}}(Y,Z) =
	\hat\delta^{n}\,C^{\sst (n)}_{\sst A_{1}A_{2}A_{3}}(Y,Z)\,.
\ee
Notice that $C_{\sst A_{1}A_{2}A_{3}}^{\sst [n]}$ and
$C_{\sst A_{1}A_{2}A_{3}}^{\sst (n)}$
are different functions for $n\ge1$\,.
The entire couplings can be finally resummed as
\be
	C_{\sst A_{1}A_{2}A_{3}}(\hat\delta;Y,Z)
	=\sum_{n=0}^{\infty}\,\hat\delta^{n}\,
	C_{\sst A_{1}A_{2}A_{3}}^{\sst (n)}(Y,Z)\,,
	\label{exp lambda}
\ee
where we have used the same notation
for both $C_{\sst A_{1}A_{2}A_{3}}(L^{-1};Y,Z)$
and $C_{\sst A_{1}A_{2}A_{3}}(\hat\delta;Y,Z)$
although they are different functions. 

In order to make contact with the standard tensor notation, let us provide an explicit example. A vertex of the form
\be
C(\hat\delta;Y,Z)=(Y_1^2\,Y_2\,Y_3\,Z_1+\text{cycl.})-\tfrac{\hat{\delta}}{L}\,(Y_1\,Y_2\,Z_1\,Z_2+\text{cycl.})+\tfrac{3}{4}\,\left(\tfrac{\hat{\delta}}{L}\right)^2\,Z_1\,Z_2\,Z_3\,,
\ee
which will turn out to be a consistent coupling involving three partially-massless spin 2 fields (see Appendix \ref{sec: pm ex}), gives
\begin{multline}
S^{\sst (3)}=\frac{1}{2}\,\int d^{d+1}X\ \delta\Big(\sqrt{X^2}-L\Big)\,\Big[\partial_{\sst P}\,\Phi^{\sst MN}\,\partial_{\sst M}\,\partial_{\sst N}\,\Phi_{\sst LQ}\,\partial^{\sst L}\,\Phi^{\sst PQ}\\
+\frac{d-6}{L^2}\,\Phi^{\sst M}_{\phantom{\sst M}\sst N}\,\partial_{\sst M}\,\Phi_{\sst LP}\,\partial^{\sst L}\,\Phi^{\sst NP}
+\frac{(d-4)(d-6)}{4\,L^4}\,\Phi^{\sst M}_{\phantom{\sst M}\sst N}\,\Phi^{\sst N}_{\phantom{\sst N}\sst P}\,\Phi^{\sst P}_{\phantom{\sst P}\sst M}\Big]\,.
\end{multline}

\subsection{Consistent cubic interactions of massive and massless HS fields}
\label{sec:Consistent cubic}

So far, we have not specified whether the fields $\Phi^{\sst A}$
are massive or massless.
In the following we use ${\st A}=\alpha$ for massive fields and ${\st A}=a$ for massless ones.
One can consider different cases
depending on the number of massless and massive fields involved in the cubic interactions.
The presence of massive fields does not impose any constraints on the vertices,
while, whenever a massless field takes part in the interactions,
the corresponding vertices must be compatible with
the gauge symmetries of that field.

Gauge consistency can be studied order by order
(in the number of fields), and
at the cubic level gives
\be
\label{noetherproc}
\delta^{\sst (1)}_{i}S^{\sst (2)}+\delta^{\sst (0)}_{i}
S^{\sst (3)}=0 \quad \Rightarrow \quad
\delta^{\sst (0)}_{i}S^{\sst (3)} \approx 0\,,
\ee
where $\approx$ means equivalence modulo the free field equations
\eqref{Amb eom} and $\delta_{i}^{\sst (0)}$ is
the linearized gauge transformation \eqref{amb gt}
associated with the massless field $\Phi^{a_{i}}$\,.
The key point of our approach
is that the TT parts of the vertices
can be determined from the Noether procedure \eqref{noetherproc}
independently from the ellipses in \eqref{cubicact1}.
This amounts to quotient the Noether equation \eqref{noetherproc}
by the Fierz systems of the fields $\Phi^{\sst A_{i}}$ and of the gauge parameters $E^{a_{i}}$\,.
In our notation, this is equivalent to impose,  for $i=1$\,,
\be
\label{gaugeconscond1}
\left[\,C_{\sst a_{1}A_{2}A_{3}}(\hat\delta;Y,Z)\,,\, U_1\cdot\partial_{X_1}\,\right]\Big|_{U_1=0}\approx 0\,,
\ee
modulo all the $\partial_{X_{i}}^{\,2}$'s\,, $\partial_{U_{i}}\!\cdot\partial_{X_{i}}$'s
and $\partial_{U_{i}}^{\,2}$'s\,.
Due to the presence of the delta function, the total derivative terms generated by
the gauge variation do not simply vanish, but contribute as
\be
\delta\Big(\sqrt{\epsilon\,X^2}-L\Big)\,\partial_{X^M}=
-\,\delta\Big(\sqrt{\epsilon\,X^2}-L\Big)\,\hat\delta\, \frac{X_{\sst M}}{L}\,.
\ee
Using the commutation relations \eqref{commrelations} together with the identity \eqref{fBcomm},
eq.~\eqref{gaugeconscond1} is equivalent to
the following differential equation:
\be
\label{recurrrelmass}
\Big[\,Y_2\partial_{Z_3}-Y_3\partial_{Z_2}
+\tfrac{\hat\delta}{L}\left(Y_2\partial_{Y_2}-Y_3\partial_{Y_3}
-\tfrac{\mu_{2}-\mu_3}2
\right)\partial_{Y_{1}}\Big]\,C_{\sst a_{1}A_{2}A_{3}}(\hat\delta;Y,Z)=0\,.
\ee
The consistent parity-invariant cubic interactions
involving massive and massless HS fields in (A)dS can be obtained as solutions of the above equations.
Since $C_{\sst a_{1}A_{2}A_{3}}$ is a polynomial in
$\hat\delta$\,,
one can solve \eqref{recurrrelmass}  iteratively starting from the lowest order in $\hat \d$.
 To begin with, the zero-th order term
 $C^{\sst (0)}_{\sst a_{1}A_{2}A_{3}}$ in \eqref{exp lambda} is given by
\be
	C^{\sst (0)}_{\sst a_{1}A_{2}A_{3}}=
	C^{\sst (0)}_{\sst a_{1}A_{2}A_{3}}(Y_{1},Y_{2},Y_{3},Z_{1}, G)\,,
	\label{1 ML}
\ee
where
\be
	G:=Y_{1}\,Z_{1}\,+ Y_{2}\,Z_{2}+Y_{3}\,Z_{3}\,.
\ee
On the other hand, when more than two massless fields are present, it becomes
\be
	C^{\sst (0)}_{\sst a_{1}a_{2}A_{3}}=
	C^{\sst (0)}_{\sst a_{1}a_{2}A_{3}}(Y_{1},Y_{2},Y_{3}, G)\,.
	\label{2 ML}
\ee
Notice that, while \eqref{1 ML} is an
arbitrary function of five arguments, the zero-th order solution \eqref{2 ML} depends on four arguments.
This is a consequence of the different number of differential equations
imposed on the vertices. On the other hand, in the case of three massless fields,
the third differential equation is redundant so that the number of
arguments do not decrease further.
Having obtained the zero-th order parts of the solution in eqs.~(\ref{1 ML}\,,\,\ref{2 ML}),
what is left is to determine their higher order completions.
Eq.~\eqref{recurrrelmass} gives
 an inhomogeneous differential equation for $C^{\sst (n\ge1)}_{\sst a_{1}A_{2}A_{3}}$\,,
  whose solutions are fixed up to a solution of the corresponding homogeneous equation.
However, ambiguities of the interactions
related to these solutions
are nothing but redundancies as discussed in \cite{Joung:2011ww}.

Before considering eq.~\eqref{recurrrelmass},
we first solve its flat limit $L\to \infty$\,.
Once again, this is achieved via \eqref{flat shift}
after the replacement:
\be
	\lim_{L\to\infty}\, \tfrac1L\,\mu=-M\,.
\ee
The end result takes the following form:
\be
\Big[\, Y_2\partial_{Z_3}-Y_3\partial_{Z_2}
+\tfrac{\hat\delta}2\,(M_{2}-M_3)\,\partial_{Y_1}\Big]\,
C_{\sst a_{1}A_{2}A_{3}}(\hat\delta;Y,Z)=0\,,
\label{recur flat}
\ee
so that the zero-th order parts of the solutions coincide with the (A)dS ones.
Moreover, in flat space, the operator
$\hat \d$ appearing in \eqref{d rel} is simply given by
\be
	\hat\delta=\hat N\cdot\partial_{X}\,.
	\label{dt flat}
\ee
Notice also that in this case,
for a given $C^{\sst (0)}_{\sst A_{1}A_{2}A_{3}}$\,,
the lower-derivative parts of the vertices
$C^{\sst (n\ge1)}_{\sst A_{1}A_{2}A_{3}}$ can be recast into total-derivative terms,
making them homogeneous in the number of derivatives.
This observation makes it possible to write generic consistent vertices as arbitrary
 functions of some fixed building blocks.

Our strategy is as follows.
We first solve the flat-space equation \eqref{recur flat}
and express the general solution in terms of homogeneous objects in the number of derivatives.
In this way, we identify the building blocks of the flat-space cubic interactions.
Then, we take as ansatz for the (A)dS building blocks the deformation of the flat-space ones
with the addition of  further total derivatives.
Finally, we fix such ansatz requiring the latter to solve \eqref{gaugeconscond1}.
In the following, we divide the analysis into four different cases:
3 massive, 1 massless and 2 massive, 2 massless and 1 massive and finally 3 massless fields.
For each of them, we provide the most general solution
as arbitrary functions of the corresponding building blocks.

\subsubsection*{A\ \ 3 massive}

This case is rather trivial since no condition on $C_{\alpha_1\alpha_2\alpha_3}$ is imposed.
Thus, the cubic interactions of three massive fields are given by
\be
	C_{\alpha_1\alpha_2\alpha_3}=
	\cK_{\alpha_1\alpha_2\alpha_3}(Y_{1}, Y_2,Y_3,Z_{1}, Z_{2}, Z_{3})\,.
	\label{mssv}
\ee
This reflects the fact that we focused only on the TT parts of the vertices.
Finding the remaining parts is in principle non-trivial but we expect that,
working within the gauge invariant formulation \`a la St\"uckelberg (see Section \ref{sec: Stuckelberg}),
those parts can be recursively determined from \eqref{mssv}.

\subsubsection*{B\ \ 1 massless and 2 massive}
\label{2 massive: equal masses d}

When one massless (${\st A}_{1}=a_{1}$) and two massive HS fields are involved in the interactions,
one needs to analyze separately the cases wherein the two fields have equal or different masses.

\paragraph{Equal mass}

When \mt{M_2=M_3=m\neq 0}\,,
the $M$-dependent term in \eqref{recur flat} vanishes and
therefore the solution in flat space is given by its zero-th order part:
\be
	C_{a_{1}\alpha_{2}\alpha_{3}} = \cK_{a_{1}\alpha_{2}\alpha_{3}}
	(Y_1,Y_2, Y_3, Z_{1}, G)\,.
\ee
Regarding the vertices in (A)dS, we make an ansatz by deforming
 the latter with total-derivative terms as
\be
\label{equalans}
C_{a_1\alpha_2\alpha_3}=
\cK_{a_1\alpha_2\alpha_3}(\tilde{Y}_1,\tilde{Y}_{2},\tilde{Y}_{3},Z_{1},\tilde{G})\,,
\ee
where $\tilde{Y}_i$'s and $\tilde{G}$ are given by
\ba
\label{YGtilde}
\tilde{Y}_i&=&Y_i+\alpha_i\,\partial_{U_i}\!\cdot\partial_X\,,\nn
\tilde{G}&=&(Y_1+\beta_1\,\partial_{U_1}\!\cdot\partial_X)Z_1+(Y_2+\beta_2\,\partial_{U_2}\!\cdot\partial_X)Z_2+(Y_3+\beta_3\,\partial_{U_3}\!\cdot\partial_X)Z_3\,.
\ea
Requiring the gauge invariance, one ends up with
\ba
&&(\alpha_1+1)\alpha_{2}+1=0\,,\nn
&&(\alpha_1+1)(\beta_{2}+1)+\alpha_1\,\beta_{3}=0\,,\nn
&&(\beta_1+1)(\beta_{2}+1)+\beta_{3}\,(\beta_1+\beta_{2}+1)=0\,,
\label{a b cond}
\ea
whose general solutions (see \cite{Joung:2011ww} for the details) are\footnote{
Notice that in the present conventions,
the definitions of the $\alpha_{i}$'s and the $\beta_{i}$'s
differ from the ones used in \cite{Joung:2011ww}.
The latter are recovered through
the replacements: $\alpha_{i}\to (\a_{i}-1)/2$ and of the
 $\b_{i}\to (\b_{i}-1)/2$.}
\ba
&&\alpha_1=\alpha\,, \qquad\alpha_2=-\tfrac{1}{\alpha+1}\,,\qquad
\alpha_3=-\tfrac{\alpha+1}{\alpha}\,,\nn
&&\beta_1=\beta\,, \qquad\beta_2=-\tfrac{\beta+1}{\alpha+1}\,,\qquad \beta_3=-\tfrac{\alpha-\beta}{\alpha}\,.
\label{sol ab}
\ea
As we have anticipated, the different values of the $\a_{i}$'s and the $\b_{i}$'s are
related to the redundancies of the solutions.

\paragraph{Different masses}

When $M_2\neq M_3\,$,
the zero-th order part of the solution $C^{\sst (0)}_{a_{1}\alpha_{2}\alpha_{3}}$
\eqref{1 ML} is
an arbitrary function of the $Y_i$'s\,, $Z_{1}$ and $G$\,.
However, not all of these arguments admit a solution for $C^{\sst (1)}_{a_{1}\alpha_{2}\alpha_{3}}$\,.
In particular, $C^{\sst (0)}_{a_{1}\a_{2}\alpha_{3}}=Y_{2}\,, Y_{3}$ and $Z_{1}$ are already consistent and do not need to be completed with $C^{\sst (n\ge1)}_{a_{1}\a_{2}\alpha_{3}}$\,, while
\be
	C^{\sst (0)}_{a_{1}\a_{2}\alpha_{3}}=\quad  Y_{3}\,Y_{1}\ ,\quad Y_{1}\,Y_{2}\,,
\ee
involve next order contributions given by
\be
	C^{\sst (1)}_{a_{1}\a_{2}\alpha_{3}}=\quad
	\tfrac{1}{2}\,(M_{2}-M_{3})\,Z_{2}\ ,\quad \tfrac{1}{2}\,(M_{3}-M_{2})\,Z_{3}\,,
\ee
respectively,
and $C^{\sst (n\ge2)}_{a_{1}\a_{2}\alpha_{3}}=0$\,.
Therefore, the flat-space solution can be written as
\be
	C_{a_1\alpha_2\alpha_3}=\cK_{a_1\alpha_2\alpha_3}(Y_2,Y_3,Z_{1},H_2,H_3)\,,
\ee
where the $H_i$'s are given by
\be
	H_{i}:= Y_{i+1}\,Y_{i-1}+
	\tfrac{1}{2}\,\hat N\cdot\partial_{X}\,(M_{i}-M_{i+1}-M_{i-1})\,Z_{i}\,.
	\label{H}
\ee
Notice that, using the properties of the delta function \eqref{dt flat},
they can be recast in the form
\ba
\label{H2nd}
	H_{i} \eq
	Y_{i+1}\,Y_{i-1}+\tfrac{1}{2}\,\big[M_{i}^{2}-(M_{i+1}+M_{i-1})^{2}\big]\,Z_{i}\nn
	\eq Y_{i+1}\,Y_{i-1}
	-\tfrac{1}{2}\,
	\partial_{X}\cdot(\partial_{X_{i}}-\partial_{X_{i+1}}-\partial_{X_{i-1}})\,Z_{i}\,.
\ea
The first expression in eq.~\eqref{H2nd} does not contain any total-derivative part, thus one can trivially reduce it to $d$ dimensions.
On the other hand, the second one does not contain any explicit mass dependence, and this makes the deformation of arbitrary functions of the latter to (A)dS easier.
Indeed, by adding proper total-derivative terms to the $Y_{i\pm1}$'s,
one gets the (A)dS counterpart of \eqref{H2nd}\,:
\be
\label{tilde H}
	\tilde H_{i} :=
	Y_{i+1}\,(Y_{i-1}-\partial_{X}\cdot\partial_{U_{i-1}})
	-\tfrac{1}{2}\,\partial_{X}\cdot(\partial_{X_{i}}-\partial_{X_{i+1}}-\partial_{X_{i-1}})\,Z_{i}\,.
\ee
Up to field redefinitions, the latter can be recast in a form where
the gauge invariance is more transparent:
\be
	\tilde H_{i}\ \approx\
	\partial_{X_{i+1}}\!\!\cdot\partial_{X_{i-1}}\,\partial_{U_{i-1}}\!\!\cdot\partial_{U_{i+1}}
	-\partial_{X_{i+1}}\!\!\cdot\partial_{U_{i-1}}\,\partial_{X_{i-1}}\!\!\cdot\partial_{U_{i+1}}\,.
\ee
Finally, the vertices in (A)dS are given by
\be
	C_{a_1\alpha_2\alpha_3}=\cK_{a_1\alpha_2\alpha_3}
	(Y_2,Y_3,Z_{1},\tilde{H}_2,\tilde{H}_3)\,.
\ee

\subsubsection*{C\ \  2 massless and 1 massive}

This case can be recovered from the previous one as the intersection between the solutions:
\be
	C_{a_1{\sst A}_2{\sst A}_3}=\cK_{a_1{\sst A}_2{\sst A}_3}
	(Y_2,Y_3,Z_{1},{\tilde H}_2,{\tilde H}_3)\,, \qquad
	C_{{\sst A}_1 a_2{\sst A}_3}=\cK_{{\sst A}_1a_2{\sst A}_3}
	(Y_1,Y_3,Z_{2},{\tilde H}_1,{\tilde H}_3)\,,
\ee
that is
\be
	C_{a_{1}a_{2}\alpha_{3}}=\cK_{a_{1}a_{2}\alpha_{3}}
	(Y_{3}, \tilde H_{1}, \tilde H_{2}, \tilde H_{3})\,.
\ee

\subsubsection*{D\ \ 3 massless}

This case is a combination of three equal mass cases:
\be
\label{massless}
	C_{a_1a_2a_3}=\cK_{a_1a_2a_3}
	(\tilde{Y}_1,\tilde{Y}_2,\tilde{Y}_3,\tilde{G})\,.
\ee
Here, the $\tilde{Y}_i$'s and $\tilde{G}$ are given by \eqref{YGtilde}
with the $\alpha_{i}$'s and the $\beta_{i}$'s satisfying
eq.~\eqref{a b cond} and cyclic permutations thereof.
Interestingly, the solutions \eqref{sol ab} of  \eqref{a b cond}
fulfill automatically also its cyclic counterparts.

\medskip

At this stage we have completed the systematic constructions of the TT parts
of the cubic interactions involving massive and massless HS fields in (A)dS.
Before considering the partially-massless cases,
let us make a few remarks.
Similarly to what happens in the (A)dS massless case \cite{Joung:2011ww},
all higher-order parts of the solutions $C^{\sst (n)}_{\sst A_{1}A_{2}A_{3}}$ are encoded via functions of simple building blocks that, being linear in $\partial_{U_{i}}$ for any $i=1,2,3$\,, describe the consistent couplings among fields of spin $1$ and $0$ only.
These results resonate with the idea that spin 1 couplings
can be used as building blocks of HS interactions \cite{Sagnotti:2010at,Taronna:2011kt}.

\subsection{Consistent cubic interactions of partially-massless HS fields}

In this section we focus on the cubic interactions in a dS background where,
besides massive and massless fields, partially-massless fields also appear.
As we have seen in Section \ref{sec: pmf},
 partially-massless fields with homogeneities
$\mu=r\in \{0\,,\,1\,,\, \ldots,\, s-1\}$ admit the gauge symmetries \eqref{pm gt}.
Then, according to eq.~\eqref{noetherproc},
the cubic interactions ought to be compatible with those gauge symmetries
leading to the following condition:
\be
\label{partiallygaugeconscond1}
\left[\,C_{\sst A_{1}A_{2}A_{3}}(\hat\delta;Y,Z)\,,
\, (U_1\!\cdot\partial_{X_1})^{r_{1}+1}\,\right]\Big|_{U_1=0}\approx 0\,.
\ee
Once again, neglecting all the $\partial_{X_{i}}^{\,2}$'s\,, $\partial_{U_{i}}\!\cdot\partial_{X_{i}}$'s
and $\partial_{U_{i}}^{\,2}$'s\,, one ends up with
\begin{multline}
\label{pmf eq}
\sum_{\sst \ell_1+\ell_2+\ell_3=r_1+1} \binom{r_1+1}{\ell_1\,\ell_2\,\ell_3}\,\Big[Y_3\,\partial_{Y_3}-Y_2\,\partial_{Y_2}-2\,Z_3\,\partial_{Z_3}+\tfrac{r_1+\m_2-\m_3}{2}\Big]_{\ell_1}\times\\
\times\left(\tfrac{\hat{\delta}}{L}\,\partial_{Y_1}\right)^{\ell_1}\left(Y_3\,\partial_{Z_2}\right)^{\ell_2}
\left(-Y_2\,\partial_{Z_3}+\tfrac{2\,\hat{\delta}}{L}\,Z_3\,\partial_{Z_3}\,\partial_{Y_1}\right)^{\ell_3}
\,  C_{\sst A_1A_2A_3}(\hat \d;Y,Z)\,=\,0\,,
\end{multline}
where \mt{[a]_{n}} is the descending Pochhammer symbol we have introduced previously.
Since \eqref{pmf eq} is an higher-order partial differential equation,
 solving it is a non-trivial task.
 However, if we restrict the attention to the $s_{1}\!-\!s_{2}\!-\!s_{3}$ couplings with fixed $s_{i}$'s, then the solution
is of the form:
\be
	C_{\sst A_{1}A_{2}A_{3}}(\hat \d;Y,Z)\,=
	\sum_{\sigma_{i}+\tau_{i+1}+\tau_{i-1}=s_{i}}
	\hspace{-10pt}
	c^{\tau_{1}\,\tau_{2}\,\tau_{3}}_{\sst A_{1}A_{2}A_{3}}(\hat\delta)\ \
	Y_{1}^{\,\sigma_{1}}\,Y_{2}^{\,\sigma_{2}}\,Y_{3}^{\,\sigma_{3}}\,
	Z_{1}^{\,\tau_{1}}\,Z_{2}^{\,\tau_{2}}\,Z_{3}^{\,\tau_{3}}\,,
\ee
where the number of the undetermined coefficients $c^{\tau_{1}\,\tau_{2}\,\tau_{3}}_{\sst A_{1}A_{2}A_{3}}$
is of the order $N\sim s_{1}\,s_{2}\,s_{3}$.
Hence, the coupling can be viewed as a vector in a $N$-dimensional space,
and eq.~\eqref{pmf eq} reduces to a set of linear equations for that vector.
Then, the consistent couplings correspond to the solution space of such linear system.
This procedure can be conveniently implemented in Mathematica.
For instance, in the case of \mt{4\!-\!4\!-\!2} couplings between two spin 4 fields
at their first partially-massless points ($\mu=1$) and a massless spin 2,
we find one ten-derivative,
two eight-derivative, two six-derivative and one four-derivative couplings:
\vspace{-10pt}

{ \footnotesize
\ba
C_{1} \eq
Y_1^4\,Y_2^4\, Y_3^2-1
2\,\hat{\delta}^2 \,Y_1^2\,Y_2^2 \left(Y_1\,Z_1+Y_2 \,Z_2\right)^2
+48\,\hat{\delta}^3\,Y_1\,Y_2 \left(Y_1\, Z_1+Y_2\, Z_2\right) Z_3
\left(2\, Y_1\, Z_1+2 \,Y_2 \,Z_2+Y_3 \,Z_3\right) \nn
&&-\,24\, \hat{\delta}^4\,Z_3^2 \left[6\, Y_1^2\, Z_1^2+6\, Y_2^2 \,Z_2^2+
4 \,Y_2 \,Y_3\, Z_2\, Z_3+Y_3^2\, Z_3^2+2 \,Y_1\, Z_1 \left(7 \,Y_2\, Z_2+2\, Y_3\, Z_3\right)\right]
+96\,\hat{\delta}^5\, Z_1\, Z_2\, Z_3^3\,,\nn
C_{2}\eq Y_1^3\,Y_2^3\,Y_3^2\,Z_3
-3\,\hat{\delta} \,Y_1^2\, Y_2^2 \left(Y_1\, Z_1+Y_2 \,Z_2\right)^2
+12\,\hat{\delta}^2\,Y_1\,Y_2 \left(Y_1\, Z_1+Y_2 \,Z_2\right) Z_3
\left(2\, Y_1\, Z_1+2\, Y_2\, Z_2+Y_3\, Z_3\right) \nn
&&-\,6\, \hat{\delta}^3\,Z_3^2 \left[6\,Y_1^2\, Z_1^2+6\, Y_2^2 \,Z_2^2+4\, Y_2\, Y_3\, Z_2\, Z_3
+Y_3^2\,Z_3^2+2\, Y_1\, Z_1 \left(7\, Y_2\, Z_2+2\, Y_3\, Z_3\right)\right]
+24\,\hat{\delta}^4\,Z_1\,Z_2\,Z_3^3\,, \nn
C_{3}\eq
Y_1^3\, Y_2^3\, Y_3 \left(Y_1\, Z_1+Y_2\, Z_2\right)
+\hat{\delta}\,Y_1^2 Y_2^2 \left(6\, Y_1^2\, Z_1^2+11\, Y_1\, Y_2\, Z_1\, Z_2+6\, Y_2^2\,Z_2^2\right)
\nn &&
-18\, \hat{\delta}^2\,
Y_1\, Y_2 \left(Y_1\,Z_1+Y_2 \,Z_2\right) Z_3 \left(2\, Y_1 \,Z_1+2 \,Y_2\, Z_2+Y_3\,  Z_3\right) \nn
&&+\,6\,\hat{\delta}^3\, Z_3^2 \left[6\,Y_1^2\, Z_1^2+2\, Y_2\, Z_2 \left(3\, Y_2\, Z_2+Y_3\, Z_3\right)+Y_1\, Z_1 \left(15\,Y_2 \,Z_2+2\, Y_3\, Z_3\right)\right]
-12\,\hat{\delta}^4\,Z_1\, Z_2\, Z_3^3\,,\nonumber
\ea
\ba
C_{4}\eq
-\,Y_1^2\,Y_2^2 \left(Y_1^2\,Z_1^2+2\,Y_1\,Y_2\,Z_1\,Z_2+Y_2^2\,Z_2^2-Y_3^2\,Z_3^2\right)\nn
&&+\,4\, \hat{\delta}\,Y_1\,Y_2 \left(Y_1\,Z_1+Y_2\,Z_2\right)Z_3 \left(2\,Y_1\,Z_1+2\,Y_2\,Z_2+Y_3\,Z_3\right)\nn
&&
-\,2\,\hat{\delta}^2\,Z_3^2 \left[6\,Y_1^2\,Z_1^2+6\,Y_2^2\,Z_2^2+4\,Y_2\,Y_3\,Z_2\,Z_3+Y_3^2\,Z_3^2+2\,Y_1\,Z_1 \left(7\,Y_2\,Z_2+2\,Y_3\,Z_3\right)\right]
+8\,\hat{\delta}^3\,Z_1\,\,Z_2\,\,Z_3^3\,,
\nn
C_{5}\eq
Y_1^2\,Y_2^2 \left(Y_1\,Z_1+Y_2\,Z_2\right) \left(Y_1\,Z_1+Y_2\,Z_2+Y_3\,Z_3\right)\nn
&&
-\,\hat{\delta}\,Y_1\,Y_2\,Z_3 \left[6\,Y_1^2\,Z_1^2+2\,Y_2\,Z_2 \left(3\,Y_2\,Z_2+2\,Y_3\,Z_3\right)+Y_1\,Z_1 \left(13\,Y_2\,Z_2+4\,Y_3\,Z_3\right)\right] \nn
&&
+\,2\,\hat{\delta}^2 \,Z_3^2 \left[3\,Y_1^2\,Z_1^2+Y_2\,Z_2 \left(3\,Y_2\,Z_2+Y_3\,Z_3\right)+Y_1\,Z_1 \left(8\,Y_2\,Z_2+Y_3\,Z_3\right)\right]
-2\,\hat{\delta}^3\,Z_1\,Z_2\,Z_3^3\,,
\nn
C_{6}\eq
Y_1\,Y_2\,Z_3 \left(Y_1\,Z_1+Y_2\,Z_2+Y_3\,Z_3\right)^2\nn
&&-\,\hat{\delta}\,Z_3^2 \left[3\,Y_1^2\,Z_1^2+3\,Y_2^2\,Z_2^2+4\,Y_2\,Y_3\,Z_2\,Z_3+Y_3^2\,Z_3^2+4\,Y_1\,Z_1 \left(2\,Y_2\,Z_2+Y_3\,Z_3\right)\right]\nn
&&+\,4\, \hat{\delta}^2\,Z_1\,Z_2\,Z_3^3\,,
\ea}
\!\!
where  for simplicity we choose $L=1$ while the $L$ dependence can be recovered replacing $\hat{\delta}$ by $\hat{\delta}/L$. 
In Appendix \ref{sec: pm ex}, we also provide
the examples of \mt{2\!-\!2\!-\!2} and \mt{3\!-\!3\!-\!2} couplings for any combinations
of the masses.

Remember that in the previous section the
solutions were obtained in a compact form
recasting the lower-derivative parts of the vertices into total derivatives.
We expect this way of simplifying couplings to work in the partially-massless cases too.
Indeed, the following class of highest-derivative couplings
\be
	C_{\sst A_{1}A_{2}A_{3}}
	=\cK_{\sst A_{1}A_{2}A_{3}}(\tilde Y_{1},\tilde Y_{2},\tilde Y_{3})\,,
	\label{pmf hd}
\ee
is also compatible with the partially-massless gauge invariance
provided the homogeneities of the fields satisfy
\be
\label{mec}
r_{i}-|\mu_{i+1}-\mu_{i-1}|\in 2\,\mathbb{N}_{\ge 0}\,.
\ee
Here the $i$-th field is at the $r_{i}$-th partially-massless
point while the other two fields have generic homogeneities
$\mu_{i+1}$ and $\mu_{i-1}$\,.
The proof of the conditions \eqref{mec} can be found in Appendix \ref{sec: hd pm}.
This implies that a partially-massless spin $s$ field can interact with two scalars
if and only if the masses of the latter satisfy \eqref{mec}.
Moreover, when all the three fields are partially-massless,
the $r_{i}$'s satisfy a triangular inequality wherein
one or three of them are even integers, $r_{i}=e_{i}$\,,
while the others
are odd, $r_{i}=o_{i}$\,:
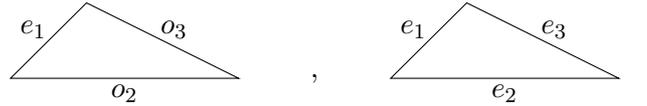
\smallskip
\begin{center}
\begin{tikzpicture}
\draw (1,1) -- (0,0) -- (3,0)-- (1,1);
\draw (0.3,0.65) node{$e_{1}$};
\draw (1.5,-0.2) node {$o_{2}$};
\draw (2.15,0.65) node {$o_{3}$};
\draw (4,0)  node {,};
\draw (6,1) -- (5,0) -- (8,0)-- (6,1);
\draw (5.3,0.65) node{$e_{1}$};
\draw (6.5,-0.2) node {$e_{2}$};
\draw (7.15,0.65) node {$e_{3}$};
\draw (8.4,0)  node {.};
\end{tikzpicture}
\end{center}
\vspace{-8pt}
Note that this triangular inequality is not imposed on the spins but on the
homogeneities $\mu_{i}$'s which are related to the masses according to eq.~\eqref{mass}.
The conditions \eqref{mec} reveal the systematics of the partially-massless vertices.
Let us recall that whenever one massless field takes part to generic massive interactions,
the vertices split into two categories
according to whether the other two fields have equal or different masses.
The condition \eqref{mec} is a
generalization of this pattern to the partially-massless cases.
We expect that, as in the massless case (see the 1 massless and 2 massive case of Section \ref{2 massive: equal masses d}),
whenever \eqref{mec} holds
we have $\tilde G$-like building blocks
on top of the $\tilde Y_{i}$'s, otherwise,
one is left with $\tilde Y_{2}$,
$\tilde Y_{3}$, $\tilde H_{2}$-like and $\tilde H_{3}$-like
building blocks.\footnote{In fact, it is even possible that the $\tilde G$ and the $\tilde H_{i}$'s
defined for the massless case still work for the partially-massless cases. However, checking it requires non-trivial computations and we postpone
this issue for future work.}

Despite at present we lack the building blocks for the interactions
involving partially-massless fields, we can do a systematic analysis of
the zero-th order parts of the solution $C^{\sst (0)}_{\sst A_{1}A_{2}A_{3}}$\,.
In this case,  eq.~\eqref{pmf eq} reduces to
\be
(Y_{2}\partial_{Z_{3}}-Y_{3}\partial_{Z_{2}})^{r_1+1}\,
C^{\sst (0)}_{\sst A_{1}A_{2}A_{3}}(Y,Z)=0\,,
\ee
whose corresponding solutions are given by
\be
C^{\sst (0)}_{\sst A_{1}A_{2}A_{3}}=
\sum_{m_{2}+m_{3}\le r_{1} }
Z_2^{\,m_2}\,Z_3^{\,m_3}\
\bar C^{\sst (0)\,m_2m_3}_{\sst A_{1}A_{2}A_{3}}(Y_1,Y_2,Y_3, Z_{1}, G)\,.
\ee
Notice that, compared to the massless case,
some factors of $Z_{2}$ and $Z_{3}$ are also allowed
increasing the number of possible ways of writing the couplings.
However, when restricted to particular
couplings with fixed spins,
the number of solutions may be smaller than in the massless case.

\section{St\"uckelberg formulation}
\label{sec: Stuckelberg}

In this section, we first
consider the free theories of massive and massless HS fields in the St\"uckelberg
formalism, and then extend the discussion to the cubic level.
Once again, we focus on the TT parts of the vertices.
It is worth stressing that, as in the massless case,
 working with a gauge invariant
description for massive fields might give us a recipe in order to fix the remaining parts of the vertices. Moreover, as mentioned in the Introduction, St\"{u}ckelberg formulation represents a convenient framework in order to study the massless limit of massive theories.

\subsection{Free St\"{u}ckelberg fields from dimensional reduction}
\label{sec: Stuckfields}

The St\"{u}ckelberg description of massive HS fields can be conveniently
obtained through dimensional reduction of a $(d+1)$-dimensional massless theory.
In the following we first provide the example of a spin 1 field
and then generalize it to arbitrary-spin fields.

\subsubsection*{Spin 1}

Let us consider the theory of a massive spin 1 field
$a_\mu$:
\be
\label{massivephot}
S=-\tfrac12\int d^d x \sqrt{\epsilon\,g}\,\left(\,\tfrac{1}{2}\,f_{\mu\nu}\,f^{\mu\nu}
+m^2\,a_\mu\, a^\mu\right)\,,
\ee
where $f_{\mu\nu}=\partial_{\mu}\,a_{\nu}-\partial_{\nu}\,a_{\mu}$\,.
Because of the mass term this theory is not gauge invariant and describes
the propagation of the DoF associated to a massive spin 1 particle. Notice that,
performing the limit $m\rightarrow 0$ at this level,
one ends up with a massless spin $1$ field, loosing one DoF.
On the other hand, before taking the massless limit, one can introduce a new scalar field $\alpha_{\sst 1}$ via the St\"uckelberg shift:
\be
	a_{\mu}=\alpha_{{\sst 0}\,\mu}+\tfrac1m\,\partial_{\mu}\,\alpha_{\sst 1}\,,
	\label{s1 shift}
\ee
in such a way that the resulting action acquires the gauge symmetries
$\delta \alpha_{{\sst 0}\,\mu}=\partial_\mu\,\varepsilon_{\sst 0}$
and $\delta\alpha_{\sst 1}=-m\,\varepsilon_{\sst 0}$\,.
Then, the action becomes
\be
\label{AdS s1}
	S=-\tfrac12\int d^d x\,\sqrt{\epsilon\,g}\,
 \left[\,\tfrac{1}{2}\,f_{{\sst 0}\,\mu\nu}\,f_{\sst 0}^{\mu\nu}
 	+ m^{2}\,\alpha_{{\sst 0}\,\mu}\,\alpha_{\sst 0}^{\mu}
	+ \partial_\mu\alpha_{\sst 1}\,\partial^\mu\alpha_{\sst 1}
	+2\,m\,\alpha_{\sst 0}^{\mu}\,\partial_{\mu}\alpha_{\sst 1}\,\right],
\ee
which, in the massless limit, describes a massless spin 1 and
spin 0 field, preserving the number of DoF.

The above discussion can be restated in the ambient space formalism.
First of all, one can obtain the St\"uckelberg action
through radial reduction of the massless ambient-space one:
\be\label{spin1 amb}
	S=-\tfrac14\int d^{d+1} X\,\delta\big(\sqrt{\epsilon\,X^{2}}-L\big)\
	F_{\sst MN}\,F^{\sst MN}\,,
\ee
where the spin-1 field is homogeneous and tangent:
\be
	(X\cdot\partial_{X}+\mu+1)\,A_{\sst M}=0\,, \qquad
	X^{\sst M}\,A_{\sst M}=0\,.
\ee
The tangentiality condition implies $A_{d}=0$ and, after the identification $A_{\mu}= a_{\mu}$\,, one recovers the action \eqref{massivephot}\,.
Remember that the gauge symmetry \mt{\d \bm A_{\sst M}\,=\,\partial_{\sst M} E}
of the action \eqref{spin1 amb}
is incompatible with the tangentiality condition when $\m$ is different from zero.
On the other hand, one can insist on a gauge invariant formulation also for $\m\neq 0$ provided the tangentiality condition is relaxed. In this case one has to promote the tangent field $A_{\sst M}$ to a generic one $\bm A_{\sst M}$ with non-vanishing radial part:
$X^{\sst M}\,\bm A_{\sst M} \neq 0$\,.
Then, after identifying
\be
	\bm A_{\mu}=\alpha_{{\sst 0}\,\mu}\,, \qquad
	\bm A_{d}=\alpha_{\sst 1}\,,
\ee
in \eqref{spin1 amb}, one recovers the St\"uckelberg action \eqref{AdS s1}.
Moreover, the usual St\"uckelberg symmetry is obtained
by decomposing \mt{\d \bm A_{\sst M}\,=\,\partial_{\sst M} E} into
its tangent and radial parts.
Such decomposition can be also carried out in terms of
 ambient-space fields as
$\bm A_{\sst M}=A_{\sst 0\,M}+ A_{\sst 1}\,L\,X^{\sst M}/X^{2}$\,,
ending up with
\be
\d A_{\sst 0\,M}=
\left({\d}_{\sst M}^{\sst N}-\tfrac{X_{\sst M}\,X^{\sst N}}{X^2}\right)
\partial_{\sst N} E\,,\qquad
\d A_{\sst 1}=\tfrac{1}{L}\,X^{\sst M}\, \partial_{\sst M} E=-\tfrac{\m}{L}\, E\,.
\ee
Finally, the St\"uckelberg shift \eqref{s1 shift} can be realized as well at the ambient space level as
\be
	A_{\sst M}=\left(\delta^{\sst N}_{\sst M}+\tfrac 1\mu\,\partial_{X^{M}}\,X^{\sst N}\right)
	\bm A_{\sst N}\,.
\ee

\subsubsection*{General spins}

In the previous sections we have discussed how
the quadratic action of massive HS fields
\eqref{AdS act} can be
obtained through radial reduction of the ambient-space massless one \eqref{freeAdS}.
In the following, we introduce St\"uckelberg fields
promoting the tangent ambient-space fields $\Phi$
to generic unconstrained ones $\bm\Phi$\,.
In this case, after the radial reduction, one is led to
\be
\label{genfuncd+1}
\bm\Phi(R,x;v,u)=\big(\tfrac RL\big)^{u\cdot\partial_{u}+v\partial_{v}-2-\mu}
{\bm \varphi}(x;v,u)\,.
\ee
The $(d+1)$-dimensional tensor fields ${\bm\varphi}$ can be expanded
into $d$-dimensional ones of different ranks as
\be
	\bm\varphi(x;v,u):=
	\sum_{r=0}^{\infty}\,\frac{v^{r}}{r!}\,\varphi_{\sst r}(x,u)\,,
\ee
where the components $\varphi_{\sst r}$ with $r=1,2, \ldots$
correspond to the St\"uckelberg fields.
Although the action and the corresponding field equations for this system
stay the same as in the unitary gauge (\mt{\varphi_{\sst r\ge 1}=0}),
having relaxed the tangentiality condition, the theory acquires the gauge symmetries:
\be
	\d^{\sst (0)}\,\bm\Phi(X,U)=U\cdot\partial_{X}\,\bm E(X,U)\,,\label{Stuck gt}
\ee
 with gauge parameters:
\be
\label{genfuncd+1}
\bm E(R,x;v,u)=\big(\tfrac RL\big)^{u\cdot\partial_{u}+v\partial_{v}-\mu}
{\bm \varepsilon}(x;v,u)\,.
\ee
The $(d+1)$-dimensional gauge parameters ${\bm\varepsilon}$
can be expanded into $d$-dimensional ones as
\be
	\bm\varepsilon(x;v,u):=
	\sum_{r=0}^{\infty}\,\frac{v^{r}}{r!}\,\varepsilon_{\sst r}(x,u)\,.
	\label{St gt}
\ee
Let us mention once again that, depending on the kind of formulation,
the gauge fields as well as the gauge parameters can have trace constraints. However,
 since we focus on the TT part of the Lagrangian,
they are not relevant for our discussion.

The radial reduction considered so far can be also
restated in terms of ambient-space quantities as
\be
	\bm\Phi
	:=\sum_{r=0}^{\infty}\,\frac1{r!}\,\big(\tfrac{L\,X\cdot U}{X^{2}}\big)^{r}\,
	\Phi_{r}\,,\qquad
	\bm E
	:=\sum_{r=0}^{\infty}\,\frac1{r!}\,\big(\tfrac{L\,X\cdot U}{X^{2}}\big)^{r}\,
	E_{r}\,,
\ee
where
\be
	\Phi_{\sst r}=\big(\tfrac RL\big)^{u\cdot\partial_{u}+2(r-1)-\mu}\,
	\varphi_{\sst r}\,,\qquad
	E_{\sst r}=\big(\tfrac RL\big)^{u\cdot\partial_{u}+2r-\mu}\,
	\varepsilon_{\sst r}\,.
\ee
Decomposing the gauge transformation \eqref{Stuck gt} into its tangent and
 normal parts, one gets
\be
	\delta^{\sst (0)}\,\Phi_{\sst r}=
 \Big[\,U\cdot\partial_{X}+(\m-2\,r)\,\tfrac{X\cdot U}{X^2}\,\Big]\,E_{\sst r}+\tfrac{L}{X^2}\,
 \Big[\,U^{2}-\tfrac{(X\cdot U)^{2}}{X^2}\Big]\,E_{\sst r+1}
	-\tfrac{r\,(\mu-r+1)}{L}\,E_{\sst r-1}\,.
\ee
From these gauge transformations, one can see that, when $\mu\neq 0$\,,
all $\Phi_{\sst r\ge1}$'s can be gauge fixed to zero, going back to the unitary gauge.
On the other hand, in the massless limit,
one can gauge fix to zero only the $\Phi_{\sst r\ge2}$'s, ending up
with a massless field $\Phi_{\sst 0}$ together with a massive one $\Phi_{\sst 1}$
(corresponding to \mt{\mu=-2})\,.
This differs from what happens in flat space
where none of the $\Phi_{\sst r}$'s can be gauged away.
In other words, if we consider the massless limit of the flat-space Lagrangian of massive HS fields
 \`a la St\"uckelberg,
it decomposes into the sum of massless ones: e.g.
a massive spin $s$ reduces to massless spin $s,s\!-\!1$, down to $0$ fields.
Therefore, the total number of physical DoF stays the same as in the massive case.

Similarly to the spin 1 case, it is possible to restate the St\"{u}ckelberg shift in terms of ambient-space quantities as
\be
\label{StuckshiftAdS}
\Phi(X,U)=\sum_{r=0}^\infty\, \frac{a_{r}}{r!}\,(U\!\cdot\partial_X)^r\,W^r\,
\bm\Phi(X,U)\,,\qquad
	W:=\tfrac1L\,X\cdot\partial_U\,.
\ee
Demanding either the compatibility with the unitary gauge, \emph{i.e.}  $\delta^{\sst (0)}\,\Phi=0$ under \eqref{Stuck gt}, or with the tangentiality condition \eqref{t con},
the coefficients $a_{r}$'s are fixed as
\be
	a_r=\frac{L^{r}}{[\mu]_r}\,.
\ee
In the flat limit one gets
\be
\label{Wflat}
	W=\hat N\cdot\partial_{U}\,,
	\qquad a_{r}=\frac{(-1)^r}{M^{r}}\,.
\ee
Notice that both the AdS and the flat-space results present a pole
in the massless limit, while in dS further (partially-massless) poles
appear at $\mu=1,\ldots s-1$.

\subsection{Cubic interactions of HS fields
with St\"{u}ckelberg symmetries}
\label{Cubic interactions Stuckelberg}

In this section we present the St\"{u}ckelberg formulation of the consistent cubic interactions of
massless and massive HS fields. Once again we restrict the attention to the TT parts of such vertices which are provided in terms of operators and fields in the ambient space formalism.
The key point is that in this case the dependence on $X^{\sst M}$ cannot be neglected anymore and
the possible $(d+1)$-dimensional cubic vertices are more general than the unitary gauge ones (\ref{cubicact1}).
In particular, as in Section~\ref{sec: cubic unitary}, we can simplify
the ansatz for the cubic couplings making use of all scalar operators:
\ba
	S^{\sst (3)} \eq \frac1{3!}\int d^{d+1}X\ \delta\Big(\sqrt{\epsilon\,X^2}-L\Big)\
	C_{\sst A_{1}A_{2}A_{3}}(\hat\delta;Y, Z, W) \times \nn
	&& \hspace{50pt} \times\ \bm\Phi^{\sst A_{1}}(X_1,U_1)\
	\bm\Phi^{\sst A_{2}}(X_2,U_2)\
	\bm\Phi^{\sst A_{3}}(X_3,U_3)\ \Big|_{^{X_i=X}_{U_i=0}}\,,
\ea
where, compared to the unitary gauge case, we have introduced the additional
scalar quantities
\be
\label{YZW var}
W_i=\tfrac1L\,X_{i}\cdot\partial_{U_i}\,.
\ee
Gauge invariance under \eqref{Stuck gt} imposes the following equation:
\be
\label{gaugeconsc2}
\big[\,C_{\sst A_{1}A_{2}A_{3}}(\hat\delta;Y,Z,W)\,,\,U_i\cdot\partial_{X_i}\,\big]\approx 0\,,
\ee
which, once again can be solved modulo the Fierz system.
However, in this case the non-commutativity between $Y_{i}$ and $W_{i+1}$
makes the analysis more involved.
On the other hand, one can get the cubic vertices for the St\"uckerberg fields
by exploiting the St\"uckelberg shift \eqref{StuckshiftAdS}. Let us stress that we have explicitly checked the equivalence between the latter approach and resolution of eq.~\eqref{gaugeconsc2}.
The non-commutativity problem arises in this approach as well,
and in order to deal with it we choose an ordering prescription
where all the $W_{i}$'s are placed
after the $Y_{i}$'s and the $Z_{i}$'s.
For this purpose, it is convenient to introduce a new variable $w$ and write the St\"uckelberg shift \eqref{StuckshiftAdS} as
\be
\label{StuckshiftAdS2}
\Phi(X,U)=e^{U\cdot\partial_{X}\,\partial_{w}}\,P(w\,W)\,\bm\Phi(X,U)\,\Big|_{w=0}\,,
\ee
where
\be
	P(z)=\sum_{r=0}^{\infty}\frac{(L\,z)^{r}}{r!\,[\mu]_{r}}=
	\,_{\sst 0} F_{\sst 1}(-\mu\,;\,-L\,z)\,.
\ee
Then, the cubic vertices in the St\"uckelberg formulation can be obtained by shifting the unitary gauge ones as
\ba
\label{STshifCV}
	C_{\sst A_{1}A_{2}A_{3}}=
	\cK_{\sst A_{1}A_{2}A_{3}}(\bm Y,\bm Z)
		\,P(w_{1}\,W_{1})\,P(w_{2}\,W_{2})\,P(w_{3}\,W_{3})\,	
		\Big|_{w_{i}=0}\,,
\ea
where the $\bm Y_{i}$'s and the $\bm Z_i$'s are given by
\ba
	&& \bm Y_{i} :=
	Y_{i}\ e^{U_{i}\cdot\partial_{X_{i}}\,\partial_{w_{i}}}\,\big|_{U_{i}=0}
	\,=\,Y_{i}+
	\partial_{X_{i}}\!\cdot\partial_{X_{i+1}}\,\partial_{w_{i}}\,, \nn
	&& \bm Z_{i}:= Z_{i}\ e^{U_{i}\cdot\partial_{X_{i}}\,\partial_{w_{i}}}\,
	\big|_{U_{i}=0}\, =\,
	Z_{i}+
	\partial_{U_{i+1}}\!\!\cdot\partial_{X_{i-1}}\,\partial_{w_{i-1}}
	+\partial_{U_{i-1}}\!\!\cdot\partial_{X_{i+1}}\,\partial_{w_{i+1}}\nn
	&&\hspace{148pt}
	+\ \partial_{X_{i+1}}\!\!\cdot\partial_{X_{i-1}}\,\partial_{w_{i+1}}\,
	\partial_{w_{i-1}}\,.\label{deformation}
\ea
 Depending on the number of massless fields involved in the interactions,
 one recovers a dependence of the vertices on the variables
$\tilde{\bm Y}_{i}$\,, $\tilde{\bm G}$ and $\tilde{\bm H}_{i}$\,,
which are defined as in eq.~\eqref{deformation}
starting from the $\tilde{Y}_{i}$'s\,, $\tilde{G}$ and the $\tilde{H}_{i}$'s\,, respectively.

\subsection{Massless limit}
\label{sec: massless}

As mentioned in the Introduction, the relation between massless and massive HS theories
is of particular interest with regards to the possibility of having a better
understanding of both ST and HS gauge theory in (A)dS.
Although it is difficult to realize a mass generation mechanism for HS fields,  one might
get some hints for that by studying the massless limit of massive theories.

In the previous section we have shown that
the cubic vertices in the St\"uckelberg formulation are
given by arbitrary functions $\cK_{\sst A_{1}A_{2}A_{3}}$ of
the $\tilde{\bm Y}_{i}$'s and of the $\tilde{\bm Z_{i}}$'s\,.
Moreover, when some of the fields are massless,
$\tilde {\bm G}$ and the $\tilde {\bm H}_{i}$'s also appear.
In order to properly study the behavior of such vertices
in the limit where some of the masses go to zero,
one should know in principle how the coupling function
$\cK_{\sst A_{1}A_{2}A_{3}}$ scales.
However, as we will see in the following,
interesting information can be also extracted considering generic
behaviors in this limit.
For simplicity, we consider the case where
all the mass parameters of the theory scale uniformly with a mass scale $\mu$\,:
\be
\label{unifscal}
\mu_i=\n_i\,\m\,.
\ee
Since the massless limit depends on the background,
we analyze the AdS and the flat-space cases  separately.

\paragraph{AdS case}

As we have seen in Section \ref{sec: Stuckfields}, in the massless limit $\mu\to 0$\,
 one can gauge fix all the lower spin components up to spin $s\!-\!2$\, ending up with:
\be
\bm\Phi=\Phi_{\sst 0}+\tfrac{L\,X\cdot U}{X^{2}}\,\Phi_{\sst 1}\,,
\ee
where $\Phi_{\sst 0}$ and $\Phi_{\sst 1}$ are a spin $s$ massless field
and a spin $s\!-\!1$ massive field, respectively.
In this way, the St\"uckelberg shift \eqref{StuckshiftAdS2} simplifies to
\be
\Phi=\left(1-\tfrac{1}{\mu}\,U\cdot\partial_{X}\,W\right)\,\bm\Phi\,.
\ee
Hence, the couplings \eqref{STshifCV}  can be expanded as
\ba
	&& C_{\sst A_{1}A_{2}A_{3}} =
	\cK_{\sst A_{1}A_{2}A_{3}}(Y,Z)+
	\frac1\mu\sum_{i=1}^{3}\,
	\cK^{\sst [i]}_{\sst A_{1}A_{2}A_{3}}(\hat\delta;Y,Z)\,W_{i}\nn
	&&\quad +\,\frac1{\mu^{2}}
	\sum_{i=1}^{3}\,
	\cK^{\sst [i+1,i-1]}_{\sst A_{1}A_{2}A_{3}}(\hat\delta;Y,Z)\,
	W_{i+1}\,W_{i-1}
	+\frac1{\mu^{3}}\,
	\cK^{\sst [1,2,3]}_{\sst A_{1}A_{2}A_{3}}(\hat\delta;Y,Z)
	\,W_{1}\,W_{2}\,W_{3}\,,
\ea
where the $\cK^{\sst [\,\ldots]}_{\sst A_{1}A_{2}A_{3}}$'s
are given by successive commutators of $\cK_{\sst A_{1}A_{2}A_{3}}$\,:
\be
	\cK^{\sst [\ldots,i]}_{\sst A_{1}A_{2}A_{3}}
	:=\big[\,\cK^{\sst [\ldots]}_{\sst A_{1}A_{2}A_{3}}\,,\,
	-\tfrac1{\nu_{i}}\,U_{i}\cdot\partial_{X_{i}}\,\big]\,.
\ee
In the $\mu\rightarrow 0$ limit, the leading terms are massive couplings
of the form $\cK^{\sst [1,2,3]}_{\sst A_{1}A_{2}A_{3}}$
involving all the massive spin $s-1$ components $\Phi_{\sst 1}=W\,\bm\Phi$\,.
On the other hand, if some of leading parts of the couplings are absent,
then the dominant ones contain less number of $W_{i}$'s
and consequently the interactions involve the corresponding massless fields.

\paragraph{Flat-space case}

The situation in flat space is rather different from the one in AdS.
First of all, in the massless limit one can not gauge fix the St\"uckelberg fields to zero
so that the latter become all massless fields.
Moreover, since the non-commutativity problem is absent, the St\"uckelberg vertices \eqref{STshifCV}
can be simplified performing the $w_{i}$-contractions as
\ba
	C_{\sst A_{1}A_{2}A_{3}}=
	\cK_{\sst A_{1}A_{2}A_{3}}(\hat{Y},\hat{Z})\,,
\ea
where
 \ba
 \label{YZredu}
\hat{Y}_i \eq
y_{i}-\tfrac{M_i^2+M_{i+1}^2-M_{i-1}^2}{2\,M_i}\,\partial_{v_i}\,,\nn
\hat{Z}_i \eq z_i+\tfrac{1}{M_{i-1}}\,y_{i+1}\,\partial_{v_{i-1}}-\tfrac{1}{M_{i+1}}\,y_{i-1}\,\partial_{v_{i+1}}+\tfrac{M_i^2+M_{i-1}^2-M_{i+1}^2}{2\,M_{i+1}\,M_{i-1}}\,
\partial_{v_{i+1}}\,\partial_{v_{i-1}}\,.
\ea
Here we have also performed the dimensional reduction
providing the building blocks $\hat Y$ and $\hat Z$
in terms of the $d$-dimensional intrinsic ones:
\be
y_i:=\partial_{u_{i}}\!\cdot\partial_{x_{i+1}}\,,\qquad
z_i:=\partial_{u_{i+1}}\!\!\cdot\partial_{u_{i-1}}\,.
\ee
Then, under the assumption \eqref{unifscal},
one can observe the following behavior:
\be
\label{dom y h}
	\hat{Y}_i=y_{i}+\cO(\mu)\,,\qquad\quad
\mu\,\hat{Z}_i=\tfrac{1}{\nu_{i-1}}\,y_{i+1}\,\partial_{v_{i-1}}-\tfrac{1}{\nu_{i+1}}\,y_{i-1}\,\partial_{v_{i-1}} +\cO(\mu)\,,
\ee
in the $\mu\rightarrow 0$ limit.
Notice that the dominant terms contained in the $\hat{Z}_i$'s lead to consistent massless interactions
and involve at least one St\"uckelberg field.
The terms proportional to the $z_i$'s\,, which can violate the gauge invariance,
 are contained in the subdominant $\cO(\mu)$ part. Similarly, the variables $\hat{G}$ and $\hat{H}_i$'s behave as
\ba
\label{tilde g limit}
	&&\hat{G}=g
	 +\tfrac{\nu_{2}^{2}+\nu_{3}^{2}-\nu_{1}^{2}}{2\,\nu_{2}\,\nu_{3}}\,y_{1}\,\partial_{v_{2}}\,\partial_{v_{3}}+\text{cyclic},\nn
		&&\hat{H}_i=y_{i+1}\,y_{i-1}+\cO(\mu)\,,
\ea
where $g:=y_{1}\,z_{1}+y_{2}\,z_{2}+y_{3}\,z_{3}$\,.
Finally, the generic leading parts of the massive cubic vertices
can be obtained by simply replacing all the variables by their leading terms (\ref{dom y h}\,,\,\ref{tilde g limit}).
The resulting vertices involve only the $y_i$'s and $g$ together with the $\partial_{v_i}$'s
which encode the contribution of the St\"uckelberg fields.
Hence, they are consistent with the gauge symmetries of the massless theory.
For the sake of completeness, one should also analyze the cases where
some of the leading parts cancel.
This analysis can be found in Appendix \ref{sec: mless limit}.

\section{Discussion}
\label{sec: discussion}

In this paper we have obtained the solutions
to the cubic-interaction problem for massive and partially-massless
HS fields in a constant-curvature background.
This has been achieved through a dimensional reduction of a $(d+1)$-dimensional massless theory with a delta function insertion in the action.\footnote{
Actually, any integrable function of the same argument is good.
In particular one can consider the insertion of an Heavyside theta function, that is tantamount to introducing a cut-off for the diverging radial integral, or similarly, a boundary for the ambient space.
Then, the total-derivative terms appearing in the interactions play the role of boundary actions
which have to be taken into account whenever the base space-time has a non-empty boundary. See \cite{Joung:2011xb}
for the recent construction of boundary actions for the free theory of massless HS fields in AdS.}
For simplicity, the entire construction has been carried out focusing on the TT part of the Lagrangian.
We expect that the completion of such vertices
can be performed within the St\"uckelberg formulation,
adding divergences and traces of the fields together with possible auxiliary fields.

Our studies are mainly motivated by ST whose very consistency rests on the presence of infinitely many HS fields.
Conversely, string interactions may provide useful information on the
systematics of the consistent HS couplings.
In \cite{Taronna:2010qq,Sagnotti:2010at},
cubic vertices of totally-symmetric tensors belonging
to the first Regge trajectory of the open bosonic string were investigated.
Those vertices are encoded in the following generating function:
\ba
\label{StringVer}
	&& \tfrac1{\sqrt{G_{N}}}\,
	\cK_{\sst A_1 A_2 A_3}=
	i\,\frac{g_o}{\alpha{\sst'}}\,{\rm Tr}\left[T_{\sst A_1}\,T_{\sst A_2}\,	
	T_{\sst A_3}\right]\,
	\exp\left(i\sqrt{2\alpha'}\,(y_1+y_2+y_3)+z_1+z_2+z_3 \right)\nn
	&&\qquad +\,i\,\frac{g_o}{\alpha{\sst'}}\,
	{\rm Tr}\left[T_{\sst A_2}\,T_{\sst A_1}\,T_{\sst A_3}\right]\,
	\exp\left(-i\sqrt{2\alpha'}\,(y_1+y_2+y_3)+z_1+z_2+z_3 \right)\,,\quad
\ea
where $G_{N}$ denotes Newton's constant, $g_o$ the open string coupling constant and
$\alpha'$ the inverse string tension related to the  masses of the string states as
\be
\label{spec st}
	M^2\,\varphi^{\sst (s)}=\frac{s-1}{\alpha'}\,\varphi^{\sst (s)}\,.
\ee
Remarkably, the Taylor coefficients of the exponential function
and the spectrum \eqref{spec st} nicely combine
to reproduce the right vertices belonging to the classification considered in Section \ref{sec:Consistent cubic} (the details can be found in Appendix \ref{sec: ST coup}).
In this respect, it would be interesting to understand how the exponential function \eqref{StringVer}
fits in with other ST properties and what its AdS counterpart may be.
In particular, we believe that the choice of the exponential is crucial for the global symmetries
as well as for the planar dualities of the theory.
Let us mention however that in AdS an exponential couplings of the form:
\be
	e^{i\,\sqrt{2\alpha'}\,(\tilde Y_{1}+\tilde Y_{2}+\tilde Y_{3})+Z_{1}+Z_{2}+Z_{3}},
\ee
where the $\tilde Y_{i}$'s are any total-derivative deformations of the $Y_{i}$'s\,,
is incompatible with any spectrum containing a massless spin 1 field, reflecting the difficulties encountered in quantizing ST on (A)dS backgrounds \cite{Tseytlin:2002gz,Bonelli:2003zu,Sagnotti:2003qa}. From this perspective it is conceivable that a better understanding of the global symmetries of ST as well as of their implementation at the interacting level may shed some light on this issue.
Moreover, coming back to flat space, St\"uckelberg fields can be also introduced into the vertices
of the first Regge trajectory \eqref{StringVer} using the $\hat Y_{i}$'s and the $\hat Z_{i}$'s in place of the $y_{i}$'s and the $z_{i}$'s. Clarifying their role is potentially interesting in view of a deeper comprehension of the states present in the lower Regge trajectories, to whom the St\"uckelberg fields may be related.

In the present paper
we also studied the massless limit of the interactions focusing on
the scaling of the masses leaving aside the behavior of the coupling functions.
However, a complete analysis should take into account such behavior,
which can depend in principle on more than one scale.
For instance, conventional symmetry breaking scenarios, where masses are
generated through interactions, need at least two mass scales:
one related to the vev of the scalars (or more generally even-spin fields) and
the other related to the coupling constants of the symmetric theory.
Therefore, in order to properly address the mass generation issue in HS theories,
it is necessary to have some control on the higher-order interactions and possibly on the full nonlinear theory.
In this respect, if ST draws its origin from the spontaneous breaking of a HS gauge symmetry,
one could expect that, besides the string tension, some new mass scales
appear in the underlying fundamental description.

Finally, in order to complete the classification
of the cubic interactions, it would be necessary to study the gauge deformations induced by
the latter. Besides allowing us to address the issues related to
the gravitational and the electromagnetic minimal couplings of HS fields,
this would possibly shed some light on HS algebras and on their implications.
Moreover, in order to get further insights into ST it would be interesting
to extend the present analysis to fermionic and mixed-symmetry fields, and eventually to higher-order interactions along the lines of \cite{Sagnotti:2010at,Taronna:2011kt,Dempster:2012vw}. Last, let us stress that the ambient-space framework has proven particularly suitable in order to deal with interactions in curved backgrounds. For this reason, it is conceivable that this approach would give new insights into the AdS/CFT correspondence in relation to HS theories.

\acknowledgments{
We are grateful to
M. Bianchi,
D. Francia
and
K. Mkrtchyan
for helpful discussions,
and especially to
A. Sagnotti for reading the manuscript and for key suggestions.
The present research was supported in part
by Scuola Normale Superiore,
by INFN and
by the MIUR-PRIN contract 2009-KHZKRX.
}

\appendix

\section{Useful identities}
\label{sec: identities}
This appendix contains some identities and mathematical tools used in our construction of the cubic vertices.
Basic commutation relations among the operators $\eqref{Y and Z}$ are
\ba
\label{commrelations}
\big[\,Y_i\,,\,U_j\!\cdot\partial_{X_j}\,\big] \eq
\delta_{ij} \
\partial_{X_{i}}\cdot\partial_{X_{i+1}}\,,\nn
\big[\,Z_i\,,\,U_{i+1}\!\cdot\partial_{X_{i+1}}\,\big]
\eq \partial_{X}\cdot\partial_{U_{i-1}}-Y_{i-1}\,\,, \nn
 \big[\,Z_i\,,\,U_{i-1}\cdot\partial_{X_{i-1}}\big]
\eq Y_{i+1}\,,\nn
\big[\,X_i\!\cdot\partial_{U_i}\,,\,F(Y,Z)\,\big] \eq -Z_{i+1}\,\partial_{Y_{i-1}}\,F(Y,Z)\,,\nn
 \big[\,X_i\!\cdot\partial_{X_i}\,,\,F(Y,Z)\,\big] \eq -Y_{i-1}\,\partial_{Y_{i-1}}\,F(Y,Z)\,,\nn
 \big[\, F(Y,Z)\,,\,U_i\!\cdot\partial_{U_i}\,\big] \eq \left(Y_i\,\partial_{Y_i}+Z_{i+1}\,\partial_{Z_{i+1}}+Z_{i-1}\,\partial_{Z_{i-1}}\right) F(Y,Z)\,.
\ea
Here $i, j$ are defined modulo 3: $(i,j)\cong (i+3,j+3)$.
Another identity used throughout all the paper concerns the commutator between an arbitrary function $f(A)$ of a linear operator $A$ and an other linear operator $B$\,:
\be
\label{fBcomm}
[\,f(A)\,,\,B\,]=\sum_{n=1}^\infty\, \frac{1}{n!}\, ({\rm ad}_A)^{n}B\, f^{(n)}(A)\,,
\ee
where ${\rm ad}_A\,B=[\,A\,,\,B\,]$ and $f^{(n)}(A)$ denotes the $n$-th derivative of $f$ with respect to $A$. In order to prove the latter formula, we represent $f(A)$ as a Fourier integral
so that the commutator appearing in \eqref{fBcomm} can be written as
\be
\label{fBcomm2}
[\,f(A)\,,\,B\,]=\int^\infty_{-\infty} dt\  [\,e^{itA}\,,\,B\,]\,f(t)\,.
\ee
Using the well-known identity
\be
e^{itA}\, B\, e^{-itA}=\sum_{n=0}^\infty\, \frac{(it)^n}{n!}\, ({\rm ad}_A)^{n}\,B\,,
\ee
eq. \eqref{fBcomm2} becomes
\be
[\,f(A)\,,\,B\,]=\sum_{n=1}^\infty\, \frac{1}{n!}\, ({\rm ad}_A)^{n}\,B\,\int^\infty_{-\infty} dt\, (it)^n\, e^{itA}\,f(t)
=\sum_{n=1}^\infty\, \frac{1}{n!}\, ({\rm ad}_A)^{n}\,B\, f^{(n)}(A)\,.
\ee
Since our vertices are arbitrary functions of commuting operators, formula \eqref{fBcomm} applies independently to each of them.

\section{\mt{2\!-\!2\!-\!2} and \mt{3\!-\!3\!-\!2} partially-massless interactions}
\label{sec: pm ex}

This appendix is devoted to the examples of  \mt{2\!-\!2\!-\!2} and \mt{3\!-\!3\!-\!2}
couplings involving at least one partially-massless field. The results are collected in the following tables in which we organized the solutions for given $(\m_1,\m_2,\m_3)$ according to the maximal number of derivatives denoted by $\partial$\,. Arbitrary linear combinations of such solutions are consistent cubic couplings. Let us mention that in all cases we have checked, the number of solutions for the interactions involving massive fields is enhanced for those mass values satisfying eq.~\eqref{mec}. For brevity, we consider such cases only in the \mt{2\!-\!2\!-\!2} table (see e.g. $(\m_1,\m_2,\m_3)=(1,1\,,2)\,,\,(1,\m_3+1,\m_3)\,,\,(1,\m_3-1,\m_3)$\,). Moreover for simplicity we choose $L=1$ while the $L$ dependence can be recovered replacing $\hat{\delta}$ by $\hat{\delta}/L$.

\begin{center}
\textbf {2-2-2 Couplings}
{\footnotesize
\begin{longtable}[c]{|c|c|c|}
\hline
\scriptsize $(\m_1\,,\,\m_2\,,\,\m_3)$ & \scriptsize $\partial$ &  \scriptsize Couplings  \\
\hline
 \scriptsize $(1\,,\,1\,,\,1)$ & \scriptsize 6 & \scriptsize $ Y_1^2\,Y_2^2\,Y_3^2-\tfrac{1}{4}\,\hat{\delta}^2\,(Y_1\,Y_2\,Z_1\,Z_2+\text{cycl.})+\tfrac{1}{4}\,\hat{\delta}^3\,Z_1\,Z_2\,Z_3$  \\
\cline{2-3}
& \scriptsize 4 & \scriptsize $(Y_1^2\,Y_2\,Y_3\,Z_1+\text{cycl.})-\hat{\delta}\,(Y_1\,Y_2\,Z_1\,Z_2+\text{cycl.})+\tfrac{3}{4}\,\hat{\delta}^2\,Z_1\,Z_2\,Z_3$
\\
\cline{1-3}
 \scriptsize $(1\,,\,1\,,\,0)$ & \scriptsize 6& \scriptsize $Y_1^2\, Y_2^2\, Y_3^2$ \\
\cline{2-3}
& \scriptsize 4 & \scriptsize $Y_1\, Y_2\, Y_3^2\, Z_3+\hat{\delta}\, (Y_1^2\, Z_1^2+ Y_2^2\, Z_2^2+2\, Y_1\, Y_2\, Z_1\, Z_2)$
\\
\cline{2-3}
& \scriptsize 4 & \scriptsize $Y_2^2\, Y_3\, Y_1\, Z_2+Y_1^2\,Y_2\,Y_3\,Z_1-\hat{\delta}
   \,Y_2\, Y_1\, Z_1\, Z_2$ \\
   \cline{2-3}
& \scriptsize 2 & \scriptsize $Y_3^2\, Z_3^2- Y_1^2\, Z_1^2- Y_2^2\,
   Z_2^2 -2\,Y_1\, Y_2\, Z_2\, Z_1 $\\
   \cline{2-3}
& \scriptsize 2   & \scriptsize $Y_1^2\, Z_1^2+Y_2^2\, Z_2^2+2\, Y_1\, Y_2\, Z_2\, Z_1+Y_1\, Y_3\, Z_3\, Z_1+Y_2\, Y_3\, Z_2\, Z_3$\\

& & \scriptsize $-\hat{\delta}\, Z_1\, Z_2\, Z_3$\\
\cline{1-3}
 \scriptsize $(1\,,\,1\,,\,\m_3)$ & \scriptsize 6& \scriptsize $ Y_1^2\, Y_2^2\, Y_3^2+\tfrac{1}{4}\, \hat{\delta}^2\, \m_3\,(\mu _3-2)\,( Y_1\, Y_3\, Z_1\, Z_3+ Y_2\, Y_3\, Z_2\, Z_3)$\\
 & &\scriptsize  $+\tfrac{1}{8}\, \hat{\delta}^3\,\m_3\,(\mu _3-2)^2\, Z_1\, Z_2\, Z_3$\\
 \cline{2-3}
& \scriptsize 4 &  \scriptsize$Y_1\,Y_2\,Y_3^2\,Z_3+\tfrac{1}{2}\,\hat{\delta}\,(\m_3-2)\,(Y_1\,Y_3\,Z_1\,Z_3+Y_2\,Y_3\,Z_2\,Z_3)$\\
& &\scriptsize$+\tfrac{1}{4}\,\hat{\delta}^2\,(\m_3-2)^2\,Z_1\,Z_2\,Z_3$\\
\cline{2-3}
& \scriptsize 4 &\scriptsize $Y_1\,Y_2^2\,Y_3\,Z_2-\tfrac{1}{2}\,\hat{\delta}\,\m_3\,Y_2\,Y_3\,Z_2\,Z_3$\\
 \cline{2-3}
& \scriptsize 4 &\scriptsize $Y_1^2\,Y_2\,Y_3\,Z_1-\tfrac{1}{2}\,\hat{\delta}\,\m_3\,Y_1\,Y_3\,Z_1\,Z_3$\\
\cline{2-3}
& \scriptsize 2&\scriptsize $Y_3^2\,Z_3^2+Y_1\,Y_3\,Z_1\,Z_3+Y_2\,Y_3\,Z_2\,Z_3+\tfrac{1}{2}\,\hat{\delta}\,(\m_3-2)\,Z_1\,Z_2\,Z_3$\\
\cline{2-3}
& \scriptsize 2& \scriptsize $Y_1\,Y_2\,Z_1\,Z_2-\tfrac{1}{2}\,\hat{\delta}\,\m_3\,Z_1\,Z_2\,Z_3$\\

\cline{1-3}
\scriptsize $(1\,,\,1\,,\,2)$
& \scriptsize 6& \scriptsize $Y_1^2\,Y_2^2\,Y_3^2$\\

\cline{2-3}
& \scriptsize 4& \scriptsize $Y_1\,Y_2\,Y_3^2 Z_3$\\

\cline{2-3}
& \scriptsize 4& \scriptsize $Y_1\,Y_2^2\,Y_3 Z_2+\hat{\delta }\,Y_2^2 Z_2^2$\\

\cline{2-3}
& \scriptsize 4& \scriptsize $Y_1^2\,Y_2\,Y_3 Z_1+\hat{\delta }\,Y_1^2 Z_1^2$\\

\cline{2-3}
& \scriptsize 2& \scriptsize $-Y_1^2 Z_1^2-Y_2^2 Z_2^2+Y_3^2 Z_3^2$\\

\cline{2-3}
& \scriptsize 2& \scriptsize $Y_2 Z_2 \left(Y_2 Z_2+Y_3 Z_3\right)$\\

\cline{2-3}
& \scriptsize 2& \scriptsize $Y_1 Z_1 \left(Y_1 Z_1+Y_3 Z_3\right)$\\

\cline{2-3}
& \scriptsize 2& \scriptsize $Y_1\,Y_2 Z_1 Z_2-\hat{\delta } Z_1 Z_2 Z_3$\\

\cline{1-3}
 \scriptsize $(1\,,\,0\,,\,0)$& \scriptsize 6 & \scriptsize $Y_1^2\,Y_2^2\,Y_3^2-\tfrac{1}{2}\,\hat{\delta}\,(3\,Y_1^2\,Y_2\,Y_3\,Z_1+Y_1\,Y_2^2\,Y_3\,Z_2+Y_1\,Y_2\,Y_3^2\,Z_3)$\\
 && \scriptsize $+\tfrac{1}{4}\,\hat{\delta}^2\,(3\,Y_1\,Y_2\,Z_1\,Z_2+3\,Y_1\,Y_3\,Z_1\,Z_3+Y_2\,Y_3\,Z_2\,Z_3)-\tfrac{3}{8}\,\hat{\delta}^3\,Z_1\,Z_2\,Z_3$\\
\cline{1-3}
 \scriptsize $(1\,,\,0\,,\,\m_3)$ & \scriptsize 6& \scriptsize $Y_1^2\,Y_2^2\,Y_3^2+\hat{\delta}\,(\m_3-1)\,Y_1^2\,Y_2\,Y_3\,Z_1$\\
& & \scriptsize$-\tfrac{1}{4}\,\hat{\delta}^2\,(\m_3^2-1)\,(Y_3^2\,Z_3^2+2\,Y_1\,Y_3\,Z_1\,Z_3)$\\
\cline{2-3}
& \scriptsize 4& \scriptsize $Y_1^2\,Y_2\,Y_3\,Z_1+Y_1\,Y_2\,Y_3^2\,Z_3-\tfrac{1}{2}\,\hat{\delta}\,(\m_3-1)\,Y_3^2\,Z_3^2$\\
& & \scriptsize $-\tfrac{1}{2}\,\hat{\delta}\,(\m_3+1)\,Y_1\,Y_3\,Z_1\,Z_3$\\
\cline{2-3}
& \scriptsize 4& \scriptsize $Y_1\,Y_2^2\,Y_3\,Z_2+\tfrac{1}{2}\,\hat{\delta}\,(\m_3-3)\,Y_1\,Y_2\,Z_1\,Z_2-\tfrac{1}{2}\,\hat{\delta}\,(\m_3+1)\,Y_2\,Y_3\,Z_2\,Z_3$\\
& & \scriptsize $-\tfrac{1}{4}\,\hat{\delta}^2\,(\m_3+1)\,(\m_3-3)\,Z_1\,Z_2\,Z_3$\\

\cline{1-3}
 \scriptsize $(1\,,\,\m_2\,,\,\m_3)$& \scriptsize 6 & \scriptsize $Y_1^2\,Y_2^2\,Y_3^2-\tfrac{1}{4}\,\hat{\delta}^2\,[(\m_2-\m_3)^2-1]\,Y_2^2\,Z_2^2$\\
\cline{2-3}
& \scriptsize 4& \scriptsize $Y_1\,Y_2\,Y_3^2\,Z_3+\tfrac{1}{2}\,\hat{\delta}\,(\m_2-\m_3+1)\,Y_2^2\,Z_2^2$\\
 \cline{2-3}
& \scriptsize 4& \scriptsize $Y_1\,Y_2^2\,Y_3\,Z_2-\tfrac{1}{2}\,\hat{\delta}(\m_2-\m_3-1)\,Y_2^2\,Z_2^2$\\
  \cline{2-3}
& \scriptsize 4& \scriptsize $Y_1^2\,Y_2\,Y_3\,Z_1+\tfrac{1}{4}\,\hat{\delta}^2\,[(\m_2-\m_3)^2-1]\,Z_1\,Z_2\,Z_3$\\
\cline{2-3}
& \scriptsize 2& \scriptsize$Y_3^2\,Z_3^2-Y_2^2\,Z_2^2$\\
 \cline{2-3}
& \scriptsize 2& \scriptsize $Y_2^2\,Z_2^2+Y_2\,Y_3\,Z_2\,Z_3$\\
 \cline{2-3}
& \scriptsize 2& \scriptsize $Y_1\,Y_3\,Z_1\,Z_3-\tfrac{1}{2}\,\hat{\delta}\,(\m_2-\m_3+1)\,Z_1\,Z_2\,Z_3$\\
\cline{2-3}
& \scriptsize 2& \scriptsize $Y_1\,Y_2\,Z_1\,Z_2+\tfrac{1}{2}\,\hat{\delta}\,(\m_2-\m_3-1)\,Z_1\,Z_2\,Z_3$\\

\cline{1-3}
\scriptsize $(1\,,\,\m_3+1\,,\,\m_3)$
& \scriptsize 6& \scriptsize $Y_1^2\,Y_2^2\,Y_3^2$\\

\cline{2-3}
& \scriptsize 4& \scriptsize $Y_1\,Y_2\,Y_3^2 Z_3+\hat{\delta }\,Y_2^2 Z_2^2$\\

\cline{2-3}
& \scriptsize 4& \scriptsize $Y_1\,Y_2^2\,Y_3 Z_2$\\

\cline{2-3}
& \scriptsize 4& \scriptsize $Y_1^2\,Y_2\,Y_3 Z_1$\\

\cline{2-3}
& \scriptsize 2& \scriptsize $-Y_2^2 Z_2^2+Y_3^2 Z_3^2$\\

\cline{2-3}
& \scriptsize 2& \scriptsize $Y_2 Z_2 \left(Y_2 Z_2+Y_3 Z_3\right)$\\

\cline{2-3}
& \scriptsize 2& \scriptsize $Y_1\,Y_3 Z_1 Z_3-\hat{\delta } Z_1 Z_2 Z_3$\\

\cline{2-3}
& \scriptsize 2& \scriptsize $Y_1\,Y_2 Z_1 Z_2$\\

\cline{2-3}
& \scriptsize 2& \scriptsize $Y_1^2 Z_1^2$\\

\cline{1-3}
\scriptsize $(1\,,\,\m_3-1\,,\,\m_3)$
& \scriptsize 6& \scriptsize $Y_1^2\,Y_2^2\,Y_3^2$\\

\cline{2-3}
& \scriptsize 4& \scriptsize $Y_1\,Y_2\,Y_3^2 Z_3$\\

\cline{2-3}
& \scriptsize 4& \scriptsize $Y_1\,Y_2^2\,Y_3 Z_2+\hat{\delta }\,Y_2^2 Z_2^2$\\

\cline{2-3}
& \scriptsize 4& \scriptsize $Y_1^2\,Y_2\,Y_3 Z_1$\\

\cline{2-3}
& \scriptsize 2& \scriptsize $-Y_2^2 Z_2^2+Y_3^2 Z_3^2$\\

\cline{2-3}
& \scriptsize 2& \scriptsize $Y_2 Z_2 \left(Y_2 Z_2+Y_3 Z_3\right)$\\

\cline{2-3}
& \scriptsize 2& \scriptsize $Y_1\,Y_3 Z_1 Z_3$\\

\cline{2-3}
& \scriptsize 2& \scriptsize $Y_1\,Y_2 Z_1 Z_2-\hat{\delta } Z_1 Z_2 Z_3$\\

\cline{2-3}
& \scriptsize 2& \scriptsize $Y_1^2 Z_1^2$\\
\hline
\end{longtable}}
\end{center}


\begin{center}

\textbf{3-3-2 Couplings}

{\footnotesize
\begin{longtable}[c]{|c|c|c|}

\hline
\scriptsize $(\m_1\,,\,\m_2\,,\,\m_3)$ &\scriptsize$\partial$& \scriptsize{Couplings}  \\

\hline
\scriptsize $(2\,,\,2\,,\,1)$& \scriptsize 8
& \scriptsize  $Y_1^3\,Y_2^3\,Y_3^2+\frac{1}{4} \hat{\delta }^2\,Y_1^3\,Y_2\,Z_1^2$\\
&& \scriptsize  $-\frac{3}{8} \hat{\delta }^3 \left(Y_1\,Z_1+Y_2\,Z_2\right)\,Z_3 \left(Y_1\,Z_1+Y_3\,Z_3\right)$\\
&& \scriptsize  $+\frac{3}{8} \hat{\delta }^4\,Z_1\,Z_2\,Z_3^2$\\

\cline{2-3}
& \scriptsize 6& \scriptsize  $Y_1^2\,Y_2^2\,Y_3^2\,Z_3-\frac{1}{4} \hat{\delta }^2\,Z_3 \left(Y_2\,Y_3\,Z_2\,Z_3\right.$\\
&& \scriptsize  $\left.+Y_1\,Z_1 \left(Y_2\,Z_2+Y_3\,Z_3\right)\right)+\frac{1}{4} \hat{\delta }^3\,Z_1\,Z_2\,Z_3^2$\\

\cline{2-3}
& \scriptsize 6& \scriptsize  $Y_1^2\,Y_2^3\,Y_3\,Z_2,\frac{1}{2} \hat{\delta }\,Y_1^3\,Y_2\,Z_1^2$\\
&& \scriptsize  $-\frac{3}{4} \hat{\delta }^2\,Z_3 \left(Y_1^2\,Z_1^2+Y_1\,Y_2\,Z_1\,Z_2+Y_2\,Y_3\,Z_2\,Z_3\right),\frac{3}{8} \hat{\delta }^3\,Z_1\,Z_2\,Z_3^2$\\

\cline{2-3}
& \scriptsize 6& \scriptsize  $Y_1^3\,Y_2^2\,Y_3\,Z_1+\frac{1}{2} \hat{\delta }\,Y_1^3\,Y_2\,Z_1^2$\\
&& \scriptsize  $-\frac{3}{4} \hat{\delta }^2\,Y_1\,Z_1\,Z_3 \left(Y_1\,Z_1+Y_2\,Z_2+Y_3\,Z_3\right)+\frac{3}{8} \hat{\delta }^3\,Z_1\,Z_2\,Z_3^2$\\

\cline{2-3}
& \scriptsize 4& \scriptsize  $Y_1\,Y_2\,Y_3^2\,Z_3^2-\frac{1}{2} \hat{\delta }\,Y_3 \left(Y_1\,Z_1+Y_2\,Z_2\right)\,Z_3^2+\frac{1}{4} \hat{\delta }^2\,Z_1\,Z_2\,Z_3^2$\\

\cline{2-3}
& \scriptsize 4& \scriptsize  $Y_1\,Y_2^2\,Y_3\,Z_2\,Z_3-\frac{1}{2} \hat{\delta }\,Y_2\,Z_2\,Z_3 \left(Y_1\,Z_1+Y_3\,Z_3\right)+\frac{1}{4} \hat{\delta }^2\,Z_1\,Z_2\,Z_3^2$\\

\cline{2-3}
& \scriptsize 4& \scriptsize  $-Y_1^3\,Y_2\,Z_1^2+Y_1\,Y_2^3\,Z_2^2+\frac{3}{2} \hat{\delta } \left(Y_1^2\,Z_1^2-Y_2^2\,Z_2^2\right)\,Z_3$\\

\cline{2-3}
& \scriptsize 4& \scriptsize  $Y_1^2\,Y_2\,Y_3\,Z_1\,Z_3-\frac{1}{2} \hat{\delta }\,Y_1\,Z_1\,Z_3 \left(Y_2\,Z_2+Y_3\,Z_3\right)+\frac{1}{4} \hat{\delta }^2\,Z_1\,Z_2\,Z_3^2$\\

\cline{2-3}
& \scriptsize 4& \scriptsize  $Y_1^2\,Y_2\,Z_1 \left(Y_1\,Z_1+Y_2\,Z_2\right)-\frac{3}{2} \hat{\delta }\,Y_1\,Z_1 \left(Y_1\,Z_1+Y_2\,Z_2\right)\,Z_3$\\

\cline{1-3}
\scriptsize $(2\,,\,2\,,\,0)$& \scriptsize 8
& \scriptsize  $Y_1^3\,Y_2^3\,Y_3^2$\\

\cline{2-3}
& \scriptsize 6& \scriptsize  $Y_1^2\,Y_2^2\,Y_3^2\,Z_3$\\

\cline{2-3}
& \scriptsize 6& \scriptsize  $Y_1^2\,Y_2^2\,Y_3 \left(Y_1\,Z_1+Y_2\,Z_2\right)-\hat{\delta }\,Y_1^2\,Y_2^2\,Z_1\,Z_2$\\

\cline{2-3}
& \scriptsize 4& \scriptsize  $Y_1\,Y_2\,Y_3^2\,Z_3^2+\hat{\delta } \left(Y_1\,Z_1+Y_2\,Z_2\right){}^2\,Z_3$\\

\cline{2-3}
& \scriptsize 4& \scriptsize  $Y_1\,Y_2\,Y_3 \left(Y_1\,Z_1+Y_2\,Z_2\right)\,Z_3-\hat{\delta }\,Y_1\,Y_2\,Z_1\,Z_2\,Z_3$\\

\cline{2-3}
& \scriptsize 4& \scriptsize  $Y_1\,Y_2 \left(Y_1\,Z_1+Y_2\,Z_2\right){}^2-\hat{\delta } \left(Y_1\,Z_1+Y_2\,Z_2\right){}^2\,Z_3$\\

\cline{2-3}
& \scriptsize 2& \scriptsize  $Z_3 \left(-Y_1^2\,Z_1^2-2\,Y_1\,Y_2\,Z_1\,Z_2-Y_2^2\,Z_2^2+Y_3^2\,Z_3^2\right)$\\

\cline{2-3}
& \scriptsize 2& \scriptsize  $\left(Y_1\,Z_1+Y_2\,Z_2\right)\,Z_3 \left(Y_1\,Z_1+Y_2\,Z_2+Y_3\,Z_3\right)-\hat{\delta }\,Z_1\,Z_2\,Z_3^2$\\

\cline{1-3}
\scriptsize $(2\,,\,1\,,\,1)$& \scriptsize 8
& \scriptsize  $Y_1^3\,Y_2^3\,Y_3^2$\\

\cline{2-3}
& \scriptsize 6& \scriptsize  $Y_1^2\,Y_2^2\,Y_3^2\,Z_3$\\

\cline{2-3}
& \scriptsize 6& \scriptsize  $Y_1^2\,Y_2^3\,Y_3\,Z_2-\hat{\delta }\,Y_1^2\,Y_2^2\,Z_1\,Z_2+2 \hat{\delta }^2\,Y_2\,Z_2 \left(Y_1\,Z_1+Y_2\,Z_2\right)\,Z_3$\\

\cline{2-3}
& \scriptsize 6& \scriptsize  $Y_1^3\,Y_2^2\,Y_3\,Z_1+2 \hat{\delta }\,Y_1^2\,Y_2\,Z_1 \left(Y_1\,Z_1+Y_2\,Z_2\right)$\\
&& \scriptsize  $-2 \hat{\delta }^2\,Y_1\,Z_1\,Z_3 \left(2\,Y_1\,Z_1+2\,Y_2\,Z_2+Y_3\,Z_3\right)$\\

\cline{2-3}
& \scriptsize 4& \scriptsize  $-Y_1\,Y_2 \left(Y_1^2\,Z_1^2+Y_1\,Y_2\,Z_1\,Z_2-Y_3^2\,Z_3^2\right)+2 \hat{\delta }\,Y_1\,Z_1 \left(Y_1\,Z_1+Y_2\,Z_2\right)\,Z_3$\\

\cline{2-3}
& \scriptsize 4& \scriptsize  $Y_1\,Y_2^2\,Y_3\,Z_2\,Z_3+\hat{\delta }\,Y_2^2\,Z_2^2\,Z_3$\\

\cline{2-3}
& \scriptsize 4& \scriptsize  $Y_1\,Y_2^2\,Z_2 \left(Y_1\,Z_1+Y_2\,Z_2\right)-2 \hat{\delta }\,Y_2\,Z_2 \left(Y_1\,Z_1+Y_2\,Z_2\right)\,Z_3$\\

\cline{2-3}
& \scriptsize 4& \scriptsize  $Y_1^2\,Y_2\,Z_1 \left(Y_1\,Z_1+Y_2\,Z_2+Y_3\,Z_3\right)$\\
&& \scriptsize  $-\hat{\delta }\,Y_1\,Z_1\,Z_3 \left(2\,Y_1\,Z_1+2\,Y_2\,Z_2+Y_3\,Z_3\right)$\\

\cline{2-3}
& \scriptsize 2& \scriptsize  $Z_3 \left(Y_1^2\,Z_1^2-Y_2^2\,Z_2^2+2\,Y_1\,Y_3\,Z_1\,Z_3+Y_3^2\,Z_3^2\right)$\\

\cline{2-3}
& \scriptsize 2& \scriptsize  $Y_2\,Z_2\,Z_3 \left(Y_1\,Z_1+Y_2\,Z_2+Y_3\,Z_3\right)-\hat{\delta }\,Z_1\,Z_2\,Z_3^2$\\

\cline{1-3}
\scriptsize $(2\,,\,1\,,\,0)$& \scriptsize 8
& \scriptsize  $Y_1^3\,Y_2^3\,Y_3^2-\hat{\delta }\,Y_1^3\,Y_2^2\,Y_3\,Z_1$\\
&& \scriptsize  $-\frac{1}{4} \hat{\delta }^2\,Y_1\,Y_2 \left(Y_1^2\,Z_1^2+3\,Y_1\,Y_3\,Z_1\,Z_3+3\,Y_3\,Z_3 \left(Y_2\,Z_2+Y_3\,Z_3\right)\right)$\\
&& \scriptsize  $\frac{3}{8} \hat{\delta }^3\,Z_3 \left(Y_1^2\,Z_1^2+Y_2\,Y_3\,Z_2\,Z_3+3\,Y_1\,Z_1 \left(Y_2\,Z_2+Y_3\,Z_3\right)\right)-\frac{9}{16} \hat{\delta }^4\,Z_1\,Z_2\,Z_3^2$\\

\cline{2-3}
& \scriptsize 6& \scriptsize  $Y_1^2\,Y_2^2\,Y_3^2\,Z_3-\frac{1}{2} \hat{\delta }\,Y_1\,Y_2\,Y_3\,Z_3 \left(3\,Y_1\,Z_1+Y_2\,Z_2+Y_3\,Z_3\right)$\\
&&\scriptsize  $\frac{1}{4} \hat{\delta }^2\,Z_3 \left(Y_2\,Y_3\,Z_2\,Z_3+3\,Y_1\,Z_1 \left(Y_2\,Z_2+Y_3\,Z_3\right)\right)-\frac{3}{8} \hat{\delta }^3\,Z_1\,Z_2\,Z_3^2$\\

\cline{2-3}
& \scriptsize 6& \scriptsize  $Y_1^2\,Y_2^2\,Y_3 \left(Y_1\,Z_1+Y_2\,Z_2\right)$\\
&& \scriptsize  $-\frac{1}{2} \hat{\delta }\,Y_1\,Y_2 \left(Y_1^2\,Z_1^2+3\,Y_2\,Y_3\,Z_2\,Z_3+3\,Y_1\,Z_1 \left(Y_2\,Z_2+Y_3\,Z_3\right)\right)$\\
&& \scriptsize  $+\frac{3}{4} \hat{\delta }^2\,Y_1\,Z_1 \left(Y_1\,Z_1+3\,Y_2\,Z_2\right)\,Z_3$\\

\cline{1-3}
\scriptsize $(2\,,\,0\,,\,1)$& \scriptsize 8
& \scriptsize  $Y_1^3\,Y_2^3\,Y_3^2+\frac{3}{2} \hat{\delta }\,Y_1^3\,Y_2^2\,Y_3\,Z_1$\\
&& \scriptsize  $-\frac{3}{4} \hat{\delta }^2\,Y_1\,Y_2 \left(2\,Y_1^2\,Z_1^2+10\,Y_1\,Y_3\,Z_1\,Z_3+5\,Y_3^2\,Z_3^2\right)$\\
&& \scriptsize  $+\frac{15}{8} \hat{\delta }^3\,Y_1\,Z_1\,Z_3 \left(2\,Y_1\,Z_1+3\,Y_3\,Z_3\right)$\\

\cline{2-3}
& \scriptsize 6& \scriptsize  $Y_1^2\,Y_2^2\,Y_3 \left(Y_1\,Z_1+Y_3\,Z_3\right)$\\
&& \scriptsize  $-\frac{1}{2} \hat{\delta }\,Y_1\,Y_2 \left(Y_1^2\,Z_1^2+8\,Y_1\,Y_3\,Z_1\,Z_3+3\,Y_3^2\,Z_3^2\right)$\\
&& \scriptsize  $+\frac{1}{4} \hat{\delta }^2\,Y_1\,Z_1\,Z_3 \left(5\,Y_1\,Z_1+9\,Y_3\,Z_3\right)$\\

\cline{2-3}
& \scriptsize 6& \scriptsize  $Y_1^2\,Y_2^3\,Y_3\,Z_2-\frac{3}{2} \hat{\delta }\,Y_1\,Y_2^2\,Z_2 \left(Y_1\,Z_1+2\,Y_3\,Z_3\right)$\\
&& \scriptsize  $\frac{3}{4} \hat{\delta }^2\,Y_2\,Z_2\,Z_3 \left(6\,Y_1\,Z_1+Y_3\,Z_3\right)-\frac{9}{8} \hat{\delta }^3\,Z_1\,Z_2\,Z_3^2$\\

\cline{1-3}
\scriptsize $(2\,,\,0\,,\,0)$& \scriptsize 8
& \scriptsize  $Y_1^3\,Y_2^3\,Y_3^2+2 \hat{\delta }\,Y_1^2\,Y_2^3\,Y_3\,Z_2+2 \hat{\delta }^2\,Y_1\,Y_2^3\,Z_2^2$\\

\cline{2-3}
& \scriptsize 6& \scriptsize  $Y_1^2\,Y_2^2\,Y_3 \left(Y_1\,Z_1+Y_2\,Z_2+Y_3\,Z_3\right)+\hat{\delta }\,Y_1\,Y_2 \left(Y_2^2\,Z_2^2-2\,Y_1\,Y_3\,Z_1\,Z_3\right)$\\

\cline{2-3}
& \scriptsize 4& \scriptsize  $Y_1\,Y_2 \left(Y_1\,Z_1+Y_2\,Z_2+Y_3\,Z_3\right){}^2$\\
&& \scriptsize  $-2 \hat{\delta }\,Y_1\,Z_1\,Z_3 \left(Y_1\,Z_1+2\,Y_2\,Z_2+Y_3\,Z_3\right)$\\

\cline{1-3}
\scriptsize $(1\,,\,1\,,\,1)$& \scriptsize 8
& \scriptsize  $Y_1^3\,Y_2^3\,Y_3^2-\frac{1}{4} \hat{\delta }^2\,Y_1\,Y_2 \left(3\,Y_3\,Z_3 \left(2\,Y_2\,Z_2+5\,Y_3\,Z_3\right)\right.$\\
&& \scriptsize  $\left.+Y_1\,Z_1 \left(Y_2\,Z_2+6\,Y_3\,Z_3\right)\right)+\frac{3}{4} \hat{\delta }^3\,Z_3 \left(3\,Y_2\,Y_3\,Z_2\,Z_3\right.$\\
&& \scriptsize  $\left.+Y_1\,Z_1 \left(2\,Y_2\,Z_2+3\,Y_3\,Z_3\right)\right)-\frac{21}{16} \hat{\delta }^4\,Z_1\,Z_2\,Z_3^2$\\

\cline{2-3}
& \scriptsize 6& \scriptsize  $Y_1^2\,Y_2^2\,Y_3^2\,Z_3-\frac{1}{2} \hat{\delta }\,Y_1\,Y_2\,Y_3\,Z_3 \left(Y_1\,Z_1+Y_2\,Z_2+3\,Y_3\,Z_3\right)$\\
&& \scriptsize  $\frac{1}{4} \hat{\delta }^2\,Z_3 \left(3\,Y_2\,Y_3\,Z_2\,Z_3+Y_1\,Z_1 \left(Y_2\,Z_2+3\,Y_3\,Z_3\right)\right)-\frac{3}{8} \hat{\delta }^3\,Z_1\,Z_2\,Z_3^2$\\

\cline{2-3}
& \scriptsize 6& \scriptsize  $Y_1^2\,Y_2^3\,Y_3\,Z_2-\frac{1}{2} \hat{\delta }\,Y_1\,Y_2^2\,Z_2 \left(Y_1\,Z_1+6\,Y_3\,Z_3\right)$\\
&& \scriptsize  $+\frac{3}{4} \hat{\delta }^2\,Y_2\,Z_2\,Z_3 \left(2\,Y_1\,Z_1+Y_3\,Z_3\right)-\frac{3}{8} \hat{\delta }^3\,Z_1\,Z_2\,Z_3^2$\\

\cline{2-3}
& \scriptsize 6& \scriptsize  $Y_1^3\,Y_2^2\,Y_3\,Z_1-\frac{1}{2} \hat{\delta }\,Y_1^2\,Y_2\,Z_1 \left(Y_2\,Z_2+6\,Y_3\,Z_3\right)$\\
&& \scriptsize  $+\frac{3}{4} \hat{\delta }^2\,Y_1\,Z_1\,Z_3 \left(2\,Y_2\,Z_2+Y_3\,Z_3\right)-\frac{3}{8} \hat{\delta }^3\,Z_1\,Z_2\,Z_3^2$\\

\cline{1-3}
\scriptsize $(1\,,\,1\,,\,0)$& \scriptsize 8
& \scriptsize  $Y_1^3\,Y_2^3\,Y_3^2$\\

\cline{2-3}
& \scriptsize 6& \scriptsize  $Y_1^2\,Y_2^2\,Y_3^2\,Z_3$\\

\cline{2-3}
& \scriptsize 6& \scriptsize  $Y_1^2\,Y_2^2\,Y_3 \left(Y_1\,Z_1+Y_2\,Z_2\right)$\\
&& \scriptsize  $+\hat{\delta }\,Y_1\,Y_2 \left(2\,Y_1^2\,Z_1^2+3\,Y_1\,Y_2\,Z_1\,Z_2+2\,Y_2^2\,Z_2^2\right)$\\
&& \scriptsize  $-2 \hat{\delta }^2 \left(Y_1\,Z_1+Y_2\,Z_2\right)\,Z_3 \left(2\,Y_1\,Z_1+2\,Y_2\,Z_2+Y_3\,Z_3\right)+2 \hat{\delta }^3\,Z_1\,Z_2\,Z_3^2$\\

\cline{2-3}
& \scriptsize 4& \scriptsize  $-Y_1\,Y_2 \left(Y_1^2\,Z_1^2+2\,Y_1\,Y_2\,Z_1\,Z_2+Y_2^2\,Z_2^2-Y_3^2\,Z_3^2\right)$\\
&& \scriptsize  $+2 \hat{\delta } \left(Y_1\,Z_1+Y_2\,Z_2\right){}^2\,Z_3$\\

\cline{2-3}
& \scriptsize 4& \scriptsize  $Y_1\,Y_2 \left(Y_1\,Z_1+Y_2\,Z_2\right) \left(Y_1\,Z_1+Y_2\,Z_2+Y_3\,Z_3\right)$\\
&& \scriptsize  $-\hat{\delta }\,Z_3 \left(2\,Y_1^2\,Z_1^2+Y_2\,Z_2 \left(2\,Y_2\,Z_2+Y_3\,Z_3\right)+Y_1\,Z_1 \left(5\,Y_2\,Z_2+Y_3\,Z_3\right)\right)+\hat{\delta }^2\,Z_1\,Z_2\,Z_3^2$\\

\cline{2-3}
& \scriptsize 2& \scriptsize  $Z_3 \left(Y_1\,Z_1+Y_2\,Z_2+Y_3\,Z_3\right){}^2-2 \hat{\delta }\,Z_1\,Z_2\,Z_3^2$\\

\cline{1-3}
\scriptsize $(1\,,\,0\,,\,1)$& \scriptsize 8
& \scriptsize  $Y_1^3\,Y_2^3\,Y_3^2+3 \hat{\delta }\,Y_1^3\,Y_2^2\,Y_3\,Z_1+6 \hat{\delta }^2 \left(Y_1^3\,Y_2\,Z_1^2-Y_1\,Y_2^3\,Z_2^2\right)$\\
&& \scriptsize  $-6 \hat{\delta }^3\,Z_3 \left(3\,Y_1^2\,Z_1^2-3\,Y_2^2\,Z_2^2+3\,Y_1\,Y_3\,Z_1\,Z_3+Y_3^2\,Z_3^2\right)$\\

\cline{2-3}
& \scriptsize 6& \scriptsize  $Y_1^2\,Y_2^2\,Y_3 \left(Y_1\,Z_1+Y_3\,Z_3\right)+2 \hat{\delta } \left(Y_1^3\,Y_2\,Z_1^2-Y_1\,Y_2^3\,Z_2^2\right)$\\
&& \scriptsize  $-2 \hat{\delta }^2\,Z_3 \left(3\,Y_1^2\,Z_1^2-3\,Y_2^2\,Z_2^2+3\,Y_1\,Y_3\,Z_1\,Z_3+Y_3^2\,Z_3^2\right)$\\

\cline{2-3}
& \scriptsize 6& \scriptsize  $Y_1^2\,Y_2^3\,Y_3\,Z_2+\hat{\delta }\,Y_1\,Y_2^2\,Z_2 \left(3\,Y_1\,Z_1+4\,Y_2\,Z_2\right)$\\
&& \scriptsize  $-6 \hat{\delta }^2\,Y_2\,Z_2\,Z_3 \left(2\,Y_1\,Z_1+2\,Y_2\,Z_2+Y_3\,Z_3\right)+6 \hat{\delta }^3\,Z_1\,Z_2\,Z_3^2$\\

\cline{2-3}
& \scriptsize 4& \scriptsize  $Y_1\,Y_2 \left(Y_1^2\,Z_1^2-Y_2^2\,Z_2^2+2\,Y_1\,Y_3\,Z_1\,Z_3+Y_3^2\,Z_3^2\right)$\\
&& \scriptsize  $-\hat{\delta }\,Z_3 \left(3\,Y_1^2\,Z_1^2-3\,Y_2^2\,Z_2^2+4\,Y_1\,Y_3\,Z_1\,Z_3+Y_3^2\,Z_3^2\right)$\\

\cline{2-3}
& \scriptsize 4& \scriptsize  $Y_1\,Y_2^2\,Z_2 \left(Y_1\,Z_1+Y_2\,Z_2+Y_3\,Z_3\right)$\\
&& \scriptsize  $-\hat{\delta }\,Y_2\,Z_2\,Z_3 \left(4\,Y_1\,Z_1+3\,Y_2\,Z_2+2\,Y_3\,Z_3\right)+2 \hat{\delta }^2\,Z_1\,Z_2\,Z_3^2$\\

\cline{1-3}
\scriptsize $(1\,,\,0\,,\,0)$& \scriptsize 8
& \scriptsize  $Y_1^3\,Y_2^3\,Y_3^2-\frac{1}{2} \hat{\delta }\,Y_1^2\,Y_2^2\,Y_3 \left(3\,Y_1\,Z_1+Y_2\,Z_2+6\,Y_3\,Z_3\right)$\\
&& \scriptsize  $+\frac{3}{4} \hat{\delta }^2\,Y_1\,Y_2 \left(Y_3\,Z_3 \left(2\,Y_2\,Z_2+Y_3\,Z_3\right)+Y_1\,Z_1 \left(Y_2\,Z_2+6\,Y_3\,Z_3\right)\right)$\\
&& \scriptsize  $-\frac{3}{8} \hat{\delta }^3\,Z_3 \left(Y_2\,Y_3\,Z_2\,Z_3+3\,Y_1\,Z_1 \left(2\,Y_2\,Z_2+Y_3\,Z_3\right)\right)+\frac{9}{16} \hat{\delta }^4\,Z_1\,Z_2\,Z_3^2$\\

\cline{1-3}
\scriptsize $(0\,,\,0\,,\,1)$& \scriptsize 8
& \scriptsize  $Y_1^3\,Y_2^3\,Y_3^2-\frac{1}{2} \hat{\delta }\,Y_1^2\,Y_2^2\,Y_3 \left(Y_1\,Z_1+Y_2\,Z_2+10\,Y_3\,Z_3\right)$\\
&& \scriptsize  $\frac{1}{4} \hat{\delta }^2\,Y_1\,Y_2 \left(5\,Y_3\,Z_3 \left(2\,Y_2\,Z_2+3\,Y_3\,Z_3\right)+Y_1\,Z_1 \left(Y_2\,Z_2+10\,Y_3\,Z_3\right)\right)$\\
&& \scriptsize  $-\frac{5}{8} \hat{\delta }^3\,Z_3 \left(3\,Y_2\,Y_3\,Z_2\,Z_3+Y_1\,Z_1 \left(2\,Y_2\,Z_2+3\,Y_3\,Z_3\right)\right)+\frac{15}{16} \hat{\delta }^4\,Z_1\,Z_2\,Z_3^2$\\


\cline{1-3}
\scriptsize $(2\,,\,2\,,\,\m_3)$& \scriptsize 8
& \scriptsize  $Y_1^3\,Y_2^3\,Y_3^2+\frac{1}{8} \hat{\delta }^3\,Y_3 \left(Y_1\,Z_1+Y_2\,Z_2\right)\,Z_3^2 \mu _3 \left(-4+\mu _3^2\right)$\\
&& \scriptsize  $+\frac{1}{16} \hat{\delta }^4\,Z_1\,Z_2\,Z_3^2 \left(-2+\mu _3\right){}^2 \mu _3 \left(2+\mu _3\right)$\\

\cline{2-3}
& \scriptsize 6& \scriptsize  $Y_1^2\,Y_2^2\,Y_3^2\,Z_3+\frac{1}{4} \hat{\delta }^2\,Y_3 \left(Y_1\,Z_1+Y_2\,Z_2\right)\,Z_3^2 \left(-2+\mu _3\right) \mu _3$\\
&& \scriptsize  $+\frac{1}{8} \hat{\delta }^3\,Z_1\,Z_2\,Z_3^2 \left(-2+\mu _3\right){}^2 \mu _3$\\

\cline{2-3}
& \scriptsize 6& \scriptsize  $Y_1^2\,Y_2^3\,Y_3\,Z_2-\frac{1}{4} \hat{\delta }^2\,Y_2\,Y_3\,Z_2\,Z_3^2 \mu _3 \left(2+\mu _3\right)$\\

\cline{2-3}
& \scriptsize 6& \scriptsize  $Y_1^3\,Y_2^2\,Y_3\,Z_1-\frac{1}{4} \hat{\delta }^2\,Y_1\,Y_3\,Z_1\,Z_3^2 \mu _3 \left(2+\mu _3\right)$\\

\cline{2-3}
& \scriptsize 4& \scriptsize  $Y_1\,Y_2\,Y_3^2\,Z_3^2+\frac{1}{2} \hat{\delta }\,Y_3 \left(Y_1\,Z_1+Y_2\,Z_2\right)\,Z_3^2 \left(-2+\mu _3\right)$\\
&& \scriptsize  $+\frac{1}{4} \hat{\delta }^2\,Z_1\,Z_2\,Z_3^2 \left(-2+\mu _3\right){}^2$\\

\cline{2-3}
& \scriptsize 4& \scriptsize  $Y_1\,Y_2^2\,Y_3\,Z_2\,Z_3-\frac{1}{2} \hat{\delta }\,Y_2\,Y_3\,Z_2\,Z_3^2 \mu _3$\\

\cline{2-3}
& \scriptsize 4& \scriptsize  $Y_1\,Y_2^3\,Z_2^2-\frac{1}{2} \hat{\delta }\,Y_2^2\,Z_2^2\,Z_3 \left(2+\mu _3\right)$\\

\cline{2-3}
& \scriptsize 4& \scriptsize  $Y_1^2\,Y_2\,Y_3\,Z_1\,Z_3-\frac{1}{2} \hat{\delta }\,Y_1\,Y_3\,Z_1\,Z_3^2 \mu _3$\\

\cline{2-3}
& \scriptsize 4& \scriptsize  $Y_1^2\,Y_2^2\,Z_1\,Z_2-\frac{1}{4} \hat{\delta }^2\,Z_1\,Z_2\,Z_3^2 \mu _3 \left(2+\mu _3\right)$\\

\cline{2-3}
& \scriptsize 4& \scriptsize  $Y_1^3\,Y_2\,Z_1^2-\frac{1}{2} \hat{\delta }\,Y_1^2\,Z_1^2\,Z_3 \left(2+\mu _3\right)$\\

\cline{2-3}
& \scriptsize 2& \scriptsize  $Y_3\,Z_3^2 \left(Y_1\,Z_1+Y_2\,Z_2+Y_3\,Z_3\right)+\frac{1}{2} \hat{\delta }\,Z_1\,Z_2\,Z_3^2 \left(-2+\mu _3\right)$\\

\cline{2-3}
& \scriptsize 2& \scriptsize  $Y_1\,Y_2\,Z_1\,Z_2\,Z_3-\frac{1}{2} \hat{\delta }\,Z_1\,Z_2\,Z_3^2 \mu _3$\\

\cline{1-3}
\scriptsize $(2\,,\,1\,,\,\m_3)$& \scriptsize 8
& \scriptsize  $Y_1^3\,Y_2^3\,Y_3^2-\frac{3}{4} \hat{\delta }^2\,Y_1^3\,Y_2\,Z_1^2 \left(-1+\mu _3^2\right)$\\
&& \scriptsize  $-\frac{1}{8} \hat{\delta }^3\,Z_3 \left(-3\,Y_1^2\,Z_1^2+Y_3^2\,Z_3^2\right) \left(-1+\mu _3\right) \left(1+\mu _3\right) \left(3+\mu _3\right)$\\

\cline{2-3}
& \scriptsize 6& \scriptsize  $Y_1^2\,Y_2^2\,Y_3^2\,Z_3-\hat{\delta }\,Y_1^3\,Y_2\,Z_1^2 \left(-1+\mu _3\right)$\\
&& \scriptsize  $+\frac{1}{4} \hat{\delta }^2\,Z_3 \left(-1+\mu _3\right) \left(-Y_3^2\,Z_3^2 \left(1+\mu _3\right)+2\,Y_1^2\,Z_1^2 \left(3+\mu _3\right)\right)$\\

\cline{2-3}
& \scriptsize 6& \scriptsize  $Y_1^2\,Y_2^3\,Y_3\,Z_2,\frac{1}{4} \hat{\delta }^2\,Y_2\,Z_2\,Z_3 \left(1+\mu _3\right) \left(2\,Y_1\,Z_1 \left(-3+\mu _3\right)-Y_3\,Z_3 \left(3+\mu _3\right)\right)$\\
&& \scriptsize  $-\frac{1}{4} \hat{\delta }^3\,Z_1\,Z_2\,Z_3^2 \left(-3+\mu _3\right) \left(1+\mu _3\right){}^2$\\

\cline{2-3}
& \scriptsize 6& \scriptsize  $Y_1^3\,Y_2^2\,Y_3\,Z_1+\hat{\delta }\,Y_1^3\,Y_2\,Z_1^2 \left(1+\mu _3\right)$\\
&& \scriptsize  $-\frac{1}{4} \hat{\delta }^2\,Y_1\,Z_1\,Z_3 \left(2\,Y_1\,Z_1+Y_3\,Z_3\right) \left(1+\mu _3\right) \left(3+\mu _3\right)$\\

\cline{2-3}
& \scriptsize 4& \scriptsize  $Y_1\,Y_2 \left(-Y_1^2\,Z_1^2+Y_3^2\,Z_3^2\right)$\\
&& \scriptsize  $+\frac{1}{2} \hat{\delta }\,Z_3 \left(-Y_3^2\,Z_3^2 \left(-1+\mu _3\right)+Y_1^2\,Z_1^2 \left(3+\mu _3\right)\right)$\\

\cline{2-3}
& \scriptsize 4& \scriptsize  $Y_1\,Y_2^2\,Y_3\,Z_2\,Z_3-\frac{1}{2} \hat{\delta }\,Y_2\,Z_2\,Z_3 \left(-Y_1\,Z_1 \left(-3+\mu _3\right)+Y_3\,Z_3 \left(1+\mu _3\right)\right)$\\
&& \scriptsize  $-\frac{1}{4} \hat{\delta }^2\,Z_1\,Z_2\,Z_3^2 \left(-3+\mu _3\right) \left(1+\mu _3\right)$\\

\cline{2-3}
& \scriptsize 4& \scriptsize  $Y_1^2\,Y_2\,Z_1 \left(Y_1\,Z_1+Y_3\,Z_3\right)$\\
&& \scriptsize  $-\frac{1}{2} \hat{\delta }\,Y_1\,Z_1\,Z_3 \left(Y_3\,Z_3 \left(1+\mu _3\right)+Y_1\,Z_1 \left(3+\mu _3\right)\right)$\\

\cline{2-3}
& \scriptsize 4& \scriptsize  $Y_1^2\,Y_2^2\,Z_1\,Z_2-\hat{\delta }\,Y_1\,Y_2\,Z_1\,Z_2\,Z_3 \left(1+\mu _3\right)$\\
&& \scriptsize  $+\frac{1}{4} \hat{\delta }^2\,Z_1\,Z_2\,Z_3^2 \left(-1+\mu _3^2\right)$\\

\cline{1-3}
\scriptsize $(2\,,\,0\,,\,\m_3)$& \scriptsize 8
& \scriptsize  $Y_1^3\,Y_2^3\,Y_3^2+\frac{3}{2} \hat{\delta }\,Y_1^3\,Y_2^2\,Y_3\,Z_1 \mu _3+\frac{3}{4} \hat{\delta }^2\,Y_1^3\,Y_2\,Z_1^2 \mu _3 \left(2+\mu _3\right)$\\
&& \scriptsize  $-\frac{1}{8} \hat{\delta }^3\,Z_3 \left(3\,Y_1^2\,Z_1^2+3\,Y_1\,Y_3\,Z_1\,Z_3+Y_3^2\,Z_3^2\right) \mu _3 \left(2+\mu _3\right) \left(4+\mu _3\right)$\\

\cline{2-3}
& \scriptsize 6& \scriptsize  $Y_1^2\,Y_2^2\,Y_3 \left(Y_1\,Z_1+Y_3\,Z_3\right)+\hat{\delta }\,Y_1^2\,Y_2\,Z_1 \left(Y_3\,Z_3 \left(-2+\mu _3\right)+Y_1\,Z_1 \mu _3\right)$\\
&& \scriptsize  $-\frac{1}{4} \hat{\delta }^2\,Z_3 \mu _3 \left(3\,Y_1\,Y_3\,Z_1\,Z_3 \left(2+\mu _3\right)+Y_3^2\,Z_3^2 \left(2+\mu _3\right)+2\,Y_1^2\,Z_1^2 \left(4+\mu _3\right)\right)$\\

\cline{2-3}
& \scriptsize 6& \scriptsize  $Y_1^2\,Y_2^3\,Y_3\,Z_2+\frac{1}{2} \hat{\delta }\,Y_1\,Y_2^2\,Z_2 \left(Y_1\,Z_1 \left(-4+\mu _3\right)-2\,Y_3\,Z_3 \left(2+\mu _3\right)\right)$\\
&& \scriptsize  $\frac{1}{4} \hat{\delta }^2\,Y_2\,Z_2\,Z_3 \left(2+\mu _3\right) \left(-2\,Y_1\,Z_1 \left(-4+\mu _3\right)+Y_3\,Z_3 \mu _3\right)$\\
&& \scriptsize  $+\frac{1}{8} \hat{\delta }^3\,Z_1\,Z_2\,Z_3^2 \left(-4+\mu _3\right) \mu _3 \left(2+\mu _3\right)$\\

\cline{2-3}
& \scriptsize 4& \scriptsize  $Y_1\,Y_2 \left(Y_1\,Z_1+Y_3\,Z_3\right){}^2$\\
&& \scriptsize  $-\frac{1}{2} \hat{\delta }\,Z_3 \left(Y_1\,Z_1+Y_3\,Z_3\right) \left(Y_3\,Z_3 \mu _3+Y_1\,Z_1 \left(4+\mu _3\right)\right)$\\

\cline{1-3}
\scriptsize $(1\,,\,1\,,\,\m_3)$& \scriptsize 8
& \scriptsize  $Y_1^3\,Y_2^3\,Y_3^2+\frac{3}{4} \hat{\delta }^2\,Y_1\,Y_2\,Y_3 \left(Y_1\,Z_1+Y_2\,Z_2\right)\,Z_3 \mu _3 \left(2+\mu _3\right)$\\
&& \scriptsize  $-\frac{1}{8} \hat{\delta }^3\,Z_3 \mu _3 \left(2+\mu _3\right) \left(3\,Y_1\,Z_1 \left(-Y_2\,Z_2 \left(-2+\mu _3\right)+Y_3\,Z_3 \left(2+\mu _3\right)\right)\right.$\\
&& \scriptsize  $\left.+Y_3\,Z_3 \left(3\,Y_2\,Z_2 \left(2+\mu _3\right)+Y_3\,Z_3 \left(4+\mu _3\right)\right)\right)$\\
&& \scriptsize  $-\frac{3}{16} \hat{\delta }^4\,Z_1\,Z_2\,Z_3^2 \left(-2+\mu _3\right) \mu _3 \left(2+\mu _3\right){}^2$\\

\cline{2-3}
& \scriptsize 6& \scriptsize  $Y_1^2\,Y_2^2\,Y_3^2\,Z_3+\hat{\delta }\,Y_1\,Y_2\,Y_3 \left(Y_1\,Z_1+Y_2\,Z_2\right)\,Z_3 \mu _3$\\
&& \scriptsize  $-\frac{1}{4} \hat{\delta }^2\,Z_3 \mu _3 \left(Y_3\,Z_3 \left(2\,Y_2\,Z_2+Y_3\,Z_3\right) \left(2+\mu _3\right)\right.$\\
&& \scriptsize  $\left.+2\,Y_1\,Z_1 \left(-Y_2\,Z_2 \left(-2+\mu _3\right)+Y_3\,Z_3 \left(2+\mu _3\right)\right)\right)-\frac{1}{4} \hat{\delta }^3\,Z_1\,Z_2\,Z_3^2 \mu _3 \left(-4+\mu _3^2\right)$\\

\cline{2-3}
& \scriptsize 6& \scriptsize  $Y_1^2\,Y_2^3\,Y_3\,Z_2-\hat{\delta }\,Y_1\,Y_2^2\,Y_3\,Z_2\,Z_3 \left(2+\mu _3\right)+\frac{1}{4} \hat{\delta }^2\,Y_2\,Y_3\,Z_2\,Z_3^2 \mu _3 \left(2+\mu _3\right)$\\

\cline{2-3}
& \scriptsize 6& \scriptsize  $Y_1^3\,Y_2^2\,Y_3\,Z_1-\hat{\delta }\,Y_1^2\,Y_2\,Y_3\,Z_1\,Z_3 \left(2+\mu _3\right)+\frac{1}{4} \hat{\delta }^2\,Y_1\,Y_3\,Z_1\,Z_3^2 \mu _3 \left(2+\mu _3\right)$\\

\cline{2-3}
& \scriptsize 4& \scriptsize  $Y_1\,Y_2\,Y_3\,Z_3 \left(Y_1\,Z_1+Y_2\,Z_2+Y_3\,Z_3\right)$\\
&& \scriptsize  $-\frac{1}{2} \hat{\delta }\,Z_3 \left(Y_3\,Z_3 \left(Y_3\,Z_3 \mu _3+Y_2\,Z_2 \left(2+\mu _3\right)\right)\right.$\\
&& \scriptsize  $\left.+Y_1\,Z_1 \left(-Y_2\,Z_2 \left(-2+\mu _3\right)+Y_3\,Z_3 \left(2+\mu _3\right)\right)\right)-\frac{1}{4} \hat{\delta }^2\,Z_1\,Z_2\,Z_3^2 \left(-4+\mu _3^2\right)$\\

\cline{2-3}
& \scriptsize 4& \scriptsize  $Y_1^2\,Y_2^2\,Z_1\,Z_2-\hat{\delta }\,Y_1\,Y_2\,Z_1\,Z_2\,Z_3 \left(2+\mu _3\right)+\frac{1}{4} \hat{\delta }^2\,Z_1\,Z_2\,Z_3^2 \mu _3 \left(2+\mu _3\right)$\\

\cline{1-3}
\scriptsize $(1\,,\,0\,,\,\m_3)$& \scriptsize 8
& \scriptsize  $Y_1^3\,Y_2^3\,Y_3^2+\frac{3}{2} \hat{\delta }\,Y_1^3\,Y_2^2\,Y_3\,Z_1 \left(1+\mu _3\right)$\\
&& \scriptsize  $-\frac{3}{4} \hat{\delta }^2\,Y_1\,Y_2\,Y_3\,Z_3 \left(2\,Y_1\,Z_1+Y_3\,Z_3\right) \left(1+\mu _3\right) \left(3+\mu _3\right)$\\
&& \scriptsize  $+\frac{1}{8} \hat{\delta }^3\,Y_3\,Z_3^2 \left(1+\mu _3\right) \left(3+\mu _3\right) \left(2\,Y_3\,Z_3 \left(-1+\mu _3\right)+3\,Y_1\,Z_1 \left(1+\mu _3\right)\right)$\\

\cline{2-3}
& \scriptsize 6& \scriptsize  $Y_1^2\,Y_2^2\,Y_3 \left(Y_1\,Z_1+Y_3\,Z_3\right)$\\
&& \scriptsize  $-\hat{\delta }\,Y_1\,Y_2\,Y_3\,Z_3 \left(Y_3\,Z_3 \left(1+\mu _3\right)+Y_1\,Z_1 \left(3+\mu _3\right)\right)$\\
&& \scriptsize  $+\frac{1}{4} \hat{\delta }^2\,Y_3\,Z_3^2 \left(1+\mu _3\right) \left(Y_3\,Z_3 \left(-1+\mu _3\right)+Y_1\,Z_1 \left(3+\mu _3\right)\right)$\\

\cline{2-3}
& \scriptsize 6& \scriptsize  $Y_1^2\,Y_2^3\,Y_3\,Z_2+\frac{1}{2} \hat{\delta }\,Y_1\,Y_2^2\,Z_2 \left(Y_1\,Z_1 \left(-3+\mu _3\right)-2\,Y_3\,Z_3 \left(3+\mu _3\right)\right)$\\
&& \scriptsize  $+\frac{1}{4} \hat{\delta }^2\,Y_2\,Z_2\,Z_3 \left(3+\mu _3\right) \left(-2\,Y_1\,Z_1 \left(-3+\mu _3\right)+Y_3\,Z_3 \left(1+\mu _3\right)\right)$\\
&& \scriptsize  $+\frac{1}{8} \hat{\delta }^3\,Z_1\,Z_2\,Z_3^2 \left(-3+\mu _3\right) \left(1+\mu _3\right) \left(3+\mu _3\right)$\\

\cline{1-3}
\scriptsize $(2\,,\,\m_2\,,\,1)$& \scriptsize 8
& \scriptsize  $Y_1^3\,Y_2^2\,Y_3^3-\frac{1}{4} \hat{\delta }^2\,Y_1^3\,Y_3\,Z_1^2 \left(-3+\mu _2\right) \left(-1+\mu _2\right)$\\
&& \scriptsize  $-\frac{1}{8} \hat{\delta }^3\,Z_2 \left(Y_1^2\,Z_1^2+Y_3 \left(Y_1\,Z_1+Y_2\,Z_2\right)\,Z_3\right) \left(-5+\mu _2\right) \left(-3+\mu _2\right) \left(-1+\mu _2\right)$\\
&& \scriptsize  $-\frac{1}{16} \hat{\delta }^4\,Z_1\,Z_2^2\,Z_3 \left(-5+\mu _2\right) \left(-3+\mu _2\right){}^2 \left(-1+\mu _2\right)$\\

\cline{2-3}
& \scriptsize 6& \scriptsize  $Y_1^2\,Y_2\,Y_3^3\,Z_3-\frac{1}{2} \hat{\delta }\,Y_1^3\,Y_3\,Z_1^2 \left(-3+\mu _2\right)$\\
&& \scriptsize  $-\frac{1}{4} \hat{\delta }^2\,Z_2 \left(Y_1^2\,Z_1^2+Y_3 \left(Y_1\,Z_1+Y_2\,Z_2\right)\,Z_3\right) \left(-5+\mu _2\right) \left(-3+\mu _2\right)$\\
&& \scriptsize  $-\frac{1}{8} \hat{\delta }^3\,Z_1\,Z_2^2\,Z_3 \left(-5+\mu _2\right) \left(-3+\mu _2\right){}^2$\\

\cline{2-3}
& \scriptsize 6& \scriptsize  $Y_1^2\,Y_2^2\,Y_3^2\,Z_2$\\
&& \scriptsize  $+\frac{1}{4} \hat{\delta }^2\,Y_3\,Z_2 \left(Y_1\,Z_1+Y_2\,Z_2\right)\,Z_3 \left(-3+\mu _2\right) \left(-1+\mu _2\right)$\\
&& \scriptsize  $+\frac{1}{8} \hat{\delta }^3\,Z_1\,Z_2^2\,Z_3 \left(-3+\mu _2\right){}^2 \left(-1+\mu _2\right)$\\

\cline{2-3}
& \scriptsize 6& \scriptsize  $Y_1^3\,Y_2\,Y_3^2\,Z_1+\frac{1}{2} \hat{\delta }\,Y_1^3\,Y_3\,Z_1^2 \left(-1+\mu _2\right)$\\
&& \scriptsize  $+\frac{1}{4} \hat{\delta }^2\,Y_1\,Z_1\,Z_2 \left(Y_1\,Z_1+Y_3\,Z_3\right) \left(-5+\mu _2\right) \left(-1+\mu _2\right)$\\

\cline{2-3}
& \scriptsize 4& \scriptsize  $-Y_1^3\,Y_3\,Z_1^2+Y_1\,Y_3^3\,Z_3^2$\\
&& \scriptsize  $-\frac{1}{2} \hat{\delta }\,Z_2 \left(Y_1^2\,Z_1^2+Y_3 \left(Y_1\,Z_1+Y_2\,Z_2\right)\,Z_3\right) \left(-5+\mu _2\right)$\\
&& \scriptsize  $-\frac{1}{4} \hat{\delta }^2\,Z_1\,Z_2^2\,Z_3 \left(-5+\mu _2\right) \left(-3+\mu _2\right)$\\

\cline{2-3}
& \scriptsize 4& \scriptsize  $Y_1\,Y_2\,Y_3^2\,Z_2\,Z_3+\frac{1}{2} \hat{\delta }\,Y_3\,Z_2 \left(Y_1\,Z_1+Y_2\,Z_2\right)\,Z_3 \left(-3+\mu _2\right)$\\
&& \scriptsize  $+\frac{1}{4} \hat{\delta }^2\,Z_1\,Z_2^2\,Z_3 \left(-3+\mu _2\right){}^2$\\

\cline{2-3}
& \scriptsize 4& \scriptsize  $Y_1\,Y_2^2\,Y_3\,Z_2^2-\frac{1}{2} \hat{\delta }\,Y_2\,Y_3\,Z_2^2\,Z_3 \left(-1+\mu _2\right)$\\

\cline{2-3}
& \scriptsize 4& \scriptsize  $Y_1^2\,Y_3\,Z_1 \left(Y_1\,Z_1+Y_3\,Z_3\right)+\frac{1}{2} \hat{\delta }\,Y_1\,Z_1\,Z_2 \left(Y_1\,Z_1+Y_3\,Z_3\right) \left(-5+\mu _2\right)$\\

\cline{2-3}
& \scriptsize 4& \scriptsize  $Y_1^2\,Y_2\,Y_3\,Z_1\,Z_2-\frac{1}{2} \hat{\delta }\,Y_1\,Y_3\,Z_1\,Z_2\,Z_3 \left(-1+\mu _2\right)$\\

\cline{2-3}
& \scriptsize 2& \scriptsize  $Y_3\,Z_2\,Z_3 \left(Y_1\,Z_1+Y_2\,Z_2+Y_3\,Z_3\right)+\frac{1}{2} \hat{\delta }\,Z_1\,Z_2^2\,Z_3 \left(-3+\mu _2\right)$\\

\cline{2-3}
& \scriptsize 2& \scriptsize  $Y_1\,Y_2\,Z_1\,Z_2^2-\frac{1}{2} \hat{\delta }\,Z_1\,Z_2^2\,Z_3 \left(-1+\mu _2\right)$\\

\cline{1-3}
\scriptsize $(1\,,\,\m_2\,,\,1)$& \scriptsize 8
& \scriptsize  $Y_1^3\,Y_2^2\,Y_3^3+\frac{1}{4} \hat{\delta }^2\,Y_1\,Y_3 \left(-2+\mu _2\right) \left(-3\,Y_2^2\,Z_2^2 \left(-4+\mu _2\right)+Y_1\,Y_3\,Z_1\,Z_3 \mu _2\right)$\\
&& \scriptsize  $+\frac{1}{4} \hat{\delta }^3\,Y_3\,Z_2\,Z_3 \left(Y_1\,Z_1+2\,Y_2\,Z_2+Y_3\,Z_3\right) \left(-4+\mu _2\right) \left(-2+\mu _2\right) \mu _2$\\
&& \scriptsize  $+\frac{1}{16} \hat{\delta }^4\,Z_1\,Z_2^2\,Z_3 \left(-4+\mu _2\right) \left(-2+\mu _2\right){}^2 \mu _2$\\

\cline{2-3}
& \scriptsize 6& \scriptsize  $Y_1^2\,Y_2\,Y_3^3\,Z_3+\frac{1}{2} \hat{\delta }\,Y_1\,Y_3 \left(-2\,Y_2^2\,Z_2^2 \left(-4+\mu _2\right)+Y_1\,Y_3\,Z_1\,Z_3 \left(-2+\mu _2\right)\right)$\\
&& \scriptsize  $+\frac{1}{4} \hat{\delta }^2\,Y_3\,Z_2\,Z_3 \left(-4+\mu _2\right) \left(2\,Y_1\,Z_1 \left(-2+\mu _2\right)\right.$\\
&& \scriptsize  $\left.+2\,Y_3\,Z_3 \left(-2+\mu _2\right)+Y_2\,Z_2 \left(-2+3 \mu _2\right)\right)$\\
&& \scriptsize  $+\frac{1}{8} \hat{\delta }^3\,Z_1\,Z_2^2\,Z_3 \left(-4+\mu _2\right) \left(-2+\mu _2\right){}^2$\\

\cline{2-3}
& \scriptsize 6& \scriptsize  $Y_1^2\,Y_2^2\,Y_3^2\,Z_2+\hat{\delta }\,Y_1\,Y_2^2\,Y_3\,Z_2^2 \left(-2+\mu _2\right)$\\
&& \scriptsize  $-\frac{1}{4} \hat{\delta }^2\,Y_3\,Z_2\,Z_3 \left(2\,Y_2\,Z_2+Y_3\,Z_3\right) \left(-2+\mu _2\right) \mu _2$\\

\cline{2-3}
& \scriptsize 6& \scriptsize  $Y_1^3\,Y_2\,Y_3^2\,Z_1-\frac{1}{2} \hat{\delta }\,Y_1^2\,Y_3^2\,Z_1\,Z_3 \mu _2$\\
&& \scriptsize  $-\frac{1}{4} \hat{\delta }^2\,Y_1\,Y_2\,Z_1\,Z_2^2 \left(-6+\mu _2\right) \left(-4+\mu _2\right)+\frac{1}{8} \hat{\delta }^3\,Z_1\,Z_2^2\,Z_3 \left(-6+\mu _2\right) \left(-4+\mu _2\right) \mu _2$\\

\cline{2-3}
& \scriptsize 4& \scriptsize  $Y_1\,Y_3 \left(-Y_2^2\,Z_2^2+Y_3\,Z_3 \left(Y_1\,Z_1+Y_3\,Z_3\right)\right)$\\
&& \scriptsize  $+\hat{\delta }\,Y_3\,Z_2\,Z_3 \left(Y_1\,Z_1 \left(-4+\mu _2\right)+Y_3\,Z_3 \left(-4+\mu _2\right)+Y_2\,Z_2 \left(-2+\mu _2\right)\right)$\\
&& \scriptsize  $+\frac{1}{4} \hat{\delta }^2\,Z_1\,Z_2^2\,Z_3 \left(-4+\mu _2\right) \left(-2+\mu _2\right)$\\

\cline{2-3}
& \scriptsize 4& \scriptsize  $Y_1\,Y_2\,Y_3\,Z_2 \left(Y_2\,Z_2+Y_3\,Z_3\right)$\\
&& \scriptsize  $-\frac{1}{2} \hat{\delta }\,Y_3\,Z_2\,Z_3 \left(Y_3\,Z_3 \left(-2+\mu _2\right)+Y_2\,Z_2 \mu _2\right)$\\

\cline{2-3}
& \scriptsize 4& \scriptsize  $Y_1^2\,Y_2\,Y_3\,Z_1\,Z_2+\frac{1}{2} \hat{\delta }\,Y_1\,Z_1\,Z_2 \left(Y_2\,Z_2 \left(-4+\mu _2\right)-Y_3\,Z_3 \mu _2\right)$\\
&& \scriptsize  $-\frac{1}{4} \hat{\delta }^2\,Z_1\,Z_2^2\,Z_3 \left(-4+\mu _2\right) \mu _2$\\

\cline{1-3}
\scriptsize $(0\,,\,\m_2\,,\,1)$& \scriptsize 8
& \scriptsize  $Y_1^3\,Y_2^2\,Y_3^3+\frac{3}{2} \hat{\delta }\,Y_1^2\,Y_2^2\,Y_3^2\,Z_2 \left(-3+\mu _2\right)$\\
&& \scriptsize  $-\frac{1}{4} \hat{\delta }^2\,Y_1\,Y_3 \left(-1+\mu _2\right) \left(-2\,Y_2^2\,Z_2^2 \left(-5+\mu _2\right)\right.$\\
&& \scriptsize  $\left.+2\,Y_2\,Y_3\,Z_2\,Z_3 \left(1+\mu _2\right)+Y_3^2\,Z_3^2 \left(1+\mu _2\right)\right)$\\
&& \scriptsize  $-\frac{1}{8} \hat{\delta }^3\,Y_3\,Z_2\,Z_3 \left(Y_3\,Z_3 \left(-7+\mu _2\right)+2\,Y_2\,Z_2 \left(-5+\mu _2\right)\right) \left(-1+\mu _2^2\right)$\\

\cline{2-3}
& \scriptsize 6& \scriptsize  $Y_1^2\,Y_2\,Y_3^2 \left(Y_2\,Z_2+Y_3\,Z_3\right)$\\
&& \scriptsize  $+\frac{1}{2} \hat{\delta }\,Y_1\,Y_3 \left(-8\,Y_2\,Y_3\,Z_2\,Z_3+Y_2^2\,Z_2^2 \left(-5+\mu _2\right)-Y_3^2\,Z_3^2 \left(-1+\mu _2\right)\right)$\\
&& \scriptsize  $-\frac{1}{4} \hat{\delta }^2\,Y_3\,Z_2\,Z_3 \left(Y_3\,Z_3 \left(-7+\mu _2\right) \left(-1+\mu _2\right)+Y_2\,Z_2 \left(-5+\mu _2\right) \left(1+\mu _2\right)\right)$\\

\cline{2-3}
& \scriptsize 6& \scriptsize  $Y_1^3\,Y_2\,Y_3^2\,Z_1-\frac{1}{2} \hat{\delta }\,Y_1^2\,Y_3\,Z_1 \left(-2\,Y_2\,Z_2 \left(-5+\mu _2\right)+Y_3\,Z_3 \left(1+\mu _2\right)\right)$\\
&& \scriptsize  $+\frac{1}{4} \hat{\delta }^2\,Y_1\,Z_1\,Z_2 \left(-5+\mu _2\right) \left(Y_2\,Z_2 \left(-3+\mu _2\right)-2\,Y_3\,Z_3 \left(1+\mu _2\right)\right)$\\
&& \scriptsize  $-\frac{1}{8} \hat{\delta }^3\,Z_1\,Z_2^2\,Z_3 \left(-5+\mu _2\right) \left(-3+\mu _2\right) \left(1+\mu _2\right)$\\

\cline{1-3}
\scriptsize $(2\,,\,\m_2\,,\,0)$& \scriptsize 8
& \scriptsize  $Y_1^3\,Y_2^2\,Y_3^3+\hat{\delta }\,Y_1^3\,Y_2\,Y_3^2\,Z_1 \left(-2+\mu _2\right)+\frac{1}{4} \hat{\delta }^2\,Y_1^3\,Y_3\,Z_1^2 \left(-2+\mu _2\right) \mu _2$\\
&& \scriptsize  $+\frac{1}{8} \hat{\delta }^3\,Z_2 \left(Y_1\,Z_1+Y_3\,Z_3\right){}^2 \left(-4+\mu _2\right) \left(-2+\mu _2\right) \mu _2$\\

\cline{2-3}
& \scriptsize 6& \scriptsize  $Y_1^2\,Y_2\,Y_3^2 \left(Y_1\,Z_1+Y_3\,Z_3\right)$\\
&& \scriptsize  $+\frac{1}{2} \hat{\delta }\,Y_1^2\,Y_3\,Z_1 \left(Y_3\,Z_3 \left(-4+\mu _2\right)+Y_1\,Z_1 \left(-2+\mu _2\right)\right)$\\
&& \scriptsize  $+\frac{1}{4} \hat{\delta }^2\,Z_2 \left(Y_1\,Z_1+Y_3\,Z_3\right){}^2 \left(-4+\mu _2\right) \left(-2+\mu _2\right)$\\

\cline{2-3}
& \scriptsize 6& \scriptsize  $Y_1^2\,Y_2^2\,Y_3^2\,Z_2+\hat{\delta }\,Y_1^2\,Y_2\,Y_3\,Z_1\,Z_2 \left(-2+\mu _2\right)$\\
&& \scriptsize  $-\frac{1}{4} \hat{\delta }^2\,Y_3\,Z_2\,Z_3 \left(2\,Y_1\,Z_1+Y_3\,Z_3\right) \left(-2+\mu _2\right) \mu _2$\\

\cline{2-3}
& \scriptsize 4& \scriptsize  $Y_1\,Y_3 \left(Y_1\,Z_1+Y_3\,Z_3\right){}^2+\frac{1}{2} \hat{\delta }\,Z_2 \left(Y_1\,Z_1+Y_3\,Z_3\right){}^2 \left(-4+\mu _2\right)$\\

\cline{2-3}
& \scriptsize 4& \scriptsize  $Y_1\,Y_2\,Y_3\,Z_2 \left(Y_1\,Z_1+Y_3\,Z_3\right)-\frac{1}{2} \hat{\delta }\,Y_3\,Z_2\,Z_3 \left(Y_3\,Z_3 \left(-2+\mu _2\right)+Y_1\,Z_1 \mu _2\right)$\\

\cline{2-3}
& \scriptsize 4& \scriptsize  $Y_1\,Y_2^2\,Y_3\,Z_2^2+\frac{1}{2} \hat{\delta }\,Y_2\,Z_2^2 \left(Y_1\,Z_1 \left(-4+\mu _2\right)-Y_3\,Z_3 \mu _2\right)-\frac{1}{4} \hat{\delta }^2\,Z_1\,Z_2^2\,Z_3 \left(-4+\mu _2\right) \mu _2$\\

\cline{1-3}
\scriptsize $(1\,,\,\m_2\,,\,0)$& \scriptsize 8
& \scriptsize  $Y_1^3\,Y_2^2\,Y_3^3+\hat{\delta }\,Y_1^3\,Y_2\,Y_3^2\,Z_1 \left(-1+\mu _2\right)$\\
&& \scriptsize  $-\frac{1}{4} \hat{\delta }^2\,Y_1\,Y_3 \left(-1+\mu _2\right) \left(2\,Y_2^2\,Z_2^2 \left(-5+\mu _2\right)+Y_3\,Z_3 \left(2\,Y_1\,Z_1+Y_3\,Z_3\right) \left(1+\mu _2\right)\right)$\\
&& \scriptsize  $-\frac{1}{4} \hat{\delta }^3\,Y_2\,Z_2^2 \left(-5+\mu _2\right) \left(-1+\mu _2\right) \left(Y_1\,Z_1 \left(-3+\mu _2\right)-Y_3\,Z_3 \left(1+\mu _2\right)\right)$\\
&& \scriptsize  $+\frac{1}{8} \hat{\delta }^4\,Z_1\,Z_2^2\,Z_3 \left(-5+\mu _2\right) \left(-3+\mu _2\right) \left(-1+\mu _2\right) \left(1+\mu _2\right)$\\

\cline{2-3}
& \scriptsize 6& \scriptsize  $Y_1^2\,Y_2\,Y_3^2 \left(Y_1\,Z_1+Y_3\,Z_3\right)$\\
&& \scriptsize  $-\frac{1}{2} \hat{\delta }\,Y_1\,Y_3 \left(Y_2^2\,Z_2^2 \left(-5+\mu _2\right)+Y_3\,Z_3 \left(Y_3\,Z_3 \left(-1+\mu _2\right)+Y_1\,Z_1 \left(1+\mu _2\right)\right)\right)$\\
&& \scriptsize  $-\frac{1}{4} \hat{\delta }^2\,Y_2\,Z_2^2 \left(-5+\mu _2\right) \left(Y_1\,Z_1 \left(-3+\mu _2\right)-Y_3\,Z_3 \left(1+\mu _2\right)\right)$\\
&& \scriptsize  $+\frac{1}{8} \hat{\delta }^3\,Z_1\,Z_2^2\,Z_3 \left(-5+\mu _2\right) \left(-3+\mu _2\right) \left(1+\mu _2\right)$\\

\cline{2-3}
& \scriptsize 6& \scriptsize  $Y_1^2\,Y_2^2\,Y_3^2\,Z_2+\hat{\delta }\,Y_1\,Y_2\,Y_3\,Z_2 \left(Y_1\,Z_1+Y_2\,Z_2\right) \left(-1+\mu _2\right)$\\
&& \scriptsize  $-\frac{1}{4} \hat{\delta }^2\,Z_2 \left(-1+\mu _2\right) \left(Y_3\,Z_3 \left(2\,Y_2\,Z_2+Y_3\,Z_3\right) \left(1+\mu _2\right)\right.$\\
&& \scriptsize  $\left.+2\,Y_1\,Z_1 \left(-Y_2\,Z_2 \left(-3+\mu _2\right)+Y_3\,Z_3 \left(1+\mu _2\right)\right)\right)$\\
&& \scriptsize  $-\frac{1}{4} \hat{\delta }^3\,Z_1\,Z_2^2\,Z_3 \left(-3+\mu _2\right) \left(-1+\mu _2\right) \left(1+\mu _2\right)$\\

\cline{2-3}
& \scriptsize 4& \scriptsize  $Y_1\,Y_2\,Y_3\,Z_2 \left(Y_1\,Z_1+Y_2\,Z_2+Y_3\,Z_3\right)$\\
&& \scriptsize  $-\frac{1}{2} \hat{\delta }\,Z_2 \left(Y_3\,Z_3 \left(Y_3\,Z_3 \left(-1+\mu _2\right)+Y_2\,Z_2 \left(1+\mu _2\right)\right)\right.$\\
&& \scriptsize  $\left.+Y_1\,Z_1 \left(-Y_2\,Z_2 \left(-3+\mu _2\right)+Y_3\,Z_3 \left(1+\mu _2\right)\right)\right)$\\
&& \scriptsize  $-\frac{1}{4} \hat{\delta }^2\,Z_1\,Z_2^2\,Z_3 \left(-3+\mu _2\right) \left(1+\mu _2\right)$\\

\cline{1-3}
\scriptsize $(2\,,\,\m_2\,,\,\m_3)$& \scriptsize 8
& \scriptsize  $Y_1^3\,Y_2^3\,Y_3^2+\frac{1}{8} \hat{\delta }^3\,Y_2^2\,Z_2^2\,Z_3 \left(-4+\mu _2-\mu _3\right) \left(-2+\mu _2-\mu _3\right) \left(\mu _2-\mu _3\right)$\\

\cline{2-3}
& \scriptsize 6& \scriptsize  $Y_1^2\,Y_2^2\,Y_3^2\,Z_3-\frac{1}{4} \hat{\delta }^2\,Y_2^2\,Z_2^2\,Z_3 \left(-2+\mu _2-\mu _3\right) \left(\mu _2-\mu _3\right)$\\

\cline{2-3}
& \scriptsize 6& \scriptsize  $Y_1^2\,Y_2^3\,Y_3\,Z_2+\frac{1}{4} \hat{\delta }^2\,Y_2^2\,Z_2^2\,Z_3 \left(-4+\mu _2-\mu _3\right) \left(-2+\mu _2-\mu _3\right)$\\

\cline{2-3}
& \scriptsize 6& \scriptsize  $Y_1^3\,Y_2^2\,Y_3\,Z_1-\frac{1}{8} \hat{\delta }^3\,Z_1\,Z_2\,Z_3^2 \left(-4+\mu _2-\mu _3\right) \left(-2+\mu _2-\mu _3\right) \left(\mu _2-\mu _3\right)$\\

\cline{2-3}
& \scriptsize 4& \scriptsize  $Y_1\,Y_2\,Y_3^2\,Z_3^2+\frac{1}{2} \hat{\delta }\,Y_2^2\,Z_2^2\,Z_3 \left(\mu _2-\mu _3\right)$\\

\cline{2-3}
& \scriptsize 4& \scriptsize  $Y_1\,Y_2^2\,Y_3\,Z_2\,Z_3+\frac{1}{2} \hat{\delta }\,Y_2^2\,Z_2^2\,Z_3 \left(2-\mu _2+\mu _3\right)$\\

\cline{2-3}
& \scriptsize 4& \scriptsize  $Y_1\,Y_2^3\,Z_2^2+\frac{1}{2} \hat{\delta }\,Y_2^2\,Z_2^2\,Z_3 \left(-4+\mu _2-\mu _3\right)$\\

\cline{2-3}
& \scriptsize 4& \scriptsize  $Y_1^2\,Y_2\,Y_3\,Z_1\,Z_3+\frac{1}{4} \hat{\delta }^2\,Z_1\,Z_2\,Z_3^2 \left(-2+\mu _2-\mu _3\right) \left(\mu _2-\mu _3\right)$\\

\cline{2-3}
& \scriptsize 4& \scriptsize  $Y_1^2\,Y_2^2\,Z_1\,Z_2-\frac{1}{4} \hat{\delta }^2\,Z_1\,Z_2\,Z_3^2 \left(-4+\mu _2-\mu _3\right) \left(-2+\mu _2-\mu _3\right)$\\

\cline{2-3}
& \scriptsize 4& \scriptsize  $Y_1^3\,Y_2\,Z_1^2+\frac{1}{2} \hat{\delta }\,Y_1^2\,Z_1^2\,Z_3 \left(-4+\mu _2-\mu _3\right)$\\

\cline{2-3}
& \scriptsize 2& \scriptsize  $-Y_2^2\,Z_2^2\,Z_3+Y_3^2\,Z_3^3$\\

\cline{2-3}
& \scriptsize 2& \scriptsize  $Y_2\,Z_2\,Z_3 \left(Y_2\,Z_2+Y_3\,Z_3\right)$\\

\cline{2-3}
& \scriptsize 2& \scriptsize  $Y_1\,Y_3\,Z_1\,Z_3^2+\frac{1}{2} \hat{\delta }\,Z_1\,Z_2\,Z_3^2 \left(-\mu _2+\mu _3\right)$\\

\cline{2-3}
& \scriptsize 2& \scriptsize  $Y_1\,Y_2\,Z_1\,Z_2\,Z_3+\frac{1}{2} \hat{\delta }\,Z_1\,Z_2\,Z_3^2 \left(-2+\mu _2-\mu _3\right)$\\

\cline{1-3}
\scriptsize $(1\,,\,\m_2\,,\,\m_3)$& \scriptsize 8
& \scriptsize  $Y_1^3\,Y_2^3\,Y_3^2-\frac{3}{4} \hat{\delta }^2\,Y_1\,Y_2^3\,Z_2^2 \left(-3+\mu _2-\mu _3\right) \left(-1+\mu _2-\mu _3\right)$\\
&& \scriptsize  $-\frac{1}{4} \hat{\delta }^3\,Y_2\,Z_2\,Z_3 \left(2\,Y_2\,Z_2+Y_3\,Z_3\right) \left(-5+\mu _2-\mu _3\right) \left(-3+\mu _2-\mu _3\right) \left(-1+\mu _2-\mu _3\right)$\\

\cline{2-3}
& \scriptsize 6& \scriptsize  $Y_1^2\,Y_2^2\,Y_3^2\,Z_3+\hat{\delta }\,Y_1\,Y_2^3\,Z_2^2 \left(-1+\mu _2-\mu _3\right)$\\
&& \scriptsize  $+\frac{1}{4} \hat{\delta }^2\,Y_2\,Z_2\,Z_3 \left(Y_2\,Z_2 \left(-13+3 \mu _2-3 \mu _3\right)+2\,Y_3\,Z_3 \left(-3+\mu _2-\mu _3\right)\right) \left(-1+\mu _2-\mu _3\right)$\\

\cline{2-3}
& \scriptsize 6& \scriptsize  $Y_1^2\,Y_2^3\,Y_3\,Z_2+\hat{\delta }\,Y_1\,Y_2^3\,Z_2^2 \left(3-\mu _2+\mu _3\right)$\\
&& \scriptsize  $-\frac{1}{4} \hat{\delta }^2\,Y_2\,Z_2\,Z_3 \left(2\,Y_2\,Z_2+Y_3\,Z_3\right) \left(-5+\mu _2-\mu _3\right) \left(-3+\mu _2-\mu _3\right)$\\

\cline{2-3}
& \scriptsize 6& \scriptsize  $Y_1^3\,Y_2^2\,Y_3\,Z_1+\frac{1}{4} \hat{\delta }^2\,Y_1\,Z_1\,Z_3 \left(-3+\mu _2-\mu _3\right) \left(2\,Y_2\,Z_2 \left(1+\mu _2-\mu _3\right)\right.$\\
&& \scriptsize  $\left.+Y_3\,Z_3 \left(5-\mu _2+\mu _3\right)\right)+\frac{1}{4} \hat{\delta }^3\,Z_1\,Z_2\,Z_3^2 \left(1+\mu _2-\mu _3\right) \left(3-\mu _2+\mu _3\right){}^2$\\

\cline{2-3}
& \scriptsize 4& \scriptsize  $Y_1\,Y_2 \left(-Y_2^2\,Z_2^2+Y_3^2\,Z_3^2\right)+\hat{\delta }\,Y_2\,Z_2\,Z_3 \left(Y_3\,Z_3 \left(1-\mu _2+\mu _3\right)+Y_2\,Z_2 \left(3-\mu _2+\mu _3\right)\right)$\\

\cline{1-3}
\scriptsize $(\m_1\,,\,\m_2\,,\,1)$& \scriptsize 8
& \scriptsize  $Y_1^2\,Y_2^3\,Y_3^3-\frac{1}{4} \hat{\delta }^2\,Y_2^3\,Y_3\,Z_2^2 \left(-1+\left(\mu _1-\mu _2\right){}^2\right)$\\

\cline{2-3}
& \scriptsize 6& \scriptsize  $Y_1\,Y_2^2\,Y_3^3\,Z_3+\frac{1}{2} \hat{\delta }\,Y_2^3\,Y_3\,Z_2^2 \left(1+\mu _1-\mu _2\right)$\\

\cline{2-3}
& \scriptsize 6& \scriptsize  $Y_1\,Y_2^3\,Y_3^2\,Z_2+\frac{1}{2} \hat{\delta }\,Y_2^3\,Y_3\,Z_2^2 \left(1-\mu _1+\mu _2\right)$\\

\cline{2-3}
& \scriptsize 6& \scriptsize  $Y_1^2\,Y_2^2\,Y_3^2\,Z_1-\frac{1}{4} \hat{\delta }^2\,Y_2^2\,Z_1\,Z_2^2 \left(-1+\left(\mu _1-\mu _2\right){}^2\right)$\\

\cline{2-3}
& \scriptsize 4& \scriptsize  $-Y_2^3\,Y_3\,Z_2^2+Y_2\,Y_3^3\,Z_3^2$\\

\cline{2-3}
& \scriptsize 4& \scriptsize  $Y_2^2\,Y_3\,Z_2 \left(Y_2\,Z_2+Y_3\,Z_3\right)$\\

\cline{2-3}
& \scriptsize 4& \scriptsize  $Y_1\,Y_2\,Y_3^2\,Z_1\,Z_3+\frac{1}{2} \hat{\delta }\,Y_2^2\,Z_1\,Z_2^2 \left(1+\mu _1-\mu _2\right)$\\

\cline{2-3}
& \scriptsize 4& \scriptsize  $Y_1\,Y_2^2\,Y_3\,Z_1\,Z_2+\frac{1}{2} \hat{\delta }\,Y_2^2\,Z_1\,Z_2^2 \left(1-\mu _1+\mu _2\right)$\\

\cline{2-3}
& \scriptsize 4& \scriptsize  $Y_1^2\,Y_2\,Y_3\,Z_1^2+\frac{1}{4} \hat{\delta }^2\,Z_1^2\,Z_2\,Z_3 \left(-1+\left(\mu _1-\mu _2\right){}^2\right)$\\

\cline{2-3}
& \scriptsize 2& \scriptsize  $Z_1 \left(-Y_2^2\,Z_2^2+Y_3^2\,Z_3^2\right)$\\

\cline{2-3}
& \scriptsize 2& \scriptsize  $Y_2\,Z_1\,Z_2 \left(Y_2\,Z_2+Y_3\,Z_3\right)$\\

\cline{2-3}
& \scriptsize 2& \scriptsize  $Y_1\,Y_3\,Z_1^2\,Z_3-\frac{1}{2} \hat{\delta }\,Z_1^2\,Z_2\,Z_3 \left(1+\mu _1-\mu _2\right)$\\

\cline{2-3}
& \scriptsize 2& \scriptsize  $Y_1\,Y_2\,Z_1^2\,Z_2+\frac{1}{2} \hat{\delta }\,Z_1^2\,Z_2\,Z_3 \left(-1+\mu _1-\mu _2\right)$\\

\hline
\end{longtable}}
\end{center}

\section{Highest-derivative partially-massless interactions}
\label{sec: hd pm}

In this appendix we prove that the function \eqref{pmf hd} generates consistent highest-derivative interactions involving
 partially-massless fields provided the condition \eqref{mec} holds.
Our starting point is the St\"uckelberg version of
\eqref{pmf hd} given in terms of the shifted variables $\bm{\tilde Y_{i}}$ of eq.~\eqref{deformation}.
Consistency of the partially-massless interactions is tantamount to the cancellation of
 the residues of the partially-massless poles.
 To this end, we need in principle to integrate by parts all the total-derivative terms contained in the shifted variables.
 However, one can simplify the computations considering the following ansatz:
\be
\mathcal K(\bm{\tilde Y_{1}},\bm{\tilde Y_{2}},\bm{\tilde Y_{3}})
=\mathcal K(\tilde{Y}_1-\hat \delta\,\partial_{w_1}\,\partial_{\eta_1}\,,\,
\tilde{Y}_2-\hat \delta\,\partial_{w_2}\,\partial_{\eta_2}\,,\,
\tilde{Y}_3-\hat \delta\,\partial_{w_3}\,\partial_{\eta_3})\, f(\eta_1,\eta_2,\eta_3)\,,
\ee
where the results of the integrations by parts
are encoded in  the function $f$\,.
At this point, what is left is to impose the gauge invariance of the ansatz.
Taking the $\a_i$'s in the $\tilde Y_i$'s to satisfy \eqref{a b cond},
one ends up with the following differential equation for the function $f$\,:
\ba
&&\Big\{\,\big[\,1-2\,(2\,\a_1+1)\,\eta_1+4\,\a_1\,(\a_1+1)\,\eta_1^2\,\big]\,\partial_{\eta_1}
+\m_1\,\big[2\,\a_{1}+1-4\,\a_1\,(\a_1+1)\,\eta_1\,\big] \nn
&&\quad-\,2\,\big[\,\eta_2-2\,(\a_2+1)\,\eta_2^2\,\big]\,\partial_{\eta_2}
+\m_2\,\big[1-4\,(\a_2+1)\, \eta_2\,\big] \nn
&&\quad+\,2\,\big[\,\eta_3-2\,\a_3\,\eta_3^2\,\big]\,\partial_{\eta_3}
-\m_3\,\big[\,1-4\,\a_{3}\, \eta_3\,\big]\,\Big\}\,f(\eta_{1},\eta_{2},\eta_{3})=0\,,
\ea
and cyclic permutations thereof. The solution of the latter differential equations is
\ba
&& f(\eta_1,\eta_2,\eta_3)\ =\
\big[\,1-2\,(\a_1+1)\,\eta_1-2\,\a_2\,\eta_2\, \big]^{\frac{1}{2}(\m_1+\m_2-\m_3)}\nn
&& \hspace{68pt} \times\,
\big[\,1-2\,(\a_2+1)\,\eta_2-2\,\a_3\,\eta_3\, \big]^{\frac{1}{2}(\m_2+\m_3-\m_1)}\nn
&&  \hspace{68pt} \times\,
\big[\,1-2\,(\a_3+1)\,\eta_3-2\,\a_1\,\eta_1\, \big]^{\frac{1}{2}(\m_3+\m_1-\m_2)}\,,
\ea
whose Taylor coefficients at the order $\eta_i^{\,r_{i}+1}$
correspond to the residues of the poles $\mu_i=r_i$ associated with $W_i^{\,r_{i}+1}$\,.
Concentrating on the gauge consistency with respect to the $i$-th field,
one can set $\eta_{i\pm 1}=0$ so that the function $f$ becomes the generating function
of the Jacobi polynomials. One can then extract the residues as
\be
	\big(\partial_{\eta_{i}}^{\,r_{i}+1}\,f\big)(0,0,0)=(-2)^{r_i+1}\,\binom{\tfrac{1}{2} (r_i+\m_{i+1}-\m_{i-1})}{r_i+1}\,,
	\label{cond pmf}
\ee
where the homogeneity of the $i$-th field is a positive integer $r_i$\,.
The highest-derivative interactions \eqref{pmf hd} become consistent
whenever \eqref{cond pmf} vanishes.
The latter requirement is equivalent to \eqref{mec}.

\section{Massless limit in flat space}
\label{sec: mless limit}

This appendix includes further details about the massless limit in flat space considered in Section \ref{sec: massless}. It is important to notice that, in the generic analysis which led to eqs~(\ref{dom y h}\,,\,\ref{tilde g limit}), $g$ appears only through $\hat{G}$.
However, in general, it can also appear whenever the leading terms cancel identically.
More precisely, if the first $n$ leading terms cancel then the $(n+1)$-th term becomes dominant and contains $n$-th powers of $g$\,.
Therefore, for the sake of completeness, we consider all the cases in which the generic dominant terms
cancel among each other.
These situations can be systematically analyzed by focusing on the particular combinations of variables giving rise to the desired cancellations.
\begin{enumerate}[i.]

\item In the 2 massless and 1 massive case, the following combination
\be
	\mu^{-2}\left(\hat H_{1}\,\hat H_{2}-\hat Y_{3}^{\,2}\,\hat H_{3}\right)
	=-\tfrac{1}{2}\,\nu_{3}^{2}\left(
	g\,y_{3} + y_{1}\,y_{2}\,\partial_{v_{3}}^{\,2}\,
	\right) + \cO(\mu)\,,
\ee
gives rise to the cancellation of the terms proportional to $\mu^{-2}\,y^{4}$\,.

\item In the 1 massless and 2 equal massive case, there is no combination leading to the cancellation
and $g$ can show up only through $\hat G$ \eqref{tilde g limit}.

\item In the 1 massless and 2 different massive case, the following combination
\ba
	&& \mu^{-2}\left[\hat Y_{2}\,\hat Y_{3}\,\hat H_{2}
	-\hat Y_{3}^{\,2}\,\hat H_{3}
	+\tfrac{1}{2}\left(M_{2}^{2}-M_{3}^{2}\right)
	\hat Z_{1}\,\hat H_{2}\right] \nn
	&&=\tfrac{1}{2}\left(\nu_{2}^{2}-\nu_{3}^{2}\right)
	\left [g\,y_{3}
	+\tfrac{\nu_{2}}{\nu_3}\,
	y_{1}\,y_{3}\,\partial_{v_{3}}\,\partial_{v_{2}}
	-\tfrac{1}{2}\,\tfrac{\nu_{2}^2-\nu_3^2}{\nu_3^2}\,
	y_{1}\,y_{2}\,\partial_{v_{3}}^{\,2}
	\right]+\cO(\mu)\,, \qquad\quad
\ea
or, equivalently
\ba
	&& \mu^{-2}\left[\hat Y_{2}\,\hat Y_{3}\,\hat H_{3}
	-\hat Y_{2}^{\,2}\,\hat H_{2}
	+\tfrac{1}{2}\left(M_{3}^{2}-M_{2}^{2}\right)
	\hat Z_{1}\,\hat H_{3}\right] \nn
	&&=\tfrac{1}{2}\left(\nu_{3}^{2}-\nu_{2}^{2}\right)
	\left [g\,y_{2}
	+\tfrac{\nu_{3}}{\nu_2}\,
	y_{1}\,y_{2}\,\partial_{v_{3}}\,\partial_{v_{2}}
	-\tfrac{1}{2}\,\tfrac{\nu_{3}^2-\nu_2^2}{\nu_2^2}\,
	y_{1}\,y_{3}\,\partial_{v_{2}}^{\,2}
	\right]+\cO(\mu)\,, \qquad\quad
\ea
allow the cancellation of the dominant term proportional to $\mu^{-2}\,y^{4}$\,.

\item In the 3 massive case, the following combination
\ba
	&&\hat Y_{1}\,\hat Z_{1}+\hat Y_{2}\,\hat Z_{2}
	+\hat Y_{3}\,\hat Z_{3}\nn
	&&=g
	 +\tfrac{\nu_{2}^{2}+\nu_{3}^{2}-\nu_{1}^{2}}{2\,\nu_{2}\,\nu_{3}}\,y_{1}\,\partial_{v_{2}}\,\partial_{v_{3}}
	 +\tfrac{\nu_{3}^{2}+\nu_{1}^{2}-\nu_{2}^{2}}{2\,\nu_{3}\,\nu_{1}}\,y_{2}\,\partial_{v_{3}}\,\partial_{v_{1}}
	 +\tfrac{\nu_{1}^{2}+\nu_{2}^{2}-\nu_{3}^{2}}{2\,\nu_{1}\,\nu_{2}}\,y_{3}\,\partial_{v_{1}}\,\partial_{v_{2}}
	+\cO(\mu)\,,
	\qquad\quad
\ea
does not contain the dominant term proportional to $\mu^{-1}\,y^{2}\,\partial_{v}$\,.

\end{enumerate}
Notice that all resulting massless vertices that involve $g$
are decorated with the contributions of the St\"uckelberg fields.

\section{Cubic interactions of open strings in the first Regge Trajectory}
\label{sec: ST coup}

In this appendix we show that the string interactions encoded by \eqref{StringVer} nicely fit in with the classification we have provided in Section \ref{sec:Consistent cubic}.
For simplicity, let us drop Chan-Paton factors as well as the constant
$i\sqrt{G_{N}}\,g_0/\alpha'$, and focus on the first term in \eqref{StringVer}. The latter can be expanded as
\be
\label{StringVerEx}
	\cK=
	\sum_{\sigma_{i},\tau_{i}}\,
	\frac{1}{\sigma_1!\,\sigma_2!\,\sigma_3!\,\tau_1!\,\tau_2!\,\tau_3!}\,
	\left(-2\alpha'\right)^{\frac{\sigma_1+\sigma_2+\sigma_3}2}
	y_{1}^{\,\sigma_{1}}\,y_{2}^{\,\sigma_{2}}\,y_{3}^{\,\sigma_3}\,
	z_{1}^{\,\tau_{1}}\,z_{2}^{\,\tau_{2}}\,z_{3}^{\,\tau_3}\,,
\ee
where the spins of the fields are
\be
 s_{1}=\sigma_{1}+\tau_{2}+\tau_3\,,\qquad
  s_{2}=\sigma_{2}+\tau_{1}+\tau_3\,,\qquad
s_{3}=\sigma_{3}+\tau_{1}+\tau_2\,.
\ee
Concentrating on particular choices of $(s_{1},s_{2},s_{3})$\,,
we can extract consistent couplings for each of the five different categories. In particular, defining the $d$-dimensional counterpart $h_i$ of $H_i$ as
\be
h_i=y_{i-1}\,y_{i+1}+\tfrac{1}{2}\,[M_i^2-(M_{i-1}+M_{i+1})^2]\,z_i\,,
\ee
one ends up with the following five cases:
\begin{enumerate}[i.]
\item
In the 3 massless case with $(s_1,s_2,s_3)=(1,1,1)$\,, one gets
\be
	\cK=\left(-2\alpha'\right)^{\frac32}
	\left(\,y_1\,y_2\,y_3-\tfrac1{2\alpha'}\,g\,\right).
\ee
\item
In the 2 massless and 1 massive case with $(s_1,s_2,s_3)=(1,1,s)$\,, one has
\be
	\cK=-\tfrac{1}{(s-1)\,s!}
	\left(-2\alpha'\right)^{\frac{s+2}2}\,
	y_3^{s-2} \left(y_3^{\,2}\,h_3-s\,h_1\,h_2\right).
\ee
\item
In the 1 massless and 2 equal massive case with $(s_1,s_2,s_3)=(1,s,s)$\,, one finds
\be
\label{1ss st}
	\cK=\sum_{k=0}^s\,
	\tfrac{1}{k!\,[(s-k)!]^2} \left(-2\alpha'\right)^{\frac{2s+1}2-k}\,
	y_2^{\,s-k-1}\,y_3^{\,s-k-1}\,z_1^{\,k}
	\left(\, y_1\,y_{2}\,y_{3} - \tfrac{(s-k)}{2\alpha'}\,(g-y_{1}\,z_{1})\,\right).
\ee
\item
In the 1 massless and 2 different massive case with $(s_1,s_2,s_3)=(1,s,s')$ with $s<s'$\,, one gets
\be
	\cK=\sum_{k=0}^{s}\tfrac{1}{k!\,(s'-k)!\,(s-k)!}
	\left(-2\alpha'\right)^{\frac{s'+s+1}2-k}\, y_2^{s-k-1}y_3^{s'-k-1}z_1^k\,
	\left(\tfrac{s'-k}{s'-s}\,y_2\,h_2+\tfrac{s-k}{s-s'}\,y_3\,h_3\right).
\ee
\item
Finally, the 3 massive case trivially fits in with the classification.
\end{enumerate}

\bibliographystyle{JHEP}
\bibliography{ref_massive}

\end{document}